\documentclass[1p, preprint]{elsarticle}
\usepackage[utf8x]{inputenc}
\usepackage[english]{babel}
\usepackage{amssymb}
\usepackage{amsmath}
\usepackage{multirow}
\usepackage{graphicx}
\usepackage{url}
\graphicspath{{figures/}}

\renewcommand{\d}{\textrm{d}}

\title{Power laws and Self-Organized Criticality \\ in Theory and Nature}
\author[add1,add2,add3]{ Dimitrije Markovi\'c}
\ead{markovic@cbs.mpg.de}
\author[add1]{Claudius Gros}
\ead{gros@itp.uni-frankfurt.de}

\address[add1]{Institute for Theoretical Physics, Goethe University Frankfurt, Germany}
\address[add2]{Max Planck Institute for Human Cognitive and Brain Sciences, Leipzig, Germany}
\address[add3]{Biomagnetic Center, Hans Berger Clinic for Neurology, University Hospital Jena, Jena, Germany}

\begin{document}
\begin{abstract}
Power laws and distributions with heavy tails 
are common features of many complex systems.
Examples are the distribution of earthquake magnitudes, 
solar flare intensities and the sizes of neuronal avalanches.
Previously, researchers surmised that a single general concept 
may act as an underlying generative mechanism, with
the theory of self organized criticality being a
weighty contender.

The power-law scaling observed in the primary statistical 
analysis is an important, but by far not the only feature 
characterizing experimental data. The scaling function,
the distribution of energy fluctuations, the distribution
of inter-event waiting times, and other higher order spatial
and temporal correlations, have seen increased consideration
over the last years. Leading to realization that basic 
models, like the original sandpile model, are often insufficient 
to adequately describe the complexity of real-world systems with 
power-law distribution.

Consequently, a substantial amount of effort has gone into 
developing new and extended models and, hitherto, three classes 
of models have emerged. The first line of models is based on a separation 
between the time scales of an external drive and a an internal 
dissipation, and 
includes the original sandpile model and its extensions,
like the dissipative earthquake model. Within this approach
the steady state is close to criticality in terms of
an absorbing phase transition. The second line of models 
is based on external drives and internal dynamics competing 
on similar time scales and includes the coherent noise model, 
which has a non-critical steady state characterized by 
heavy-tailed distributions. The third line of models
proposes a non-critical self-organizing state, being 
guided by an optimization principle, such as the concept of 
highly optimized tolerance. 

We present a comparative overview regarding distinct
modeling approaches together with a discussion of their
potential relevance as underlying generative models for 
real-world phenomena. The complexity of physical and 
biological scaling phenomena has been found to transcend 
the explanatory power of individual paradigmal concepts. 
The interaction between theoretical development and 
experimental observations has been very fruitful, 
leading to a series of novel concepts and insights.

\end{abstract}
\maketitle
\pagenumbering{roman}
\tableofcontents
%\listoffigures
%\listoftables

\pagenumbering{arabic}

\section{Introduction}
\label{sec:crit_introduction}

Experimental and technological advancements, like the steady 
increase in computing power, makes the study of natural and 
man-made complex systems progressively popular and conceptually 
rewarding. Typically, a complex system 
contains a large number of various, potentially non-identical 
components, which often have an internal complex structure of 
their own. Complex systems may exhibit novel and emergent dynamics 
arising from local and nonlinear interactions of the constituting
elements. A prominent example for an emergent property, and possibly
the phenomenon observed most frequently in real-world complex systems, 
is the heavy-tailed scaling behavior of variables describing a 
structural feature or a dynamical characteristic of the system. 
An observable is considered to be heavy-tailed if the probability of 
observing extremely large values is more likely than it would be 
for an exponentially distributed variable \citep{feldman1998practical}. 

Heavy-tailed scaling has been observed in a large variety 
of real-world phenomena, such as the distribution of earthquake
magnitudes \citep{pisarenko2003characterization}, solar flare 
intensities \citep{dennis1985solar}, the sizes of wildfires 
\citep{newman2005power}, the sizes of neuronal avalanches 
\citep{klaus2011statistical}, wealth distribution 
\citep{levy1997new}, city population distribution \citep{newman2005power}, 
the distribution of computer file sizes \citep{douceur1999large, 
gros2012neuropsychological}, and various other examples 
\citep{bak1997nature, jensen1998self, newman2005power, newman1996self, 
clauset2009power, broder2000graph, adamic2000power}.

\begin{figure}[t]
\centerline{
\includegraphics[width=0.8\textwidth]{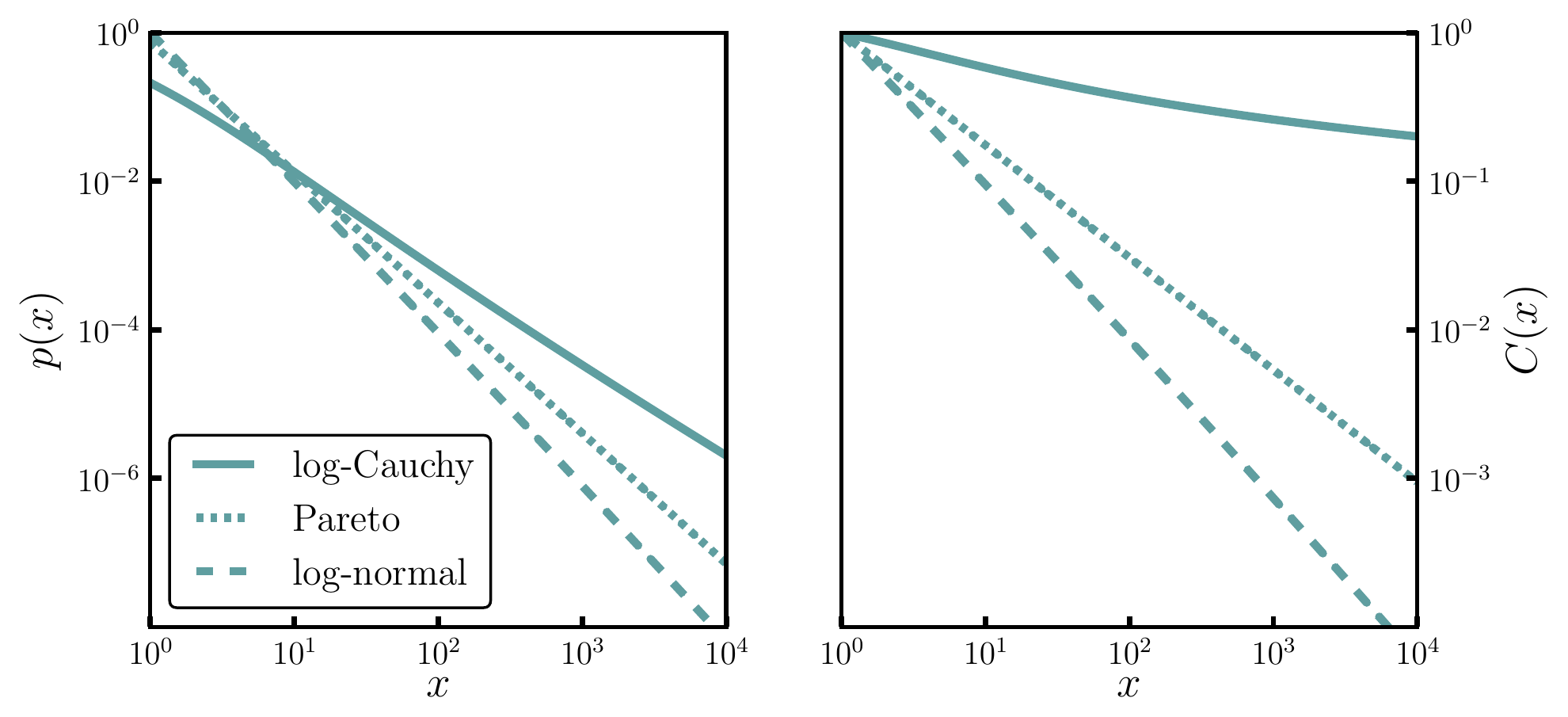}
}
 \caption{Comparison of different types of heavy-tailed distributions.
Log--Cauchy ($\sigma = 3, \mu = 0$), Log--normal distribution ($\sigma = 10, \mu = -100$) 
and Pareto distribution ($a = 1.75$, see Table \ref{tbl_heavy_tails}).
Left: The probability distribution function $p(x)$. 
Right: The corresponding complementary cumulative 
probability distribution $C(x)=\int_x^\infty p(x')dx'$. The 
distributions $p(x)$ were normalized on the range $x \in [1,\infty)$.} 
\label{fig_heavytail}
\end{figure}

Notably there are many types of distributions considered to be 
heavy-tailed, such as the L\'evy distribution, the Cauchy distribution, and 
the Weibull distribution. Still, investigations often focus
on heavy-tailed scaling in its simplest form, the form of a pure 
power law ({\it viz} the Pareto distribution). In fact, it is difficult 
to differentiate between various functional types of heavy tails
on a finite interval, especially if the data have a large variance 
and if the sample size is relatively small. In 
Fig.~\ref{fig_heavytail} we illustrate the behavior of three 
distribution functions characterized by heavy tails, the Pareto, 
the log--normal and the log--Cauchy probability 
distributions $p(x)$ (left panel), and their corresponding 
complementary cumulative probability distributions (CCDF) 
$C(x)=\int_x^{\infty}p(x')\d x'$ (right panel). The respective 
functional forms are given in Table \ref{tbl_heavy_tails}.
In spite of having more complex scaling properties, log--normal and 
log--Cauchy distributions can be approximated on a finite interval 
by a power law, that is by a straight line on a log--log plot. Note 
that the difference between log--Cauchy and Pareto distribution is 
more evident when $C(x)$ is compared.

%%%%%%%%%%%%%%%%%%%%%%%%%%%%%%%%%%%%%%%%%%%%%%%%%%%%%%%%%%%%%%%%%%%
%%%%%%%%%%%%%%%%%%%%%%%%%%%%%%%%%%%%%%%%%%%%%%%%%%%%%%%%%%%%%%%%%%%   

\citet{clauset2009power} have argued, that statistical 
methods traditionally used for data analysis ({\it e.g.}\ least-square 
fits) often misestimate the parameters describing heavy-tailed data 
sets, and consequently the actual scaling behavior. For a more 
reliable investigation of the scaling behavior one should employ methods
going beyond visually fitting data sets with power laws, such as maximum 
likelihood estimates and cross-model validation techniques. Additionally,
one should take into account the fact that most empirical data need to
be binned \citep{virkar2012power}, a procedure that reduces the 
available data resolution.

Large data sets, spanning several orders of magnitudes, are needed 
to single out the model which best fits the data and reproduces the 
heavy tail; even when advanced statistical techniques are applied. 
The collection of significantly larger data sets is however often 
difficult to achieve through experimental studies of large-scale complex 
systems, which often deal with slowly changing phenomena 
in noisy environments. Using rigorous statistical methods,
\citet{clauset2009power} re-analyzed data sets for which a least-square fit did indicate 
power-law scaling. They found that in some cases the empirical data 
actually exhibit exponential or log--normal scaling, whereas in other 
cases a power law, or a power law with an exponential cutoff, remains a
viable description---as none of the alternative distributions could be
singled out with statistical significance. Thus, in the absence of 
additional evidence, it is best to assume the simplest scaling 
of the observed phenomena, adequately described with the Pareto distribution. 

Over the past decades various models have been developed in order 
to explain the abundance of power-law scaling found in complex systems. 
Some of these power-law generating models were developed for describing 
specific systems, and have hence only a restricted applicability. 
Other models, however, aim to explain universal properties of a range of 
complex systems. They have enjoyed significant success and contributed 
to the development of the paradigm that power laws emerge naturally in 
real-world and man-made complex systems.

%%%%%%%%%%%%%%%%%%%%%%%%%%
\begin{table}[b]
\centering
\begin{tabular}{l|ll}
 name  & $\displaystyle p(x)$ & $\displaystyle C(x)$ \\ \hline
 Pareto  & $\displaystyle x^{-\alpha}$ & $\displaystyle x^{-\alpha+1}$ \\
 Log--normal & $\displaystyle \frac{1}{x}e^{-\frac{(\ln(x) - \mu)^2}{2\sigma^2}}$ & 
 $\displaystyle\frac{1}{2}\textrm{erfc}\left(\frac{\ln(x) - \mu}{2\sigma^2} \right)$ \\ 
 Log--Cauchy & $\displaystyle\frac{1}{x\left(1+\left(\frac{\ln x - \mu}{\sigma}\right)^2\right)}$ & 
 $\displaystyle\frac{1}{\pi}\textrm{arccot}\left(\frac{\ln(x)-\mu}{\sigma}\right)$ \\
 \hline
\end{tabular}
\caption{ \label{tbl_heavy_tails}
Functional form of the Pareto, Log--normal and Log--Cauchy 
distribution $p(x)$ and the corresponding 
complementary cumulative distribution, 
$C(x) = \int_x^\infty p(x')dx'$.}
\end{table}
%%%%%%%%%%%%%%%%%%%%%%%%%%

The seminal work of \citet{bak1987criticality} developed into 
an influential theory which unifies the origins of the 
power-law behavior observed in different complex systems---the 
so called theory of self-organized criticality ({\it SOC}). 
An important role for the success of {\it SOC} is the connection 
to the well-established theory of second order phase transitions 
in equilibrium statistical mechanics, for which the origin of scale-free 
behavior is well understood. The basic idea of {\it SOC} 
is that a complex system will spontaneously organize, under 
quite general conditions, into a state which is at the 
transition between two different regimes, that is at a 
critical point, without the need for external intervention or tuning. 
At such spontaneously maintained phase transition a model {\it SOC} system 
exhibits power-law scaling of event sizes, event durations and, in 
some cases, the $1/f$ scaling of the power spectra. These properties 
were also observed, to a certain extent, in natural phenomena such as 
earthquakes, solar flares, forest fires, and, more recently, 
neuronal avalanches. 

\begin{figure}[t]
 \centerline{
\includegraphics[width=0.8\textwidth ]{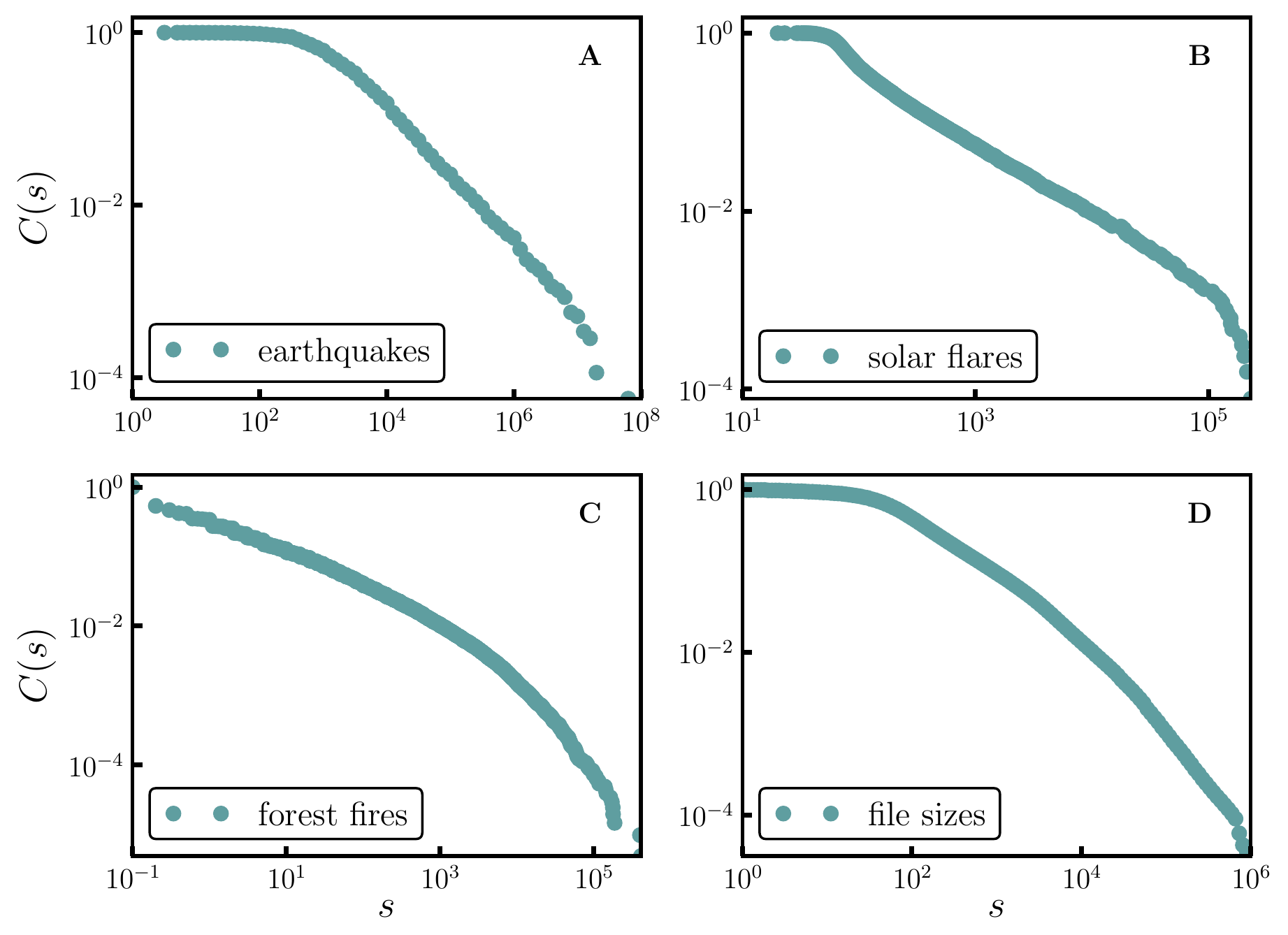}
}
\caption{
Log-log plots of the CCDF $C(s)$---a probability of observing an 
event equal to or larger than $s$---from the following empirical 
data sets: 
(A) the intensities of earthquakes occurring in 
California between 1910 and 1992, (B) peak gamma-ray intensity 
of solar flares between 1980 and 1989, 
(C) the sizes in acres of wildfires occurring on US federal land 
between 1986 and 1996 (data provided on-line by 
\citet{clauset2009power}), (D) the sizes in Kbytes of publicly 
available files on the Internet \citep{gros2012neuropsychological}.}
\label{fig_empirical}
\end{figure}

In the following chapters we will discuss in more detail the pros and cons 
of the {\it SOC} theory and its application to real-world 
phenomena. In Figure \ref{fig_empirical} we show the CCDF of some of 
the empirical data sets analyzed in \citep{clauset2009power}. Note, 
that none of the shown quantities exhibit power-law-like scaling 
across the entire range of observations.   

%%%%%%%%%%%%%%%%%%%%%%%%%%%%%%%%%%%%%%%%%%%%%%%%%%%%%%%%%%%%%
%%%%%%%%%%%%%%%%%%%%%%%%%%%%%%%%%%%%%%%%%%%%%%%%%%%%%%%%%%%%%%%%%% 

{\it SOC} is observed in a range of theoretical models.
However, several additional features characterize real-world 
complex systems and these features are mostly not captured 
by the standard modeling approach within the {\it SOC} framework. 
For example, power-law scaling 
in heterogeneous or noisy environments, or complex dynamics 
with dissipative components \citep{janosi1993self}, are 
common features of real-world systems.
As an alternative to {\it SOC}, \citet{carlson99a} proposed a 
mechanism called highly optimized tolerance ({\it HOT}) and 
argued that power-law distributions can manifest themselves
in systems with heterogeneous structures, as a consequence 
of being designed to operate optimally in uncertain environments; 
either by human design in the case of man-made systems, or by 
natural selection in the case of living organisms. The {\it HOT} 
mechanism does not require critical dynamics for the emergence 
of heavy-tailed scaling. 

In the following chapters we will describe in more details the
main concepts of {\it SOC} and {\it HOT}, together with several
other proposals for power-law generating mechanisms, and we will discuss 
their successes and limitations in predicting and explaining 
the dynamical behavior and the structure of real-world complex 
systems. In this context we will provide an assessment, 
in comparison with theory predictions, of reported statistical 
properties of the empirical time series of earthquake magnitudes, 
solar flares intensities and sizes of neuronal avalanches.
In addition we will discuss the theory of branching 
processes and the application of critical branching to the 
characterization of the dynamical regime of physical systems. 
Another important question---that we will address and discuss within 
the framework of vertex routing models---is to which extent critical 
dynamical systems actually show power-law scaling and how the process 
of experimentally observing a critical system influences the scaling 
of the collected data.

\section{Theory of Self-Organized Criticality}
\label{sec:soc}

In their seminal work \citet{bak1987criticality} provided
one of the first principles unifying the origins of 
the power law behavior observed in many natural systems. 
The core hypotheses was that systems consisting of 
many interacting components will, under certain conditions, 
spontaneously organize into a state with properties akin 
to the ones observed in a equilibrium thermodynamic 
system near a second-order phase transition. As this 
complex behavior arises spontaneously without the need for 
external tuning this phenomena was named {\it Self-organized Criticality} ({\it SOC}). 

The highly appealing feature of the {\it SOC}
theory is its relation to the well established field 
of the phase transitions and the notion of universality. 
The universality hypothesis \citep{kadanoff1990scaling} 
groups critical phenomena, as observed for many different 
physical phase transitions, into a small number of 
universality classes. Systems belonging to the same 
universality class share the values of critical exponents 
and follow equivalent scaling functions \citep{stanley1999scaling}. 
This universal behavior near a critical point is caused by 
a diverging correlation length. The correlation length becomes 
much larger than the range of the microscopic interactions, 
thus the collective behavior of the system and its components 
becomes independent of its microscopic details. This also implies 
that even the simplest model captures all the aspects of critical 
behavior of the corresponding universality class. 

Physical systems which are believed to exhibit {\it SOC} 
behavior are also characterized by a constant flux of 
matter and energy from and to the environment. Thus, they 
are intrinsically non-equilibrium systems. The concept of 
universality is still applicable to non-equilibrium phase 
transitions. However, an universal classification scheme 
is still missing for non-equilibrium phase transitions and
the full spectrum of universality classes is unknown; it 
may be large or even infinite \citep{lubeck2005universal,hinrichsen2000non}. 
The properties of non-equilibrium transitions depend not 
only on the interactions but also on the dynamics. In contrast, 
detailed balance -- a necessary precondition for a steady 
state \citep{racz2002nonequilibrium} -- constrains the dynamics 
in equilibrium phase transitions. 

Classification methods of non-equilibrium phase transition 
are diverse and phenomenologically motivated. They have to 
be checked for each model separately and, as analytic solutions 
are in most cases missing, one uses numerical simulations or 
renormalization group approaches to describe the behavior at 
the critical point. Still, as 
\citet{lubeck2005universal} pointed out, a common mistake 
is the focus on critical exponents and the neglect of scaling 
functions, which are more informative. Determining 
the functional behavior of scaling functions is a precise method 
for the classification of a given systems into a certain
universality class. The reason for this is that the variations of 
scaling exponents between different universality classes are often 
small, whereas the respective scaling functions may show significant 
differences. Thus, to properly determine the corresponding universality 
class, one should extract both scaling functions and scaling exponents.

%%%%%%%%%%%%%%%%%%%%%%%%%%
\begin{table}[b]
\centering
\begin{tabular}{rl}
\hline \hline
{\it AST} &  absorbing state transition \\
{\it SOqC} & self organized quasi criticality \\
{\it BTW} sandpile model & the original sandpile model
\\ &
proposed by \citet{bak1987criticality} \\
Manna sandpile model & a variation of the {\it BTW} model with a 
stochastic \\ & 
distribution of grains, proposed by
\citet{manna1991two} \\
{\it OFC} earthquake model & a dissipative sandpile model,
\\ &
 proposed by \citet{olami1992self} \\
Zhang sandpile model & a non-abelian variation of the {\it BTW} model
\\ &
with continuous energy, proposed by \citet{zhang1989scaling}\\
\hline \hline
\end{tabular}
\caption{A list of widely used acronyms and popular models for
self organized criticality ({\it SOC}).}
\label{tbl_abbreviations}
\end{table}
%%%%%%%%%%%%%%%%%%%%%%%%%%

\subsection{Sandpile models}
The archetypical model of a {\it SOC} system is the sandpile 
model \citep{bak1987criticality}. We will start with a general 
description. Sandpile models are often defined on a $d$ dimensional grid of 
a linear size $L$, containing $N = L^d$ intersecting points. A point of a grid 
or a lattice is called a node and to each node one relates a real or 
integer positive variable $h$. This variable can be seen as the local 
energy level, the local stress or the local height level of the sandpile
(the number of grains of sand or some other particles at that location 
on the lattice). To mimic an external drive, that is the interaction 
of the system with the environment, a single node is randomly selected 
at each time step $t$ and some small amount of energy $\delta h$ is added 
to its local energy level,
%%%%%%%%%%%%%%%%%%%%%%%%%%%%%%%%%%%%%%%%%%%
\begin{figure}[t]
\centerline{
\includegraphics[width=0.8\textwidth ]{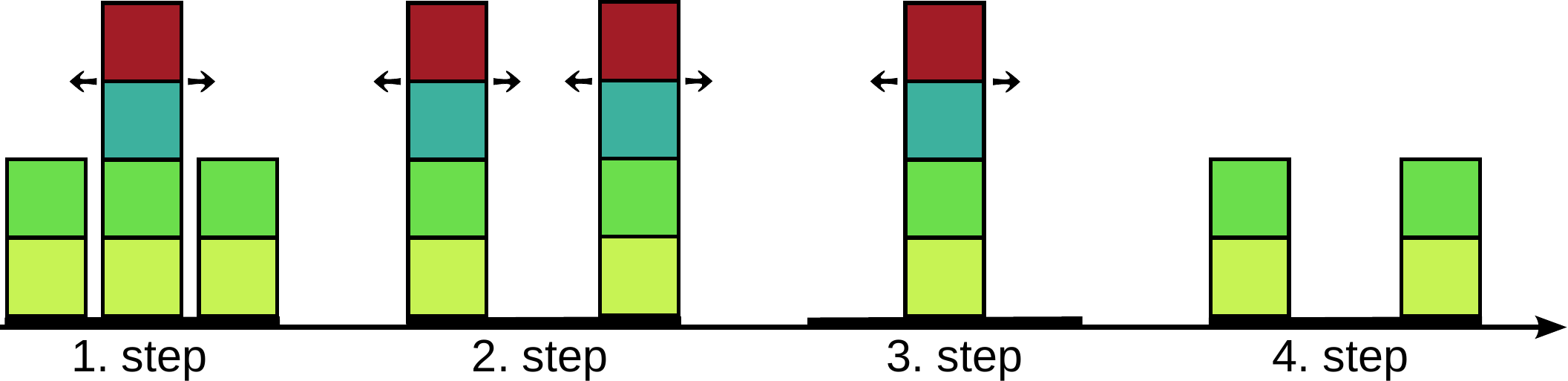}
}
\caption{An illustration of particle redistribution 
during an avalanche for a sandpile model with 
three sites. Once the local height $h$ reaches 
the activation threshold $h_T = 4$ the two neighboring 
nodes receive two particles each. Note that particles 
dissipate (disappear) only at the edge of the system.}
\label{fig_avalanche}
\end{figure}
%%%%%%%%%%%%%%%%%%%%%%%%%%%%%%%%%%%%%%%%%%%%
%
\begin{equation}
 h_{\vec{r}}(t+1) = h_{\vec{r}}(t) + \delta h~,
\end{equation}
where the index $\vec{r}=(r_1,\ldots, r_d)$, $r_i \in {1,\ldots,L}$ 
represents the location of a node on a $d$-dimensional lattice. 
If $h$ is a positive integer variable, then the increase of the local 
height proceeds in discrete steps, usually setting $\delta h = 1$. 
Once the energy at some node reaches a predefined threshold value $h_T$, 
the energy configuration of the system becomes unstable, the external 
drive is stopped, and the local energy is redistributed in the following way:
\begin{itemize}
 \item first, the energy level of the active node,
       for which $h_{\vec{r}} \geq h_T$, 
       is reduced by an amount $\Delta h$, {\it viz.}  
\begin{equation}
 h_{\vec{r}} \rightarrow h_{\vec{r}} - \Delta h~.
\end{equation}
 \item second, the nearest neighbors of the active node, 
       receive a fraction $\alpha$ of the lost energy 
       $\Delta h$. Denoting with $\vec{e}_n$ the relative location 
       of nearest neighbors with respect to location of active 
       node $\vec{r}$, we can write   
\begin{equation}
\label{eq_soc_transfer}
 h_{\vec{r}+\vec{e}_n}\rightarrow h_{\vec{r}+\vec{e}_n} + \beta\Delta h~.
\end{equation}
       For example, in the case of two dimensional ($d=2$) lattice we have 
       $\vec{e}_n = (\pm1,0)$, $(0,\pm1)$.

\item the update is repeated as long as at least one active node remains, 
      that is, until the energy configuration becomes stable.   
\end{itemize}
%%%%%%%%%%%%%%%%%%%%%%%%%%%%%%%%%%%%%%%%%%%
\begin{figure}[t]
\centerline{
\includegraphics[width=0.8\textwidth ]{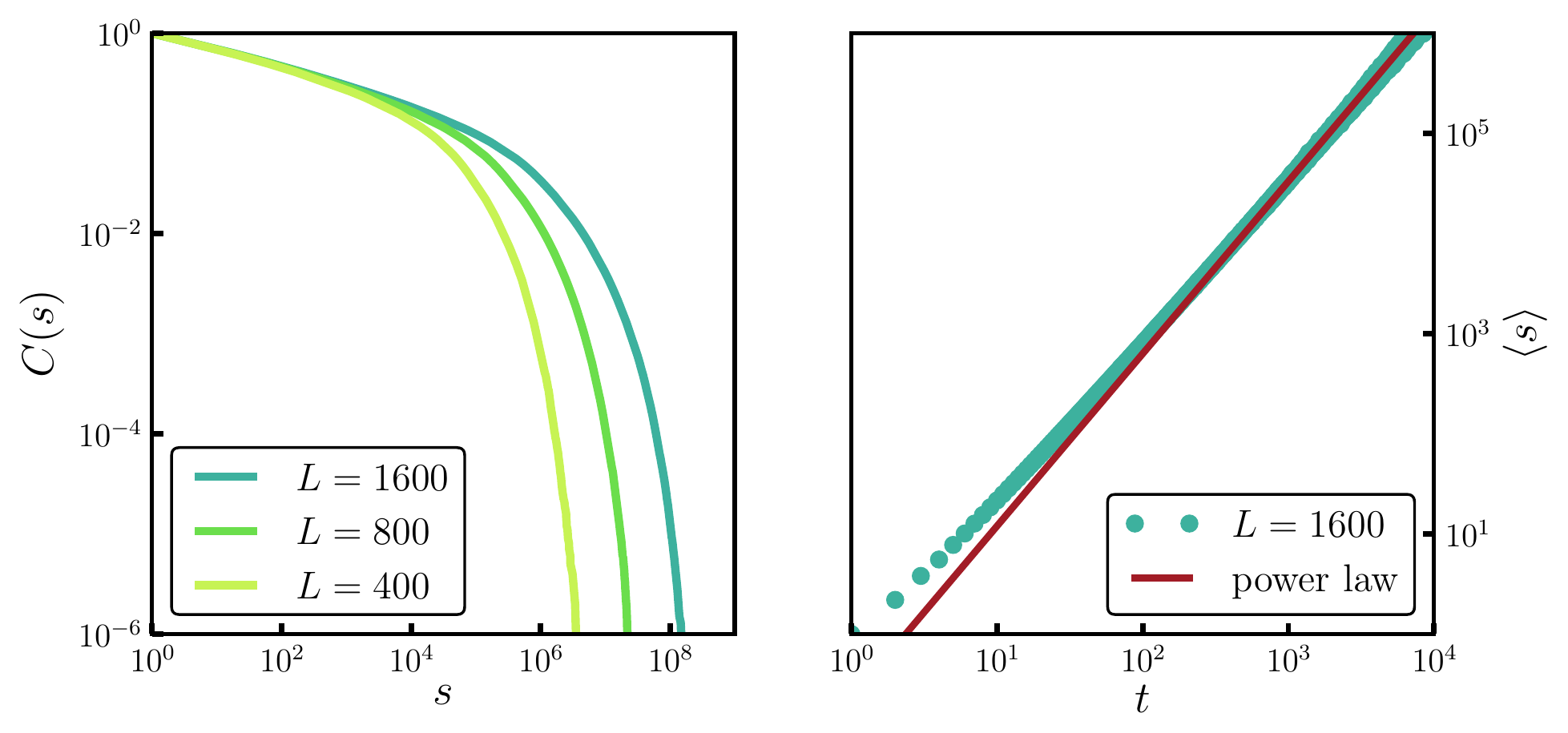}
}
\caption{Left: The complementary cumulative distribution 
$C(s|L) = \sum_{k=s}^{s_{max}} P(k|L)$ of avalanche 
sizes $s$ for the BTW sandpile model on a regular lattice 
of linear size $L$. Right: Average size $\langle s \rangle$ 
of avalanches, as a function of duration $t$, compared with the 
power-law dependence expected from the finite size scaling 
Ansatz $s\sim t^{\gamma_{ST}}$ (see Eq.~\ref{eq_scalinglaw}) 
with $\gamma_{ST} = 1.46$.}
\label{fig_distrib}
\end{figure}
%%%%%%%%%%%%%%%%%%%%%%%%%%%%%%%%%%%%%%%%%%%%
In Fig.~\ref{fig_avalanche} we illustrated the process of particle 
transport among nearest neighbors, also called an avalanche. Setting 
$$
\beta = \frac{1}{2d}
$$ 
assures local conservation of energy during an avalanche;
a necessary condition for a true {\it SOC} behavior of the 
sandpile models, as we will discuss later. However, the energy 
is conserved only locally; it is important to allow the energy to
dissipate at the lattice boundaries (grains falling off the
table), which is achieved by keeping the boundary nodes empty. 
If the amount of transferred energy $\Delta h$ -- which is 
transfered upon site activation -- equals the threshold value 
$h_T$, one calls the model an Abelian {\it SOC} model, because 
in this case the order of the energy redistribution does not 
influence the stable state configuration reached in the end
of the toppling process. The Abelian realization of the discrete 
height {\it SOC} model is better known as Bak-Tang-Wiesenfeld 
({\it BTW}) sandpile model \citep{bak1987criticality}. In 
addition, setting $\Delta h = \epsilon h$, where 
$\epsilon \in (0,1]$ leads to a non-Abelian {\it SOC} model which was 
-- in its continuous energy form -- first analyzed by 
\citet{zhang1989scaling}, thus named Zhang sandpile model 
(see Table \ref{tbl_abbreviations}).

Beside the {\it BTW} and the Zhang sandpile models, other variations of 
toppling rules exist. One possibility is a stochastic sandpile model 
proposed by \citet{manna1991two}, which was intensively studied as 
it is solvable analytically. Toppling rules can be divided into 
Abelian vs.\ non-Abelian, deterministic vs.\ stochastic and 
directed vs.\ undirected \citep{milshtein1998universality}. 
Modifications of the toppling rules employed often results in 
a change of the universality class to which the model belongs 
\citep{ben1996universality, giacometti1998dynamical}. 

Hitherto we described the critical height model, where the start 
of a toppling process solely depends on the height $h_{\vec r}$. 
Alternatively, in the critical slope model the avalanche initiation 
depends on the first derivative of the height function $h_{\vec{r}}$, 
or in the critical Laplacian model on the second derivative of 
the height function. These alternative stability criteria lead 
either to a different universality class, or to a complete absence 
of {\it SOC} behavior \cite{manna1991critical}.

%%%%%%%%%%%%%%%%%%%%%%%%%%%%%%%%%%%%%%%%%%%
\begin{figure}[t]
\centerline{
\includegraphics[width=0.8\textwidth ]{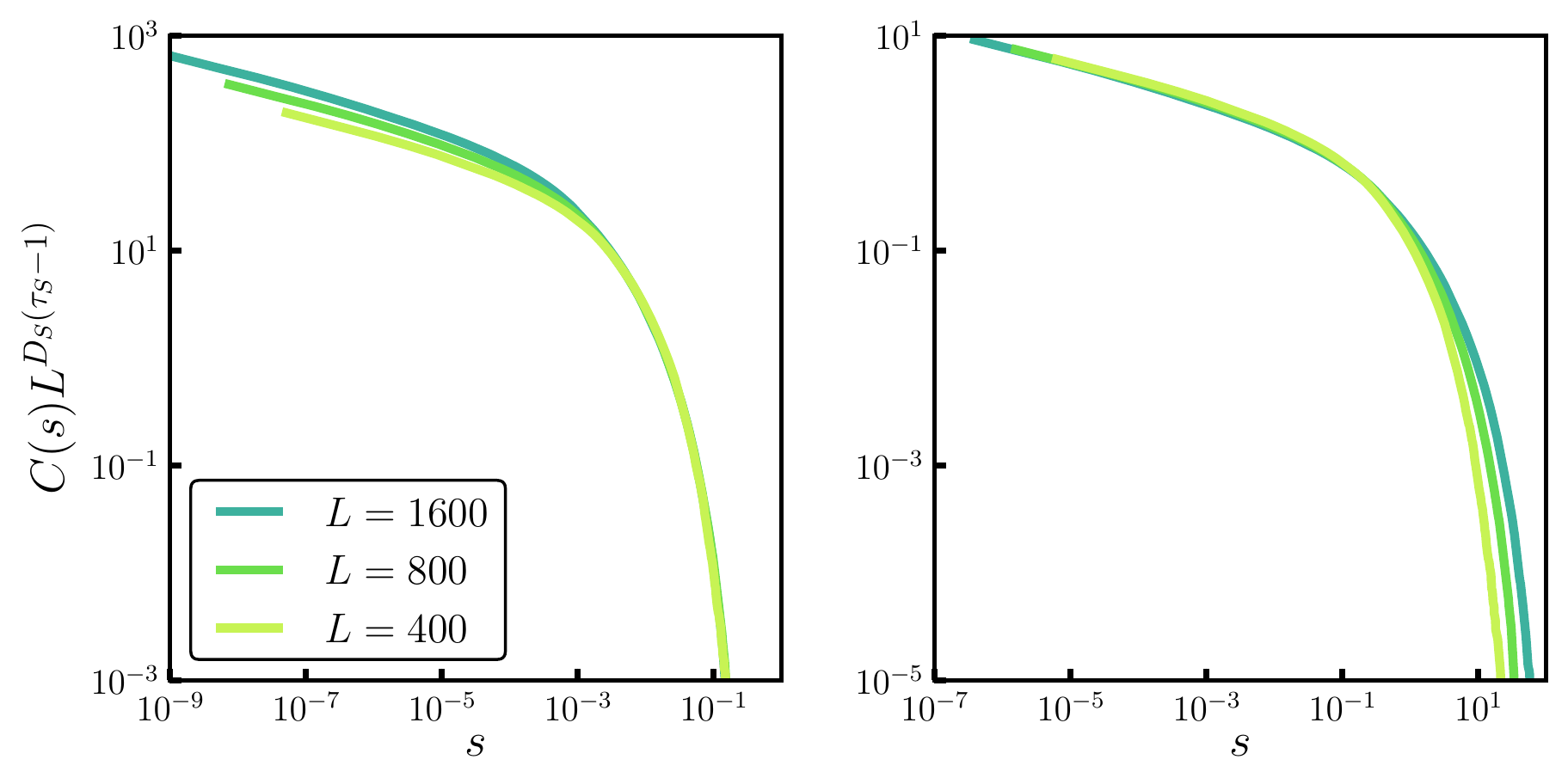}
}
\caption{Finite size scaling for the two dimensional 
{\it BTW} sandpile model, for the data shown in 
Fig.~\ref{fig_distrib}. Shown is the rescaled complementary
cumulative avalanche sized distribution {\it CCDF},
using the finite-size scaling Ansatz (\ref{eq_fss}),
appropriately integrated. The scaling parameters are
$\tau_S = 1.31$ and $D_S = 2.8$ (left)
and $\tau_S = 1.15$ and $D_S = 2$ (right). Note that the
first set of scaling exponents describes large avalanches
well, with the second set of exponents being appropriate for 
small avalanches.}
\label{fig_fsscaling}
\end{figure}
%%%%%%%%%%%%%%%%%%%%%%%%%%%%%%%%%%%%%%%%%%%%

%------------------------------
\subsection{Finite size scaling}
%------------------------------

The scaling behavior of avalanches can be extracted from the
statistical properties of several quantities: {\it e.g.} 
the size $s$ of the avalanche (the total number of activations
during an avalanche), the area $a$ of an avalanche (the 
number of distinct activated nodes), the avalanche duration 
$t$ (the number of parallel updates until a stable configuration 
is reached) and the linear size of the avalanche $r$ (usually 
estimated as the radius of gyration). In Fig.~\ref{fig_distrib} 
we show distribution of avalanche sizes obtained from the 
simulation of the {\it BTW} sandpile on a regular two dimensional 
lattice. In this review we discuss the scaling of observables -- 
like the results for the sandpile model shown in Fig.~\ref{fig_distrib} --
which result from uniform dynamics devoid of a hierarchical
organization. Scaling exponents may become complex in the
presence of underlying hierarchies \citep{sornette1998discrete} 
or specific interplay of dissipative and driving forces \citep{lee2000novel}. 
Hence, in such cases one needs to adopt the analysis of the 
scaling behavior corresponding to the discrete scale invariance 
\citep{huang2000artifactual, zhou2009numerical}, characterized by 
complex scaling exponents.
 
The theory of equilibrium critical phenomena implies that 
the scaling behavior of this quantities -- whenever the system 
is near a second-order phase transition -- follows the 
finite-size scaling ({\it FSS}) ansatz. In other words, 
one expects to find a scaling function for each observable
uniquely defining their respective scaling behavior, 
independently of the system size. Under {\it FSS} assumption 
probability distributions should have the following 
functional form \citep{cardy1996scaling}
\begin{equation}
\label{eq_fss}
P_X(x|L) = x^{-\tau_X} F_X(x/x_c),
\qquad\quad x_c = L^{D_X}~.
\end{equation}
Here $\tau_X$ and $D_X$ are the critical exponents 
for $x \in \{s,a,t,r\}$ and $L$ the linear system size.
The scaling function $F_X$ describes the finite size 
correction to the power law. Event sizes $x$ substantially
smaller than the system size follow a power law,
$F_X \to const.$ for $x << L^D$, with the fractional dimension
$D_X$ cutting off large fluctuations, $F \rightarrow 0$ for
$x \rightarrow x_c=L^{D_X}$. 

When the quantities (the size, the area, {\it etc.}) all follow 
{\it FSS}, then they will also scale as a power of each 
other in the limit $L \rightarrow \infty$, that is
the conditional probability $P_{X^\prime X}(x^\prime|x)$
of measuring $x^\prime$ given $x$ is diagonal,
\begin{equation}
\label{eq_relscale}
P_{X^\prime X}(x^\prime|x) \propto 
\delta(x^\prime-{x}^{\gamma_{X^\prime X}})~,
\end{equation} 
which arises from the requirement that 
$P_{X^\prime}(x^\prime) = \int P_{X^\prime X}(x^\prime,x)\d x$ 
is satisfied for any $x,x^\prime \in \{s,a,t,r\}$. From the 
same condition one obtains the scaling laws 
\begin{equation}
\label{eq_scalinglaw}
\gamma_{X^\prime X} = \frac{\tau_{X}-1}{\tau_{X^\prime}-1}~.
\end{equation}

Early studies of {\it SOC} behavior have demonstrated that 
certain models deviate from the expected {\it FSS} Ansatz. 
Reason for this deviation can be found in several premises
behind the {\it FSS} Ansatz: (1) boundaries should not have a 
special role in the behavior of the system; (2) a small finite 
system should behave the same as a small part of a large 
system. However, these conditions do not hold for most 
sandpile models. First, energy is dissipated at the boundaries, 
and their shape influences the scaling behavior. Second, the
average number of activations per site increases, during 
large avalanches, with the size of the system \citep{drossel2000scaling},
since energy dissipation is a boundary effect.

As an illustrative example we present in 
Fig.~\ref{fig_fsscaling} the rescaled {\it CCDF} 
of the avalanche size $s$ for the {\it BTW} sandpile 
model under the {\it FSS} assumption, 
that is rescaling $s \rightarrow s/L^{D_S}$ and 
$C_S(s) \rightarrow C_S(s)L^{D_S(\tau_S-1)}$, with
linear dimensions $L$. Depending 
on the value selected for the critical exponents,
$\tau_S$ and $D_S$, one finds nice collapse of the data 
for either large or small avalanches, though not for 
the entire range of avalanche sizes. This behavior is 
consistent with the deviation from a pure power-law scaling
for the time-dependent average avalanche size, as shown in 
Fig.~\ref{fig_distrib}, which may be approximated asymptotically
by a power law for either short or long avalanche durations,
but not for the entire range. Still, one can argue that
scaling, as described by Eq.~(\ref{eq_fss}), is expected to 
hold anyhow only asymptotically in the thermodynamic limit, 
that is, for large avalanche sizes or durations. Hence, it 
is of interest to examine whether these results indicate
to the presence of several distinct scaling regimes.

%%%%%%%%%%%%%%%%%%%%%%%%%%%%%%%%%%%%%%%%%%%
\begin{figure}[t]
\centerline{
\includegraphics[width=0.8\textwidth ]{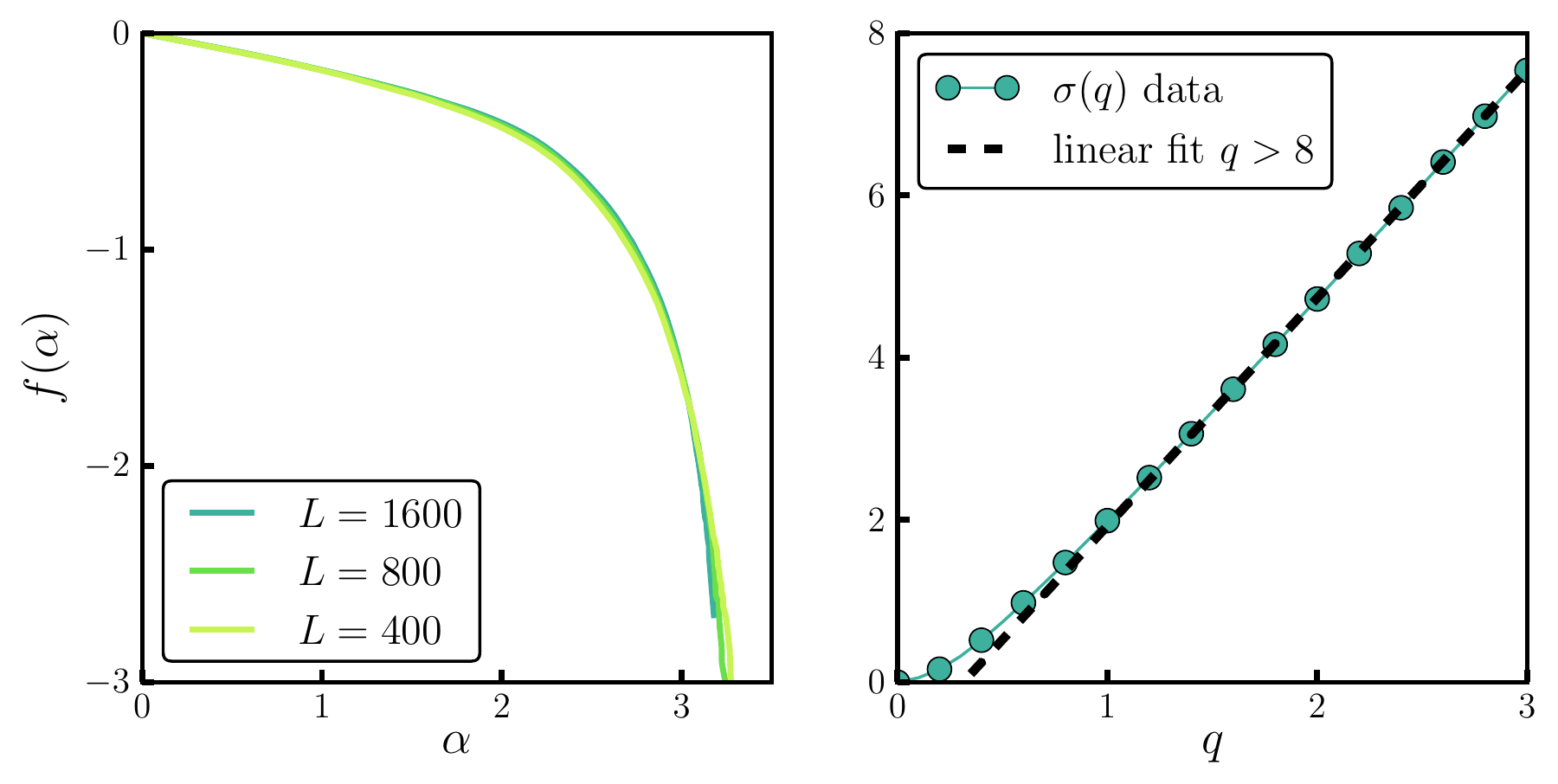}
}
\caption{Avalanche scaling properties of the two dimensional 
{\it BTW} sandpile model. (left) Multifractal spectrum $f_s(\alpha)$ 
of the avalanche size distribution $P_s$ for varying liner sizes 
$L$; (right) Scaling function $\sigma_s(q)$ of the 
$q$th moment of $P_s$, $\langle s^q \rangle \sim L^{\sigma(q)}$ 
obtained as linear fit of $\ln \langle s^q \rangle(L)$. Dashed line 
represents the fit of the region where $\sigma_S$ has linear dependence. 
The slope of the linear fit is then used to estimate 
$D_S=2.8$ and $\tau_S=1.31$.}
\label{fig_btwscaling}
\end{figure}
%%%%%%%%%%%%%%%%%%%%%%%%%%%%%%%%%%%%%%%%%%%%

%---------------------------------
\subsubsection{Multiscaling Ansatz}
%---------------------------------

It is well known, for a thermodynamic phase transition, 
that distinct scaling regime may exists. Somewhat further 
away from the critical point one normally observes scaling 
with meanfield exponents, and close to the transition 
(where the degree of closeness is given by the Ginzburg criterion) 
the scaling exponents are determined by the underlying universality class.
A possible approach in discriminating distinct scaling regimes is to
perform a rescaling transformation of the observable of interest,
an venue taken by the multifractal scaling Ansatz 
\citep{kadanoff1989scaling, de1998rare, tebaldi1999multifractal}.
Rescaling the {\it CCFF} 
\begin{equation}
\label{eq_multifractal}
 f_X(\alpha) = \frac{\log\left(C_X(\alpha|L)\right)}{\log(L)},
\qquad\quad
C_X(\alpha|L) = \int^\infty_{L^{\alpha}}P_X(x|L)dx~,
\end{equation}
one obtains with $f_X(\alpha)$ the so-called multifractal 
spectrum \citep{peitgen2004chaos}. One can furthermore
define via
\begin{equation}
\langle x^q \rangle_L = \int P_X(x|L)x^qdx \sim L^{\sigma_X(q)}~,
\end{equation}
the scaling exponents $\sigma_X(q)$ to the $q$th moment of the
distribution $P_X(x|L)$, which are related to the multifractal
spectrum $f_X(\alpha)$ through a Legendre transform,
\begin{equation}
\label{eq_legendre_tranform}
 \sigma_X(q) = \textrm{sup}_\alpha\big[f_X(\alpha) + q\alpha\big]~.
\end{equation}
If {\it FSS} is a valid assumption, viz when $P_X(x|L)$ 
follows a simple power law with a sharp cutoff given by $L^{D_X}$,
then the following form for $f_X(\alpha)$ is expected:
\begin{equation}
f_X(\alpha) = 
\begin{cases} 
\alpha(1-\tau_X) & \mbox{for\ \ \ } 0 < \alpha \leq D_x \\
-\infty & \mbox{for\ \ \ } \alpha_x > D_x
\end{cases}~.
\end{equation}
The jump to $-\infty$ is replaced by a continuous
downturn whenever the upper cutoff is not sharp, viz
if events of arbitrary large size $x$ are allowed but 
exponentially unlikely. The Legendre transform $\sigma_X(q)$ 
is given, for {\it FSS}, by
\begin{equation}
\label{eq_multi_transform}
\sigma_X(q) = 
\begin{cases} 
D_X(q - \tau_X +1)& \mbox{for\ \ \ } q > \tau_X -1 \\ 
\sigma_X(q) = 0& \mbox{for\ \ \ } q < \tau_X -1
\end{cases}~.
\end{equation}
The fractal spectrum $f_X(\alpha)$ will be piecewise linear 
for distributions having well defined and well separated scale 
regimes. On says that a fractal spectrum shows ``multifractal 
scaling'' when linear regimes are not discernible.

In Fig.~\ref{fig_btwscaling} we show the multifractal spectrum
$f_S(\alpha)$ for different system sizes $L$, and the corresponding 
moment scaling function $\sigma_s(q)$, which was obtained as the 
slope of the linear fit of $\ln \langle s^q \rangle(L)$ for a fixed 
moment $q$. The continuous downturn for large $\alpha$ seen for
$f_S(\alpha)$ results from the absence of a hard cutoff, the 
number of activated sites during an avalanche may be arbitrary 
large (in contrast to the area, which is bounded by $L^d$). One 
notes that data collapse is achieved and that $f_s(\alpha)$ and 
$\sigma_s(q)$ are not piece-wise linear,
implying multiscaling behavior of the {\it BTW} sandpile model. 

So far we have discussed methods typically used to characterize a 
scaling behavior of various {\it SOC} models, which provide a way 
to estimate both scaling exponents and scaling functions. In the 
next subsection we will discuss the underlying mechanism leading
to the emergence of the critical behavior observed in various 
sandpile models. For this purpose we introduce a general concept 
well known in the theory of non-equilibrium phase transitions,
the so called ``absorbing phase transitions''.  

%%%%%%%%%%%%%%%%%%%%%%%%%%%%%%%%%%%%%%%%%%%%%%%%%%%%%%%%%
\begin{figure}[t]
\centerline{
\includegraphics[width=0.7\textwidth ]{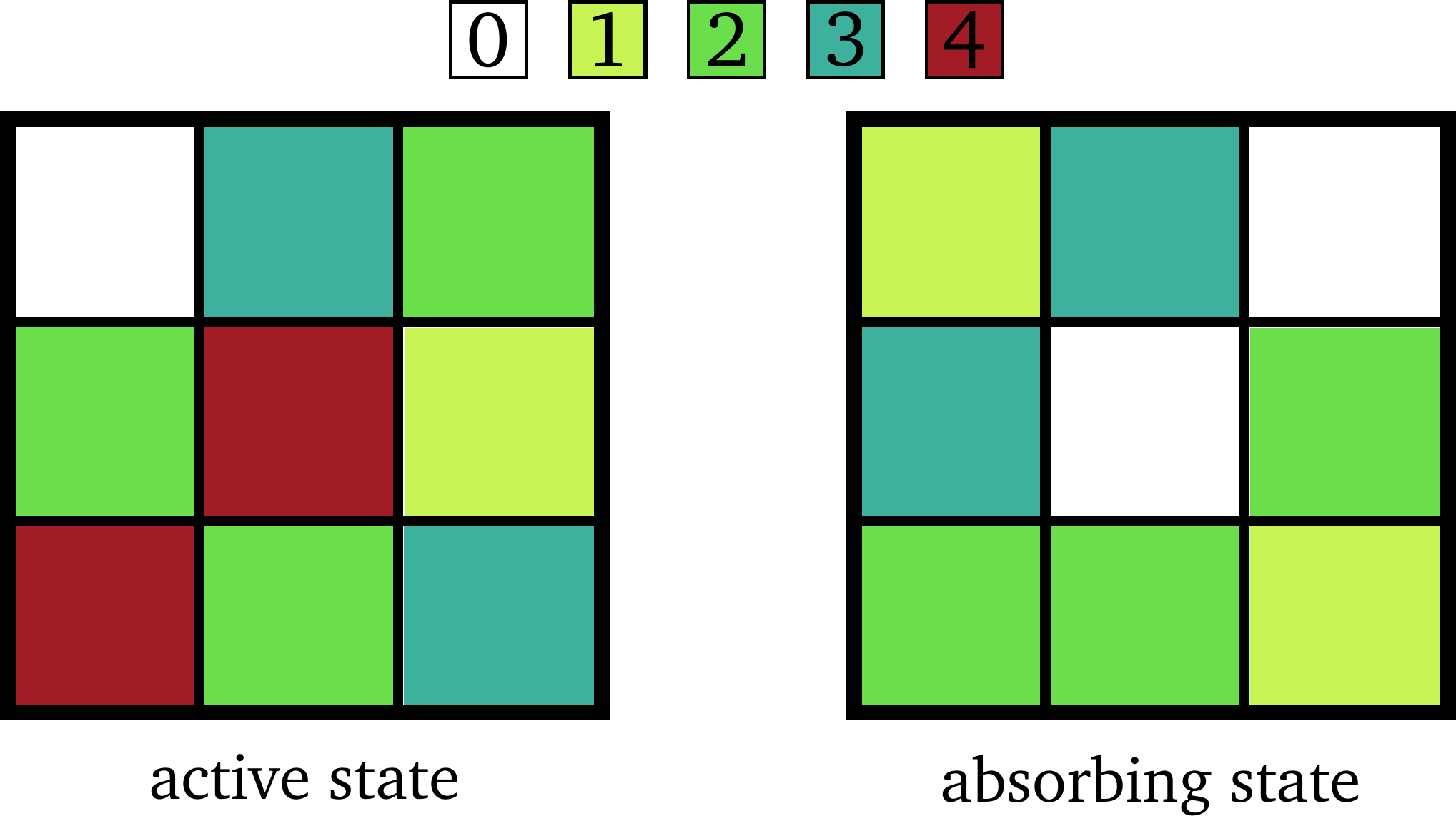}
}
\caption{Random configurations of particles on $3\times 3$ 
periodic lattice for a fixed energy sandpile model, where 
the activation threshold $h_T = 4$. Left: An active state
with a large number of particles per site, $\rho > \rho_c$. 
Right: An absorbing state with a low number of particles per
site, $\rho < \rho_c$, which is inactive.}
\label{fig_absorbingpahse}
\end{figure}
%%%%%%%%%%%%%%%%%%%%%%%%%%%%%%%%%%%%%%%%%%%%%%%%%%%%%%%%%
 
\subsection{Absorbing phase transitions and separation of time scales}
\label{subsec:absorbing}

Absorbing phase transitions exist in various forms in physical, 
chemical and biological systems that are operating far from 
equilibrium. They are considered without a counterpart in 
equilibrium systems and are studied intensively. 
For an absorbing phase transition to occur it is necessary that 
a dynamical system has at least one configuration in which the 
system is trapped forever, the so-called absorbing state. The 
opposite state is the active phase in which the time evolution 
of the configuration would never come to a stop, that is, the 
consecutive changes are autonomously ongoing.

A possible modeling venue for a dynamical system with an absorbing 
phase transition is given by the proliferation and the annihilation of
particles, where particles are seen as abstract representation of some quantity
of interest. A simple example for this picture would be a contact process on a
$d$-dimensional 
lattice \citep{marro2005nonequilibrium}, which is defined
in the following way: A lattice node can be either 
empty or occupied by a single particle; a particle may 
disappear with probability $1-p$ or create an offspring with 
probability $p$, at a randomly chosen nearest neighbor node. 
This contact process has a single absorbing state 
(with zero particles present) and one can show, in the mean 
field approximation, that this absorbing state becomes unstable 
for $p > p_c = 1/2$. For a broader discussion and a general overview 
of absorbing phase transitions we refer the reader to the recent 
review articles 
\citep{hinrichsen2000non,lubeck2005universal,marro2005nonequilibrium,racz2002nonequilibrium} and books 
\citep{henkel2009non, henkel2010non}. Here we will focus 
on the connection between the absorbing phase transitions and {\it SOC}. 

To understand the nature of {\it SOC} behavior arising in 
sandpile models we consider a fixed energy sandpile model. 
This model is obtained from the standard sandpile model by 
removing the external drive (the random addition of particles) 
and the dissipation (the removal of the particles at the boundary). 
Still, if the number of particles on a single lattice node 
exceeds some threshold value $h_T$ the particles at that node 
are redistributed to neighboring nodes as given by 
Eq.~(\ref{eq_soc_transfer}). This redistribution process continues 
as long as there are active nodes, at some position $\vec{r}$, with 
$h_{\vec r} \geq h_T$. If the initial particle density $\rho$ 
is smaller than some critical value $\rho_c$ any initial configuration of
particles will, in long-time limit, relax into a stable configuration,
corresponding to an absorbing state.
In a stable configuration there are no active nodes and each 
node can be in $h_T$ possible state (from $0$ to $h_T-1$). 
Hence, in the thermodynamic limit exist infinitely many absorbing states. For
$\rho>\rho_c$ there is always at least one active site and the redistribution
process continues forever. An illustration of absorbing and active states is
shown in Fig.~\ref{fig_absorbingpahse}. 

%%%%%%%%%%%%%%%%%%%%%%%%%%%%%%%%%%%%%%%%%%%%%%%%%%%%%%
\begin{figure}[t]
\centerline{
\includegraphics[width=0.7\textwidth ]{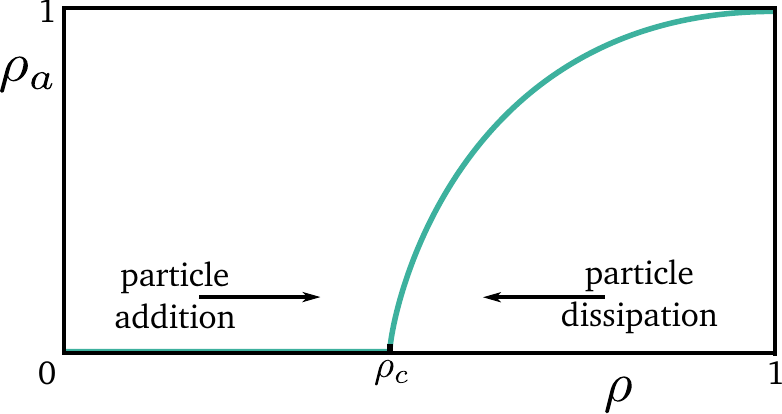}
}
\caption{The average density of active particles $\rho_a$,
the order parameter for an absorbing phase transition, 
as a function of particle density $\rho$.
The balance between the (very slow) addition of particles 
and the (relatively fast) dissipation during the active phase 
can maintain $\rho$ at the critical value $\rho_c$. This
separation of time scales is a defining property of
processes self-organizing towards criticality.
}
\label{fig_transition}
\end{figure} 
%%%%%%%%%%%%%%%%%%%%%%%%%%%%%%%%%%%%%%%%%%%%%%%%%%%%%%
 
Using the average density of active states $\rho_a$ as 
an order parameter, one usually finds that the absorbing 
to active phase transition is of second order, with 
$\rho_a$ changing continuously as $\rho$ goes through 
the $\rho_c$, as illustrated in Fig.~\ref{fig_transition}.
Thus, having a mechanism which slowly increases the amount of 
particles when $\rho < \rho_c$ (external drive) and which is 
stopped once the active state is reached, where fast dissipative 
effects take over (dissipation at the boundaries), will lead to 
the kind of self-organized critical phenomena as they are observed 
in sandpile models (Fig.~\ref{fig_transition}). Hence, we can 
relate criticality in sandpile models to the separation of 
timescales between external driving process and intrinsic 
dissipation process in systems with absorbing phase 
transitions. Thus, any non-equilibrium system, 
exhibiting an transition from an absorbing to an active 
phase, can be driven to a critical point by including 
a driving and a dissipating mechanisms with infinite 
separation of time scales \cite{dickman2000paths}. The 
separated time scales ensure the balancing of the system 
at the point of transition.
  
\subsection{{\it SOC} models on different network topologies}
Unlike regular structures or lattices, typically used 
in sandpile models, real-world complex systems mostly have 
non-regular structures, characterized often by a small 
world topology and scale-free connectivity. 
Thus, it is important to understand the influence of 
different network topologies on the scaling behavior of 
sandpile and other {\it SOC} models. 

The studies of the sandpile dynamics on Erdős–Rényi 
random graphs \citep{erdHos1959random}, have shown that 
the scaling exponents correspond to the ones obtained for 
high-dimensional lattices 
\citep{christensen1993sandpile, bonabeau1995sandpile}, 
thus belonging to the same universality class in the 
thermodynamic limit. Similar conclusions have been reached 
for the {\it BTW} sandpile on the Watts-Strogats type 
small-world networks \citep{watts1998collective}. This kind
of networks are constructed from an usual $d$-dimensional lattice by 
randomly rewiring a certain fraction of links $p$. Importantly, 
the rewiring is performed in a way such that the number 
of nearest neighbors is unchanged. This introduces 
long range interaction for $p>0$, yielding small-world 
structures for small $p$ and random structures for large $p$. 
On these networks it is simple to implement the classical 
{\it BTW} model without any modification for the toppling 
dynamics. \citet{de2002self,pan2007sandpile} concluded that 
the avalanche behavior, in the thermodynamic limit 
$L\rightarrow\infty$, corresponds to the mean field behavior 
for any $p>0$. Thus, the introduction of shortcuts to 
regular lattice structures is effectively increasing the 
dimensionality of the lattice, with the scaling behavior 
corresponding to the one observed for high dimensional lattices 
\citep{lahtinen2005sandpiles}.  

\subsubsection{Scale-free networks}
Investigations of the {\it BTW} sandpile model on 
uncorrelated scale-free networks \citep{barabasi1999emergence} 
have shown an interesting scaling behavior dependent on 
the network parameters 
\citep{goh2001universal, goh2003sandpile, lee2004sandpile, goh2005cascading}.
Scale-free graphs are graphs with a power-law distributed 
degree, that is $p_d(k)\sim k^{-\gamma}$, where the degree 
$k$ of a node is the number of its nearest neighbors. As each 
node has a variable number of neighbors, the activation threshold 
of each node is set proportional to the local vertex degree and 
defined as $h_T^{(i)} = k_i^{1-\eta}$, where $k_i$ is the out-degree 
of the $i$th node, and $0\leq \eta \leq 1$ such that 
$h_T^{(i)} \leq k_i$. Grains of sand are again added to randomly 
chosen nodes, until the activation threshold $h_T^{(i)}$ of the
selected node is surpassed. Once a node gets activated the external 
drive is stopped, and the toppling of grains proceeds until a stable 
state is reached. Dissipation is introduced either by removing 
small fraction $f$ of grains during the avalanche, or by mapping 
the network to a lattice and removing some small amount of grains at 
the boundary, which sets the maximal size of the avalanche. 
Active sites transfer a single grain to each of the 
$n = \lceil h_T^{(i)} \rceil$ randomly chosen nearest 
neighbors, where $\lceil h_T^{(i)} \rceil$ denotes
the smallest integer greater or equal to $h_T^{(i)}$.
The height of the $i$th active node $h_i$ is then 
decreased by $\lceil h_T^{(i)} \rceil$. Note that 
for $\eta > 0$ the grains are stochastically redistributed 
to nearest neighbors as the number of available grains 
$n$ is smaller then the out-degree $k_i$. 

In addition to numerical simulation, the scaling exponents 
for the avalanche size $\tau_s$ and the avalanche duration $\tau_t$ 
have been obtained analytically by taking into account the
tree like structure of the uncorrelated network and by mapping 
an avalanche to a branching processes \citep{lee2004branching},
a procedure we will discuss in Sect.~\ref{sec:branching}. 
Using the formalism of branching processes one finds that the 
scaling exponents of the avalanche distributions depend on the 
network scaling exponent $\gamma$ and threshold proportionality 
exponent $\eta$ in the following manner:
\begin{equation}
\begin{array}{llcr}
\tau_s=3/2,&
\quad\tau_t = 2 &\quad\textrm{when}\quad&
\gamma > 3-\eta \\
\tau_s=\frac{\gamma-2\eta}{\gamma -1 -\eta},&
\quad\tau_t = \frac{\gamma - 1 - \eta}{\gamma - 2}&
 \quad\textrm{when}\quad &
2 < \gamma < 3-\eta
\end{array}
\label{chap2:scalingExponents_scaleFreeGraphs}
\end{equation}
Hence, there are two separate scaling regimes dependent on the value 
of the parameter $\gamma$, which defines the network connectivity. 
At the transition of this two regimes---that is, for $\gamma = 3-\eta$---the 
avalanche scaling has a logarithmic correction  
\begin{equation}
 p_s(s)\sim s^{-3/2}(\ln s)^{-1/2},\qquad
p_t(t) \sim t^{-2}(\ln t)^{-1}~.
\end{equation}
These logarithmic corrections correspond to the scaling properties 
of critical systems at the upper critical dimension, above which 
the mean-field approximation yields the correct scaling exponents. 

The analytic results (\ref{chap2:scalingExponents_scaleFreeGraphs})
for uncorrelated graphs are well reproduced by numerical
simulations \citep{goh2005cascading}. However, real-world
networks having scale-free degree distributions, 
contain additional topological structures, such as
degree-degree correlations. Simulating the sandpile dynamics 
at the autonomous system level for the Internet, and
for the co-authorship network in the neurosciences,
one observes deviations to the random branching predictions
\citep{goh2005cascading}. The higher order structures of 
scale-free networks do therefore influence the values of the scaling
exponents. In addition, separate studies of BTW sandpile models 
on Barab\'asi-Albert scale-free networks have demonstrated that 
scaling also depends on the average ratio of the incoming 
and the outgoing links \citep{karmakar2005sandpile}, further 
demonstrating the dependence of scaling behavior on the details
of the topological structure of the underlying complex network. 

Topological changes in the structure of the network generally
do not disrupt the power-law scaling of the {\it BTW} model, 
it is however still worrisome that the scaling exponents
generically depend on the network fine structure. Such 
dependencies suggests that the number of the universality 
classes is at least very large, and may possibly even be 
infinite. With so many close-by universality classes, 
a large database and very good statistics is hence necessary, 
for a reliable classification of real-world complex system 
through experimental observation.

In the following subsection we will consider {\it SOC} models 
supplemented by dissipative terms---which are essential for 
many real-world applications---thus contrasting the {\it SOC} models 
with conserved toppling dynamics which we did discuss hitherto.

\subsection{SOC models with dissipation}
\label{subsec:dissipativesoc}
Conventional {\it SOC} models such as {\it BTW}, Zhang or 
Manna sandpile models (see Table \ref{tbl_abbreviations}),
require---to show critical scaling behavior---that 
the energy (the number of sand grains) is locally conserved. 
The introduction of local dissipation during an avalanche 
({\it e.g.} by randomly removing one or more grains during 
the toppling) leads to a subcritical avalanche behavior 
and to a characteristic event size which is independent 
of the system size. To recover self-organized critical 
behavior---or at least quasi-critical behavior, as we will discuss 
later---a modification is required for the external driving. 
Besides the stochastic addition of particles or energy, a 
loading mechanism has to be introduced. This mechanism 
increases the total energy within the system, bringing it 
closer to the critical point, but without starting an avalanche 
\citep{bonachela2009self}. From now on we will only consider 
models where the lattice nodes are represented by a continuous 
variable representing local energy levels, as defining dissipation 
under such setup is quite straightforward. 

In recent years, {\it SOC} models without energy 
conservation have raised some controversy regarding 
the statistical properties of the generated avalanche 
dynamics, and with regard to their relation to the 
critical behavior observed in conserved {\it SOC} 
models, such as the {\it BTW} model. A solvable version of a
non-conserving model of {\it SOC} was introduced by 
\citet{pruessner2002solvable}. The dissipation is controlled 
by a parameter $\beta$ (compare Eq.~\ref{eq_soc_transfer})  
which determines the fraction of energy transmitted,
by an activated node, to each neighbor. The toppling dynamics
is conserving for $\beta=1/k$, where $k$ denotes the 
number of nearest neighbors, and dissipative for $0<\beta < 1/k$. 
For the external driving one classifies the sites into
three categories. A site with energy $z_i$ is said to be 
stable for $z_i<h_T(1-\beta)$, susceptible for 
$h_T(1-\beta)\le z_i<h_T$ and active for and $1\le z_i$.
The actual external driving is then divided into a loading 
and a triggering part. 
\begin{itemize} 
\item The loading part of the external drive consists of randomly 
      selecting $n$ nodes. If the selected sites are stable, 
      having an energy level below $h_T(1-\beta)$, their respective
      energies are set to $h_T(1-\beta)$, they become susceptible.
\item For the triggering part of the external driving a single 
      node is selected randomly. Nothing happens if the site
      is stable. If the site is susceptible, its energy level 
      is set to $h_T$ and the toppling dynamics starts.
\end{itemize}
Interestingly, depending on the loading intensity, that is on the 
value of the loading parameter $n$, the avalanche dynamics will 
be in a subcritical, critical or supercritical regime, for a given
system size $N = L^d$. The critical loading parameter $n_c$ scales 
as a power of the system size $N$ and diverges in the thermodynamic 
limit. This need for fine tuning of the load, which 
can be generalized to other non-conservative {\it SOC} models, 
implies that dissipative systems exhibit just apparent self-organization. 
Furthermore even with tuned loading parameter $n = n_c$, the dynamics 
will only hover close to the critical state, without ever reaching it 
exactly. This behavior was denoted self-organized quasi-criticality 
({\it SOqC}) by \citet{bonachela2009self}.

\subsubsection{The OFC earthquake model}
%%%%%%%%%%%%%%%%%%%%%%%%%%%%%%%%%%%%%%%%%%%
\begin{figure}[t]
\centerline{
\includegraphics[width=0.8\textwidth ]{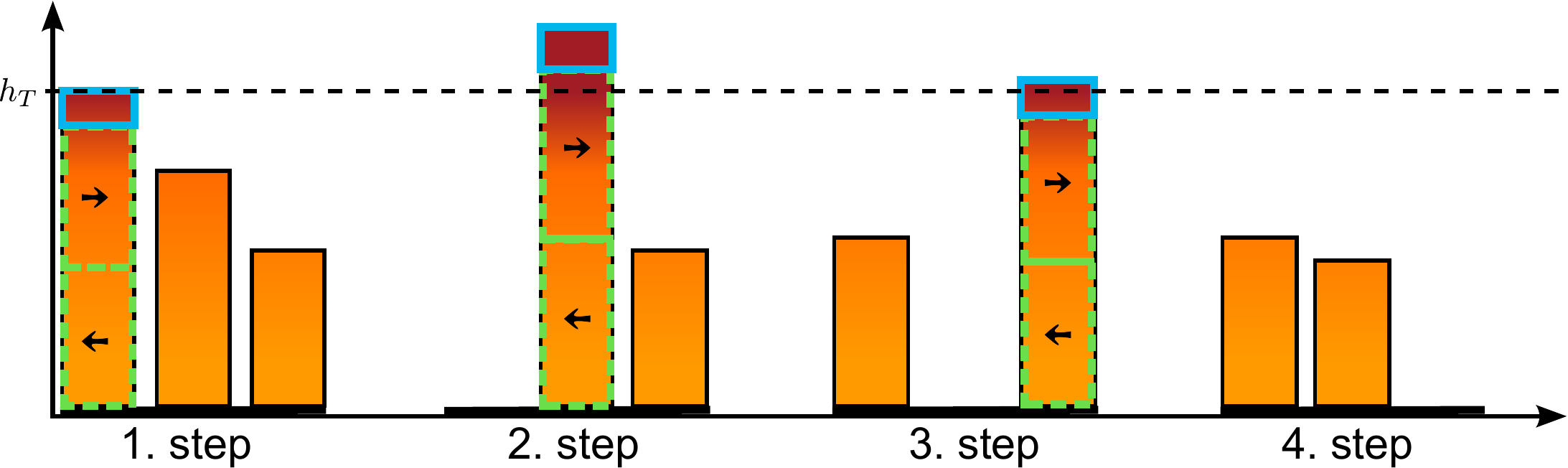}
}
\caption{An illustration of local dissipation during an 
avalanche, introduced in the OFC model. Besides the dissipation 
at the boundary, some fraction of energy (denoted by blue rectangles) 
is lost locally at each active site. The remaining energy is distributed 
equally between the neighboring nodes or dissipated at the 
open boundary.}
\label{fig_dissip_avalanche}
\end{figure}
%%%%%%%%%%%%%%%%%%%%%%%%%%%%%%%%%%%%%%%%%%%% 

Perhaps one of the most studied dissipative {\it SOC} models is the 
Olami-Feder-Christensen ({\it OFC}) model \citep{olami1992self}. 
The OFC model is an earthquake model, 
as it was originally derived as a simplified version of the
Burridge-Knopoff model \citep{burridge1967model}, which was 
designed to mirror essential features of earthquakes and tectonic 
plates dynamics. In this model the local height parameter
$h_{\vec{r}}$ is continuous and corresponds to local forces. 
The external driving, thought to be induced by slipping rigid 
tectonic plates, is global in the {\it OFC} model, whereas it 
would be local in most other sandpile models. The global 
driving force is infinitesimally slow and acts at the same time 
on all sites. Thus, the driving process can be simplified as 
following:
\begin{itemize}
\item One determines the location $\vec r^*$ with the largest stress, with
      $h_{\vec r^*}>h_{\vec r}$, for every position $r\ne \vec r^*$.
\item All forces are then increased by 
      $h_{\vec{r}}(t+1) = h_{\vec{r}}(t) + \delta h$,  where
      $\delta h=h_T-h_{\vec r^*}$.
\item The toppling dynamics then starts at $\vec r^*$,
      following Eq.~(\ref{eq_soc_transfer}), with a
      dissipation parameter $\beta$ and $\Delta h = h_{\vec{r}}$, 
      that is after activation $h_{\vec{r}}(t+1) \rightarrow 0$.
\end{itemize}
The model becomes, as usual, conservative for $\beta={1}/{2d}$. 
In addition to the local dissipation there is still dissipation at the 
boundaries (see Fig. \ref{fig_dissip_avalanche}), when assuming 
fixed zero boundary forces $h_{\vec r}$. In fact dissipative boundaries are essential for {\it SOqC} behavior to emerge. 
%%%%%%%%%%%%%%%%%%%%%%%%%%%%%%%%%%%%%%%%%%%%%%%%%%%%%%%%%%%%%%
\begin{figure}[t]
\centerline{
\includegraphics[width=0.8\textwidth ]{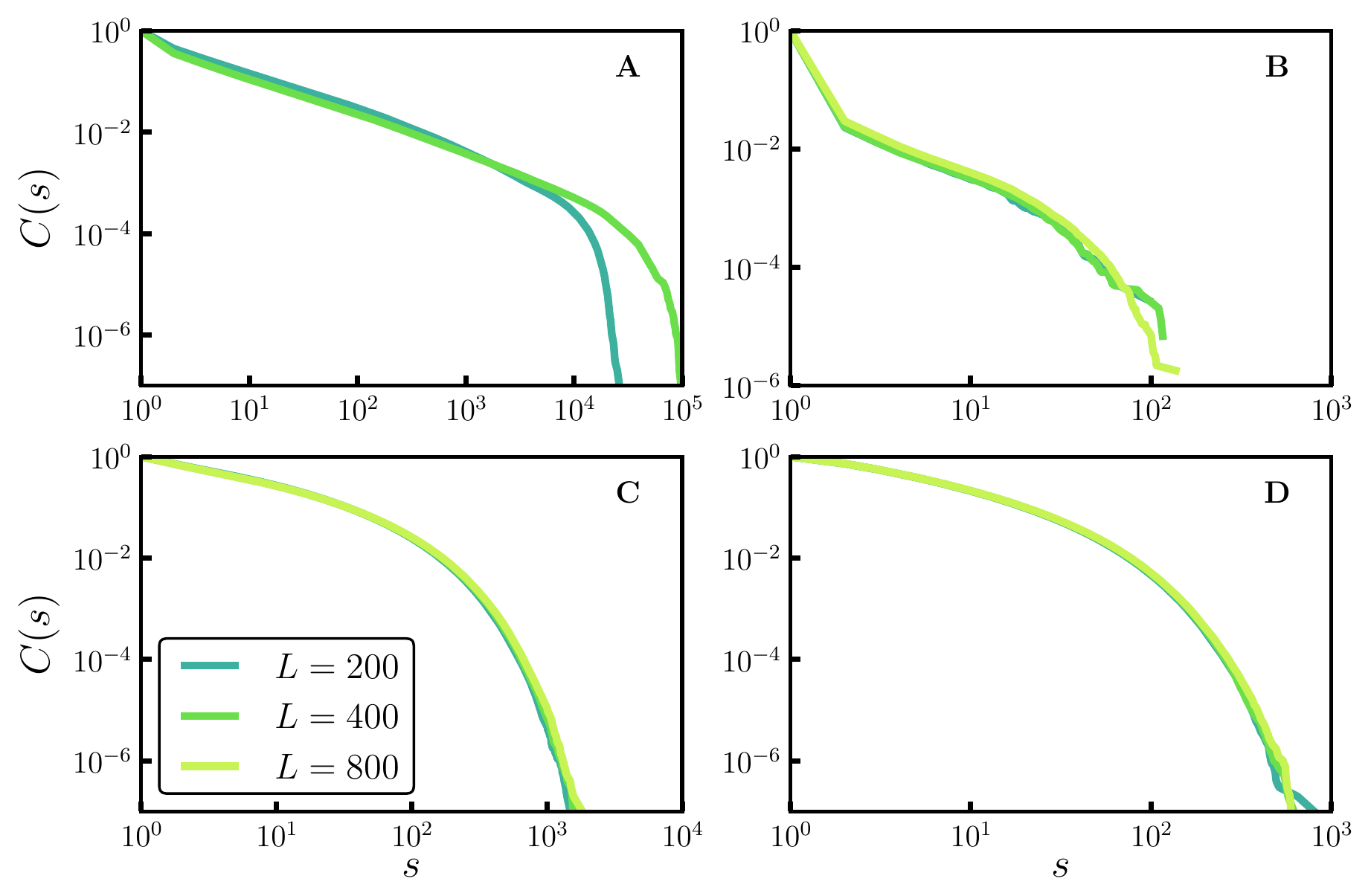}
}
\caption{Complementary cumulative avalanche size distribution 
for dissipative {\it OFC} model on different network topologies,
all having $N=L^2$ nodes. 
(A) Regular lattice with open boundaries, 
(B) regular lattice with periodic boundary conditions, 
(C) small-world network, 
(D) scale-free network.}
\label{fig_networks}
\end{figure} 
%%%%%%%%%%%%%%%%%%%%%%%%%%%%%%%%%%%%%%%%%%%%%%%%%%%%%%%%%%%%%%
%

Although initial studies of {\it OFC} models showed 
indications of critical behavior 
\citep{olami1992self,janosi1993self,jensen1998self, 
lise2001self}, later numerical studies on much larger 
system sizes found little evidence for the critical 
scaling of avalanche sizes. For dissipation rates 
$\beta > 0.18$ the scaling is very close to 
a power law and the behavior may be considered as almost 
critical that is quasi-critical \citep{miller2002measurements, 
boulter2003nonuniversality, miller2003crossover}. The difficulty 
with simulating the {\it OFC} model is that system goes through a 
transient period, which grows rapidly with system size, before it 
reaches the self-organized stationary state, thus increasing significantly 
the computational power and time needed to simulate large lattices 
\citep{wissel2006transient}. Furthermore, in the same work, 
\citet{wissel2006transient} showed that the size distribution 
of the avalanche is not of a power law form but rather of 
a log-Poisson distribution. Nevertheless, it is still 
considered that dissipative systems with loading 
mechanism are much closer to criticality than it 
would be the case in the absence of such mechanism 
\citep{bonachela2009self}. Still, although 
the {\it OFC} model is not strictly critical, it 
is somewhat more successful then other similar models 
in fitting the Omori scaling of
aftershocks \citep{hergarten2002foreshocks,wissel2006transient}. 

\begin{figure}[t]
\centerline{
\includegraphics[width=0.8\textwidth]{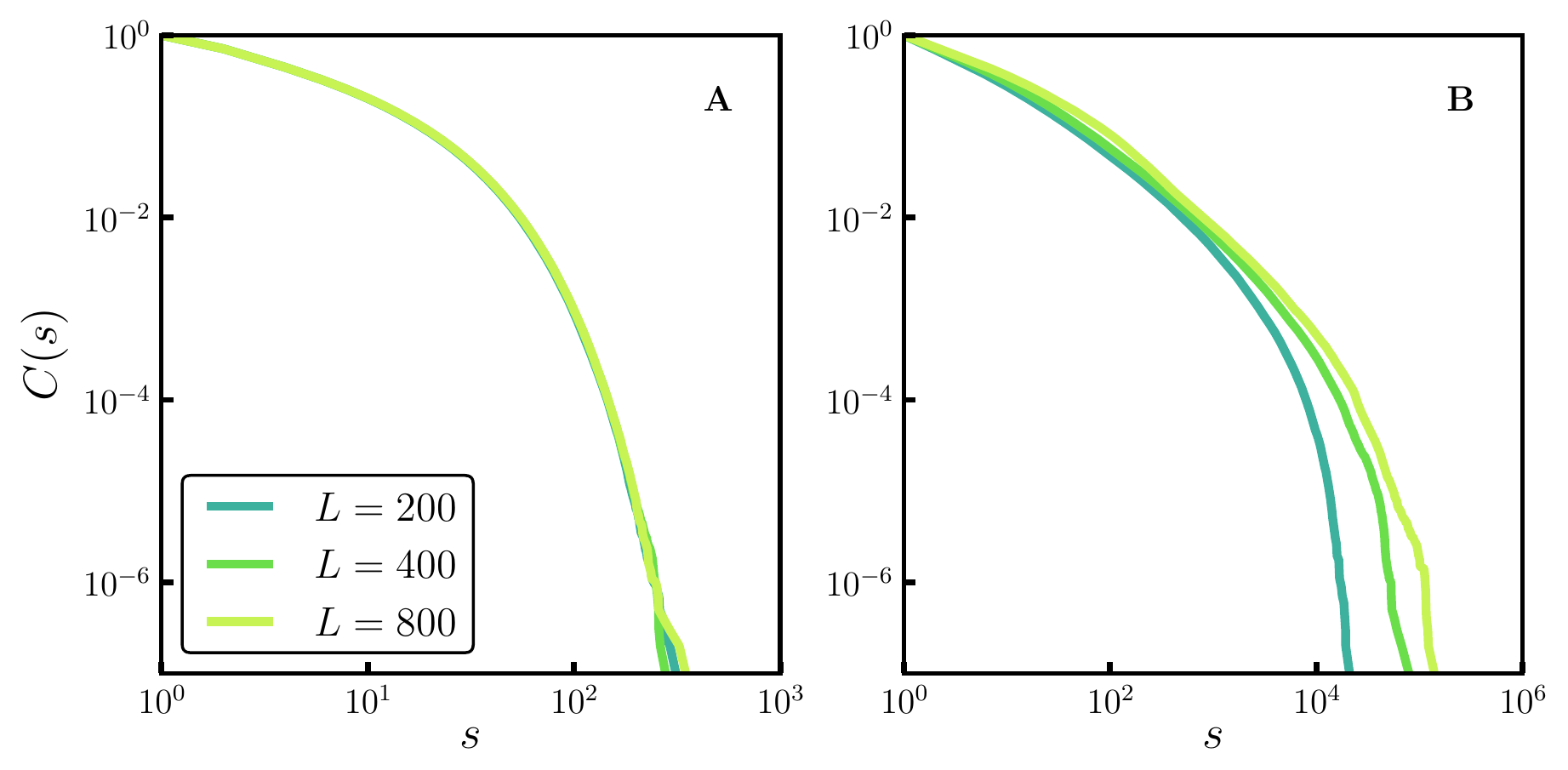}
}
\caption{Cumulative avalanche size distribution for 
dissipative {\it OFC} model on a regular lattice 
in the case of (A) non-uniform threshold, 
(B) asymmetric and random interactions.}
\label{fig_iregularities}
\end{figure}

The {\it OFC} model, which has seen several successful applications 
\citep{helmstetter2004properties, hergarten2002foreshocks, 
caruso2007analysis}, does neglect heterogeneities
as they occur in the structure of the real-world complex 
systems. Within the {\it OFC} model one assumes that the 
site activation threshold is uniform across all nodes, that 
the avalanches are undirected, that the elements have symmetric 
interactions and that the network has a regular structure and 
regular dissipative boundaries. Adding local variations, 
expected to exist in natural systems, in any of the mentioned 
properties of the model, leads to the disappearance 
of any similarity to critical scaling behavior. For example, 
introducing local variations in the threshold values 
\citep{janosi1993self}, or in the local degree of dissipation 
\citep{mousseau1996synchronization}, results in subcritical scaling 
behavior, although {\it SOqC} is preserved for very small variations. 
The change in the network structure to more irregular topology has 
a similar effect, although exceptions exist. For the case of 
quenched random networks, only finite avalanches are observed 
for any non-zero dissipation level, while power-law scaling
is retained for annealed networks 
\citep{chabanol1997analysis, lise2002nonconservative}. The 
disappearance of power-law scaling has also been observed
for the {\it OFC} model on scale-free networks 
\citep{caruso2006olami} and regular lattice with periodic boundary 
conditions \citep{grassberger1994efficient} 
(see Fig.~\ref{fig_networks}). Interestingly, {\it OFC} model 
on small-world topology, with a small rewiring probability 
and undirected connections, shows properties similar to the 
ones obtained on regular lattices \citep{caruso2006olami}. 
Examples for the  scaling of avalanche sizes in the presence 
of various site dependent irregularities for the {\it OFC} model 
are shown in Fig.~\ref{fig_iregularities}. 

%%%%%%%%%%%%%%%%%%%%%%%%%%%%%%%%%%%%%%%%%%%%%%%%%%%%%%%%
%%%%%%%%%%%%%%%%%%%%%%%%%%%%%%%%%%%%%%%%%%%%%%%%%%%%%%%%

Non-conserving {\it SOC} models are able to reproduce 
certain aspects of scaling exhibited by real-world phenomena. 
The incorporation of structural variations, which are common 
features of natural and man made systems, results however 
in qualitative changes for the observed
scaling. This circumstance is quite worrying, as pointed 
by \citet{jensen1998self}. If a model is applicable to 
real physical systems, it should also exhibit some 
robustness to disorder. In section \ref{sec:exp} we will discuss 
in more details empirically observed properties of earthquakes and 
solar flares, which will also reveal additional differences 
between real-world phenomena and both conserved and non-conserved 
{\it SOC} models. The implications of {\it SOC} theory on the 
observed power-law behavior of neuronal avalanches, and possible 
extensions of {\it SOC} theory or alternative explanation 
of their origin, will also be discussed.

\section{Alternative models for generating heavy-tailed distributions}
\label{sec:alternative}

The quest for explaining and understanding the 
abundance of power-law scaling in complex systems 
has produced, in the past several decades, a range of
of models and mechanisms for the generation of
power laws and related heavy-tailed distributions.

Some among these models provide relatively simple generating 
mechanisms \citep{newman2005power}, e.g.\ many 
properties of random walks are characterized by 
power laws, while others are based on more intricate 
principles, such as the previously described 
{\it SOC} mechanism. We will now shortly describe three
classes of basic generating mechanism, and then 
discuss in more detail a recently proposed heavy-tail 
generating mechanism, the so 
called principle of {\it highly optimized tolerance}. 
The emphasis of our discussion will be on general
underlying generating principles, and not on the details
of the various models. For additional information with respect
to several alternative mechanisms, not mentioned here, we refer 
the reader to several sources \citep{mitzenmacher2004brief, newman2005power, sornette2004critical, schwab2013zipf}.
     
%%%%%%%%%%%%%%%%%%%%%%%%%%%%%%%%%%%%%%%%%
\subsection{Variable selection and power laws}
%%%%%%%%%%%%%%%%%%%%%%%%%%%%%%%%%%%%%%%%%

One can generate power laws when selecting the
quantity of interest appropriately
\citep{sornette2002mechanism, newman2005power}. 
This procedure is, however, in many cases not an 
artifact but the most natural choice.
Consider an exponentially distributed variable 
$y$, being logarithmically dependent on a
quantity of interest $x$,
\begin{equation}
p(y)\sim\mathrm{e}^{ay}, \quad\qquad y=b\log(x),
\quad\qquad \frac{dy}{dx}=\frac{b}{x}~.
\label{chap3:y-log-x}
\end{equation}
The distribution $p(x)$ 
\begin{equation}
p(x) = p(y)\frac{dy}{dx}\sim \frac{b}{x}\mathrm{e}^{ab\log(x)}
\sim x^{ab-1}
\label{chap3:p_x-p_y}
\end{equation}
then has a power-law tail. Exponential distributions are
ubiquitous, any quantity having a characteristic length
scale, a characteristic time scale, etc.\ is exponentially
distributed. A logarithmic dependence $y\sim\log(x)$
does also appear frequently; {\it e.g.} the information content,
the Shannon information, has this functional form
\citep{gros2010complex}. Power laws may hence quite
naturally arise in systems, like the human language, 
governed by information theo\-re\-tical principles 
\citep{newman2005power}.

For another example consider two variables being the
inverse of each other,
\begin{equation}
x\sim \frac{1}{y},\qquad\quad
p(x) \sim \frac{p(y)}{x^2}~.
\label{chap:x-inverse-y}
\end{equation}
The distribution $p(x)$ has hence a power-law tail for large 
$x$, whenever the limit $\lim_{y\to0}p(y)$ is well 
behaved. E.g.\ for finite $p(y=0)$ the tail
is $p(x)\sim 1/x^2$. Whether or not a relation akin
to (\ref{chap:x-inverse-y}) is physically or biologically
correct depends on the problem at hand. It is important,
when examining real-world data, to keep in mind that
straightforward explanations for power-law dependencies---like 
the ones discuss above---may be viable, before 
jumping to elaborated schemes and fancy explanations.

%%%%%%%%%%%%%%%%%%%%%%%%%%%%%%%%%%%%%%%%%%%
\begin{figure}[t]
\centerline{
\includegraphics[width=0.8\textwidth ]{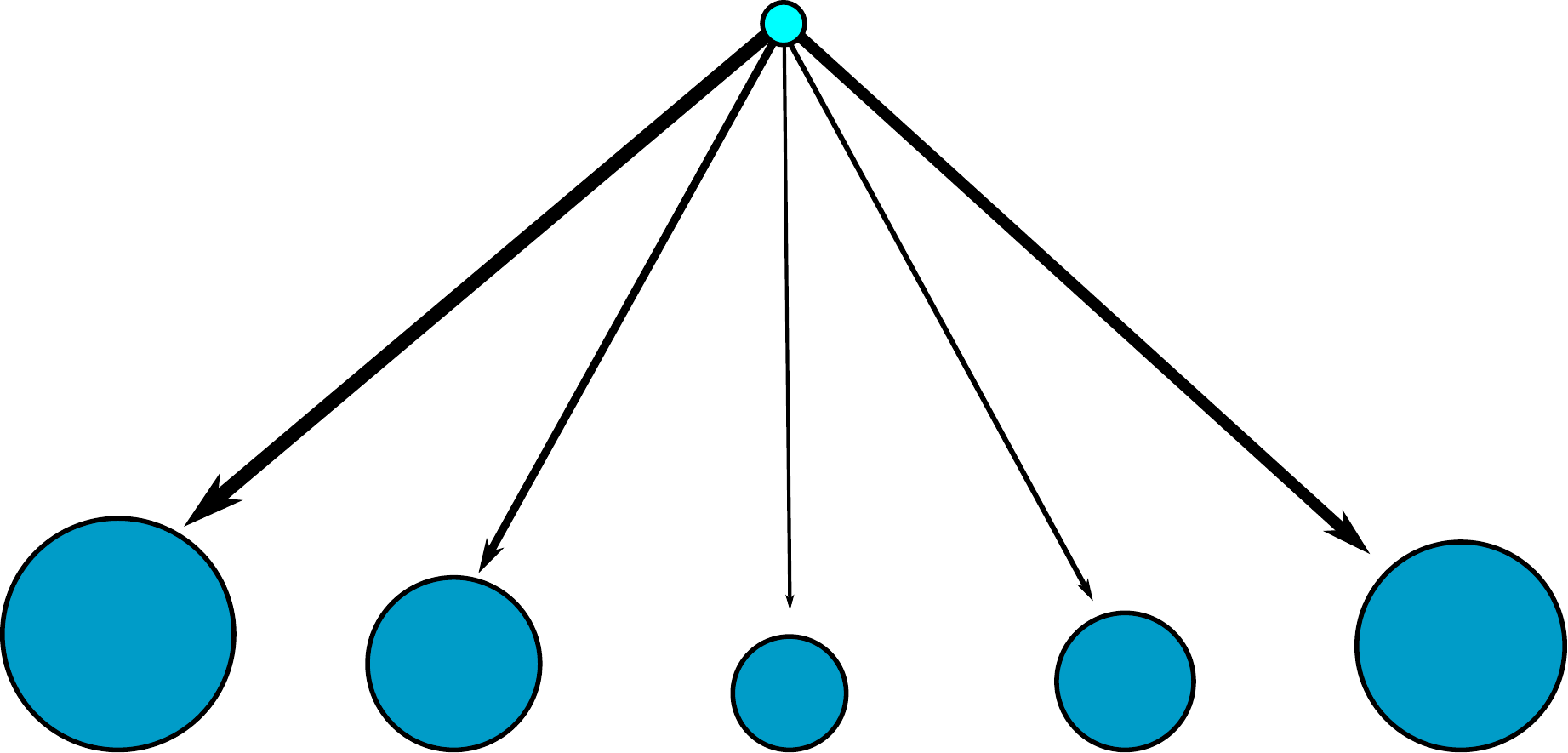}
}
\caption{An illustration of the Yule process. A probability 
that a newly created unit (top node) joins one of the existing 
communities (lower nodes) is proportional to the size of that 
community, indicated by the respective sizes of the nodes.  
}
\label{fig_yule}
\end{figure}
%%%%%%%%%%%%%%%%%%%%%%%%%%%%%%%%%%%%%%%%%

\subsection{Growth processes directed by importance measures}

One of the most applied principle, comparable to the success 
of {\it SOC} theory, is the {\it Yule Process} \citep{yule1925mathematical} 
or the ``rich-gets-richer'' mechanism, which was originally 
introduced to explain the power-law distribution of sizes of biological 
taxa. Later other researchers adapted and generalized the {\it Yule process} 
for the power-law scaling observed in various other systems. 
Today the {\it Yule process} goes by different names, 
for example it is known as {\it Gibrat's law} when applied to 
the distribution of city sizes \citep{eeckhout2004gibrat}, 
the {\it cumulative advantage} for the distribution of 
paper citations \citep{price1976general,redner1998popular}, 
the {\it preferential attachment} when modeling the scale-free 
structure of real-world networks 
\citep{newman2001clustering, dorogovtsev2000structure},
such as number of links to pages on the world wide web 
\citep{barabasi1999emergence, gros2012neuropsychological}. 

These models describe the dynamic growth of a system
which is biased by the size of existing units, as 
illustrated in Fig.~\ref{fig_yule}. The
system being a collection of interacting objects ({\it e.g.} 
cities, articles, web pages, people, etc.), where 
new objects are created from time to time, the number
of objects thus increasing continuously. To each object 
one relates a quantity representing its importance,
for example city sizes, the number of citations (for 
scientific articles), the number of links (for webpages), etc. 
It can be shown that the tail of this quantity follows a 
power-law distribution if the growth rate of this importance
measure is assumed to be proportional to its current value 
\citep{newman2005power,gros2010complex}. 
For example, the probability that a paper gets cited is 
higher if that paper has already many citations, the probability
of adding a link to a webpage is high if the webpage is well known,
{\it i.e.} if it has already many incoming links. Thus, this principle 
can be used to explain the scaling behavior of any system which 
seems to incorporate such a growth process, where the growth 
rate is biased locally by the importance of the respective node.

%%%%%%%%%%%%%%%%%%%%%%%%%%%%%%%%%%%%%%%%%
\subsection{Balancing competing driving forces, the coherent noise model}
%%%%%%%%%%%%%%%%%%%%%%%%%%%%%%%%%%%%%%%%%
%%%%%%%%%%%%%%%%%%%%%%%%%%%%%%%%%%%%%%%%%%%
\begin{figure}[t]
\centerline{
\includegraphics[width=0.7\textwidth ]{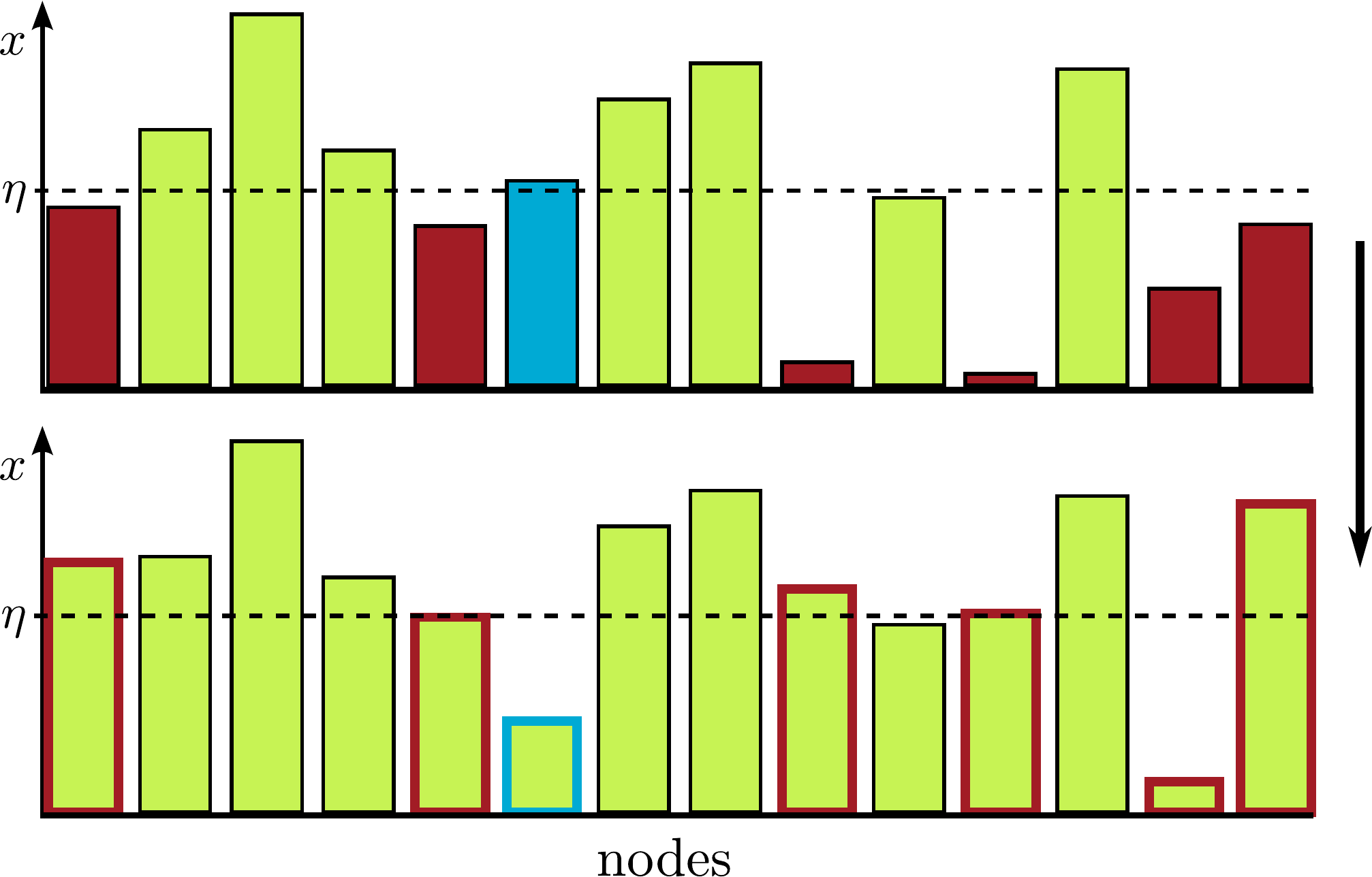}
}
\caption{An illustration of the {\it coherent noise} 
mechanism. An external stress $\eta$ affects the 
nodes whose thresholds are smaller then $\eta$ 
(red, top diagram).  The thresholds of the affected nodes 
are uniformly re-assigned (stroked red, bottom diagram). 
A small fraction of randomly drawn nodes (blue, top diagram)
receives in addition a new, randomly selected, threshold
(stroked blue, bottom diagram).  
}
\label{fig_coherent_noise}
\end{figure}
%%%%%%%%%%%%%%%%%%%%%%%%%%%%%%%%%%%%%%%%%

A dynamical system may organize itself towards criticality
as the result of balancing competing driving forces,
as discussed in the context of absorbing state transitions
in Sect.~\ref{subsec:absorbing}. Generalizing this concept 
one can consider the effect of competing driving forces on 
the dynamics of the resulting state.

An interesting class of models with competing drives are
random barrier models. An example is the Bak and Sneppen model
\citep{bak1993punctuated}, which is a model for co-evolutionary 
avalanches. In this model one has barriers $x_i\in[0,1]$ which represent
obstacles to evolutionary changes. At every time step
the lowest barrier is removed, corresponding to an
evolutionary process of species $i$ and reset to a
random value. The barriers $x_j$ of certain number of other 
species will also change and their barrier values will be 
reset randomly. The resulting state is critical and it can 
be related to a critical branching process \citep{gros2010complex}
(see \ref{sec:branching}).

In the Bak and Sneppen model there are two competing 
driving forces, the removal of low barriers and the
homogeneous redistribution of barrier levels. Another
model with an equivalent set of driving forces,
which we will now discuss briefly, has been termed 
``coherent noise model'' \citep{newman1996avalanches}.
The two steps of the time evolution, illustrated in
Fig.~\ref{fig_coherent_noise}, correspond to
an external driving and an internal dissipative
process respectively.
\begin{itemize}
\item All barriers below a randomly drawn stress 
      level $\eta$ are removed and uniformly 
      re-assigned (external forcing).
\item A certain fraction $f\in[0,1]$ of barriers 
      is removed anyhow and uniformly re-assigned
      (internal dissipation).
\end{itemize}
The coherent noise model has a functional degree
of freedom, the distribution $\rho_s(\eta)$ for
the stress levels, which is generally assumed to
be monotonically decreasing, with low stress levels
being more likely than larger ones. The distribution
$p(x,t)$ of barrier levels $x\in [0,1]$ will reach
a steady state, resulting from the competition of above 
two driving forces. The time evolution is given by
\begin{eqnarray*}
p(x,t+1) &=& \int_0^1  \rho_s(\eta) p(x,t)\Theta(x-\eta) d\eta
          - fp(x,t)
         \\ &+& 
\int_0^1  dx\int_0^1 \rho_s(\eta)p(x,t)
                \big[1-\Theta(x-\eta)\big]d\eta
          + f
\label{chap3:coherent_noise_time}
\end{eqnarray*}
where the terms in the second line enforce the conservation
of the number of barriers, $\int p(x,t+1)dx=\int p(x,t)dx$,
and where $\Theta$ is the Heaviside step function. The
equilibrium barrier distribution $p(x)\equiv p(x,t)=p(x,t+1)$ 
is then given by
\begin{equation}
p(x) = \frac{c}{1+f-P_s(x)},
\qquad\quad
P_s(x) = \int_0^x\rho_s(\eta)\,d\eta~,
\label{chap3:coherent-noise-solution}
\end{equation}
where $c$ is an appropriate normalization constant.
All barriers would pile up at the maximal barrier 
level in the absence of dissipation $f\to0$. A 
non-trivial distribution results only when both
external forcing and internal dissipation are active,
the steady-state solution is structureless if only
the internal redistribution of barriers $\propto f$ would
be active, the reason why one can consider this process 
to be analogous to friction in physics.
The steady-state barrier distribution 
(\ref{chap3:coherent-noise-solution}) looks otherwise 
unsuspicious, not showing any obvious signs of criticality. 
A phase transition, and an eventual self-organization towards
criticality, is in any case not expected for the coherent
noise model due to the absence of agent-agent interactions.
However, the resulting distribution of event sizes $s=\int_0^\eta p(x)dx$ 
shows an intermediate region of power-law scaling, 
and a large event is followed by a series of 
smaller aftershocks with power-law scaling \citep{sneppen1997coherent}.

The coherent noise model was used initially to describe the 
properties of mass extinctions observed in fossil records 
\citep{newman1997model}. It was also considered as a model 
of earthquakes, describing the properties of aftershocks 
\citep{wilke1998aftershocks,celikoglu2012earthquakes}, 
and used for the prediction of aftershocks 
\citep{sarlis2012predictability}. Recently, 
\citet{melatos2009superfluid} applied the coherent noise model 
in a study of pulsar glitches. Interestingly, the model is
quite sensitive to initial conditions 
\citep{ergun2005sensitivity}; a property in 
common with the Bak-Sneppen model.

%%%%%%%%%%%%%%%%%%%%%%%%%%%%%%%%%%%%%%%%%
\subsection{Highly optimized tolerance}
\label{subsec:hot}
%%%%%%%%%%%%%%%%%%%%%%%%%%%%%%%%%%%%%%%%%

The mechanism of {\it highly optimized tolerance} 
({\it HOT}) is motivated by the fact that most complex 
systems, either biological or man-made, consist of many 
heterogeneous components, which often have a complex 
structure and behavior of their own \citep{carlson99a}. 
Thus, real complex systems often exhibit self-dissimilarity 
of the internal structure rather then self-similarity, 
which would be expected if the self-organization 
toward a critical state would be the sole organizational 
principle \citep{carlson00a, carlson2002complexity}. 

Self-similarity is a property of systems which have similar 
structures at different scales, a defining property 
of fractals. It is not uncommon to find fractal
features in living organisms, in specific cells or tissue 
structures \citep{weibel1991fractal}. Self-similarity does 
however exist, for real-world systems, 
only within a finite range of scales. Cell shapes and functions 
differ substantially from one organ to another and there are 
highly specialized non-similar units within individual cells.
Analogous statements can be made for the case of 
artificial systems, such as the Internet or computers. 
Actually, the diversity in the components of complex systems 
is needed to provide a robust performance in the presence of 
uncertainties, either arising from changes in the behavior of the 
system components or from changes in the environment. The 
balance between self-similarity and diversity hence comes not from 
a generic generating principle, but from the driving design process. 
Optimal design is achieved, for the case of living organisms, 
through natural selection and for the man-made complex systems, 
through human intervention.

Both man-made and biological complex systems can show a surprising 
sensitivity to unexpected small perturbations, if they had not been 
designed or evolved to deal with them. To give an example, the 
the network of Internet servers is very robust against the variations in 
internet traffic volume, nevertheless highly sensitive to bugs 
in the network software. Likewise, complex organisms may be highly robust 
with respect to environmental variations and yet may easily die if 
the regulatory mechanism, which maintains this robustness, is 
attacked and damaged by microscopic pathogens, toxins or injury. 
A substantial variety of complex systems is hence characterized by
a property one may denote as ``robust-yet-fragile'' 
\citep{carlson99a, carlson00a, carlson2002complexity}. 

\citet{carlson99a} have argued, using simple 
models, that optimization of a design objective, in the 
presence of uncertainty and specified constrains, might 
lead to features such as high robustness and resilience 
to "known" system failures,
high sensitivity to design flaws and unanticipated 
perturbations, structured and self-dissimilar configurations, 
and heavy-tail distributions \citep{doyle00a}.
Depending on the specific objectives which are optimized,
and their relation to the system constrains, the exact 
scaling can follow a power law or some other heavy-tailed 
distribution \citep{carlson2002complexity}. The 
main difference between the {\it SOC} and the {\it HOT} 
mechanism is their explanation of large, possibly
catastrophic events. Large events arise, within {\it SOC},
as a consequence of random fluctuations which get amplified
by chance. As for {\it HOT}, large events are caused by a 
design which favors small, frequent losses, having
rather predictable statistics, over large losses 
resulting from rare perturbations.
%%%%%%%%%%%%%%%%%%%%%%%%%%%%%%%%%%%%%%%%%%%%%%%%%%
\begin{figure}
\centerline{
\includegraphics[width=0.7\textwidth]{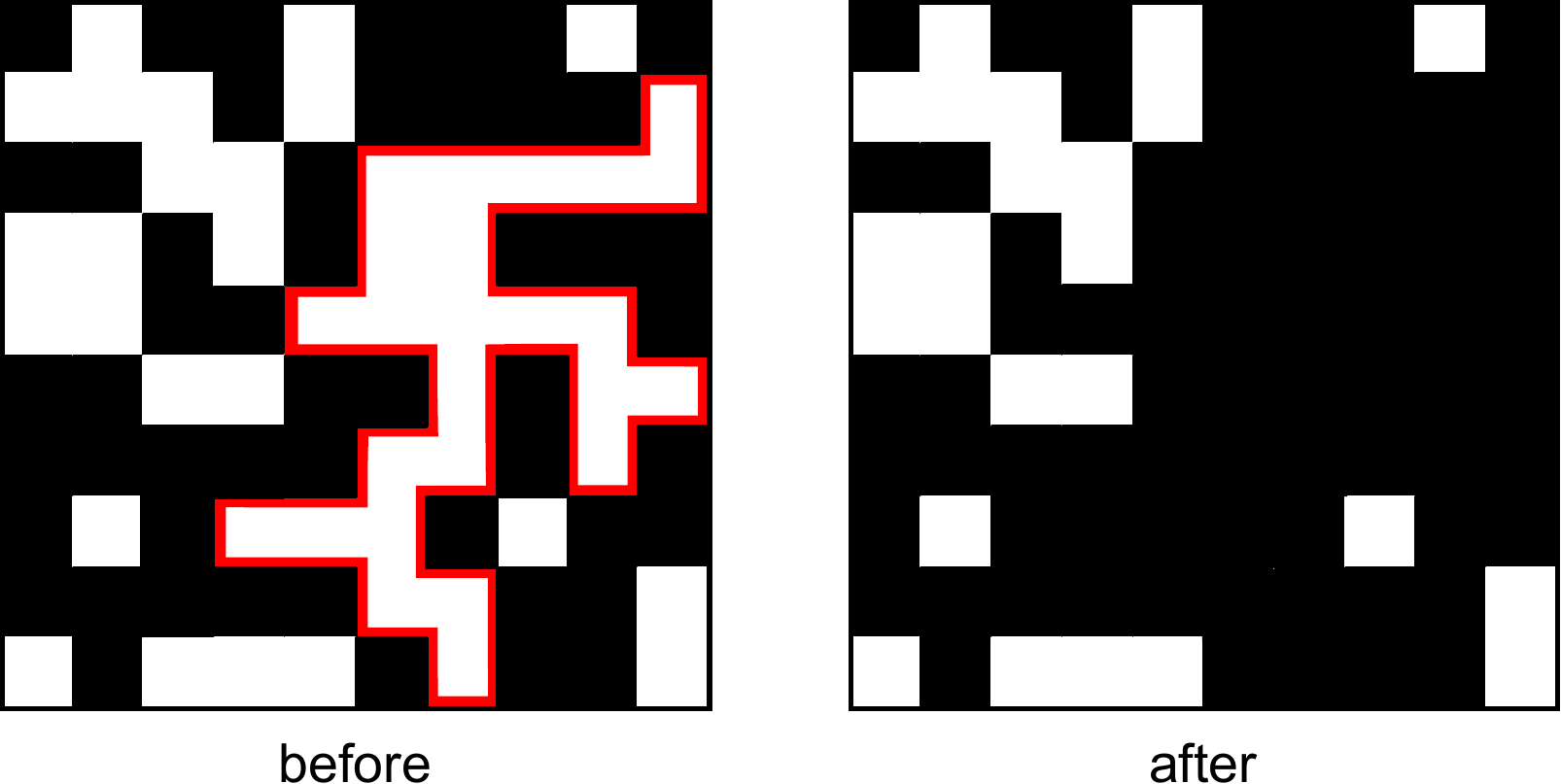}
}
\caption{Illustration of a site percolation process on a 
$10\times 10$ regular lattice for $p = 1/2$; occupied nodes 
are colored white. Before the perturbation targeting the 
largest cluster, which is shaded in red (left) and after the 
perturbation leading to the removal of all occupied nodes 
within the perturbed cluster (right).}
\label{fig_clusters}
\end{figure}
%%%%%%%%%%%%%%%%%%%%%%%%%%%%%%%%%%%%%%%%%%%%%%%%%% 

\subsubsection{HOT site percolation}
The {\it HOT} mechanism can be illustrated with a model based 
on two dimensional site percolation \citep{carlson00a}. This type 
of model is often taken as a starting point for describing the spreading 
of fire in forest patches or the spreading of epidemics through social 
cliques. It also serves, more generally, as a model for energy dissipation. 
Considering the reaction of the system under a disruption, one
is interested in these cases in the number of trees surviving 
a fire outbreak, in the number of individuals unaffected by an epidemic, 
and in the amount of energy preserved within the system. For 
{\it HOT} one considers optimized percolation processes reducing
to the classical {\it Bernoulli} percolation when no optimization 
is performed.

For the classical percolation problem, in the absence of any
optimization procedure, lattice sites are occupied (with a particle, 
a tree, etc.) with probability $\rho$ and empty with probability 
$1-\rho$. Two sites are connected, on a square lattice
with linear dimensions $L$ \citep{carlson00a},
if they are nearest neighbors of each others and a 
group of sites is connected whenever 
there is a path of nearest neighbors between all sites
of the cluster (see Fig.~\ref{fig_clusters}).
The cluster sizes are exponentially distributed if the average 
density $\rho$ of occupied nodes is below the critical density 
$\rho_c$. At criticality the characteristic cluster 
size diverges and the cluster size distribution follows a 
power law. For densities above criticality there is a finite 
probability of forming an infinite cluster covering
a finite fraction of the system, even in the thermodynamic
limit. The probability that a given occupied site is connected to an 
infinite cluster is the percolation order parameter $P_{\infty}(\rho)$, 
which is zero for $\rho \leq \rho_c$, and monotonically increasing 
from zero to one for $\rho>\rho_c$.

One now considers clusters of occupied sites to be subject to 
perturbations (e.g.\ a spark when considering forest fires) 
that are spatially distributed with probability $f(\vec{r})$. 
When a perturbation is initiated 
at the location $\vec{r}$ of the lattice, the perturbation 
spreads over the entire cluster containing the site 
originally targeted by the attack, changing the status of all
sites of the cluster from occupied to unoccupied (the trees burn 
down), as illustrated in Fig.~\ref{fig_clusters}. On the other hand, if 
the perturbed site is empty, nothing happens. The system
is most robust if, on average, as few sites as possible are affected
by the perturbation. The aim of the optimization process
is then to optimally distribute particles onto the 
lattice, for a given average density of occupied sites.
One hence defines the yield $Y$ of the optimization process as the
average fraction of sites surviving an attack. Optimization 
of the yield can be achieved, through an evolutionary process, 
by increasing continuously the density of particles.
\begin{itemize}
\item Starting with a configuration of $N_p$ particles
      one considers a number $D$ of possible states of
      $N_p+1$ particles generated by adding a single
      particle to the present state.
\item One evaluates the yield $Y$ for all $D$ prospective 
      new states by simulating disruptions, distributed by
      $f(\vec{r})$. The state with the highest yield is
      then selected.
\end{itemize}
The optimization parameter, for this algorithm, is 
in the range $0\le D\le(N-N_p)$, where $D=0$ corresponds
to no optimization, {\it i.e.} to classical percolation. Increasing
the number $D$ of trial states will, in general, lead to
an increase in performance.

%%%%%%%%%%%%%%%%%%%%%%%%%%%%%%%%%%%%%%%%%%%%%%%%%%
\begin{figure}
\centerline{
\includegraphics[width=0.8\textwidth]{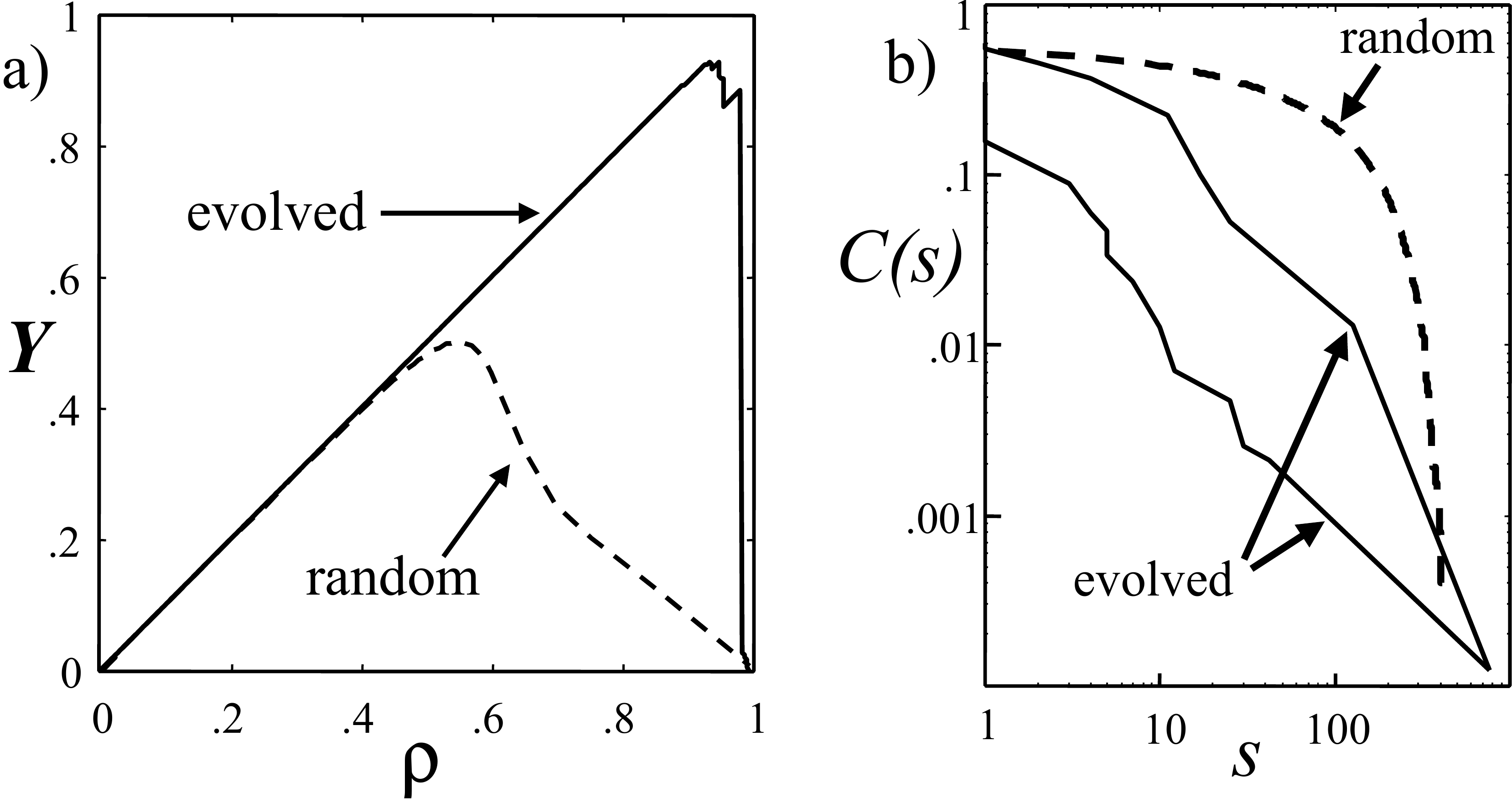}
}
\caption{Comparison between evolved {\it HOT} states and 
random percolation. (a) Yield {\it vs.} density in the 
case of random percolation and evolved lattice configuration.
(b) Cumulative distribution of event sizes $C(s)$ at the point 
of maximum yield of the evolved lattice configuration
(log-log plot), for the case of random percolation and 
for two evolved configurations.
[Courtesy of J.~Doyle \citep{carlson99a}.]}
\label{fig_yield}
\end{figure}
%%%%%%%%%%%%%%%%%%%%%%%%%%%%%%%%%%%%%%%%%%%%%%%%%%  

In Fig.~\ref{fig_yield} the yield $Y$ is shown as a function of the 
mean density $\rho$, both for the case of random percolation and for
the state evolved through the {\it HOT} process.
The yield peaks near the percolation threshold $\rho_c = 1/2$, for 
random percolation, decreasing monotonically to zero for $\rho>\rho_c$,
a behavior easily understood when considering the thermodynamic limit 
$L \rightarrow \infty$. In the thermodynamic limit there are two
possible outcome for an perturbation. Either the perturbation hits,
with probability $P_\infty(\rho)$, the infinite clusters,
or, with probability $1-P_\infty(\rho)$, some other finite 
cluster or an empty site. In the first case a finite 
fraction $P_\infty(\rho)$ of occupied sites are removed,
in the second case only an intensive number of sites:
\begin{equation}
Y(\rho) =  P_{\infty}(\rho)(\rho - P_\infty(\rho))
+ (1-P_{\infty}(\rho))\rho = \rho - P_{\infty}^2(\rho)~,
\end{equation}     
the yield is directly related to the order parameter
when no optimization is performed. A yield close to
the maximally achievable value $\rho$ can, on the other
side, be achieved when performing optimization with
an optimization parameter $D$ close to its maximal value.
The resulting distribution of occupied sites is highly
inhomogeneous, many small clusters arise in regions
of high attack rates $f(\vec{r})$, regions with low 
disruption rates are, on the other side, characterized
by a smaller number of large clusters. The {\it HOT} 
state reflects the properties of the distribution 
$f(\vec{r})$ and is hence highly sensitive to changes of 
$f(\vec{r})$. The distribution of clusters is, in contrast,
translationally invariant in critical state $\rho = \rho_c$ 
when no optimization is performed, and independent from $f(\vec{r})$. 
This model of optimized percolation hence illustrates 
the ``robust-yet-fragile'' principle.

%%%%%%%%%%%%%%%%%%%%%%%%%%%%%%%%%%%%%%%%%%%%%%%%%%
\begin{figure}
\centerline{
\includegraphics[width=0.8\textwidth]{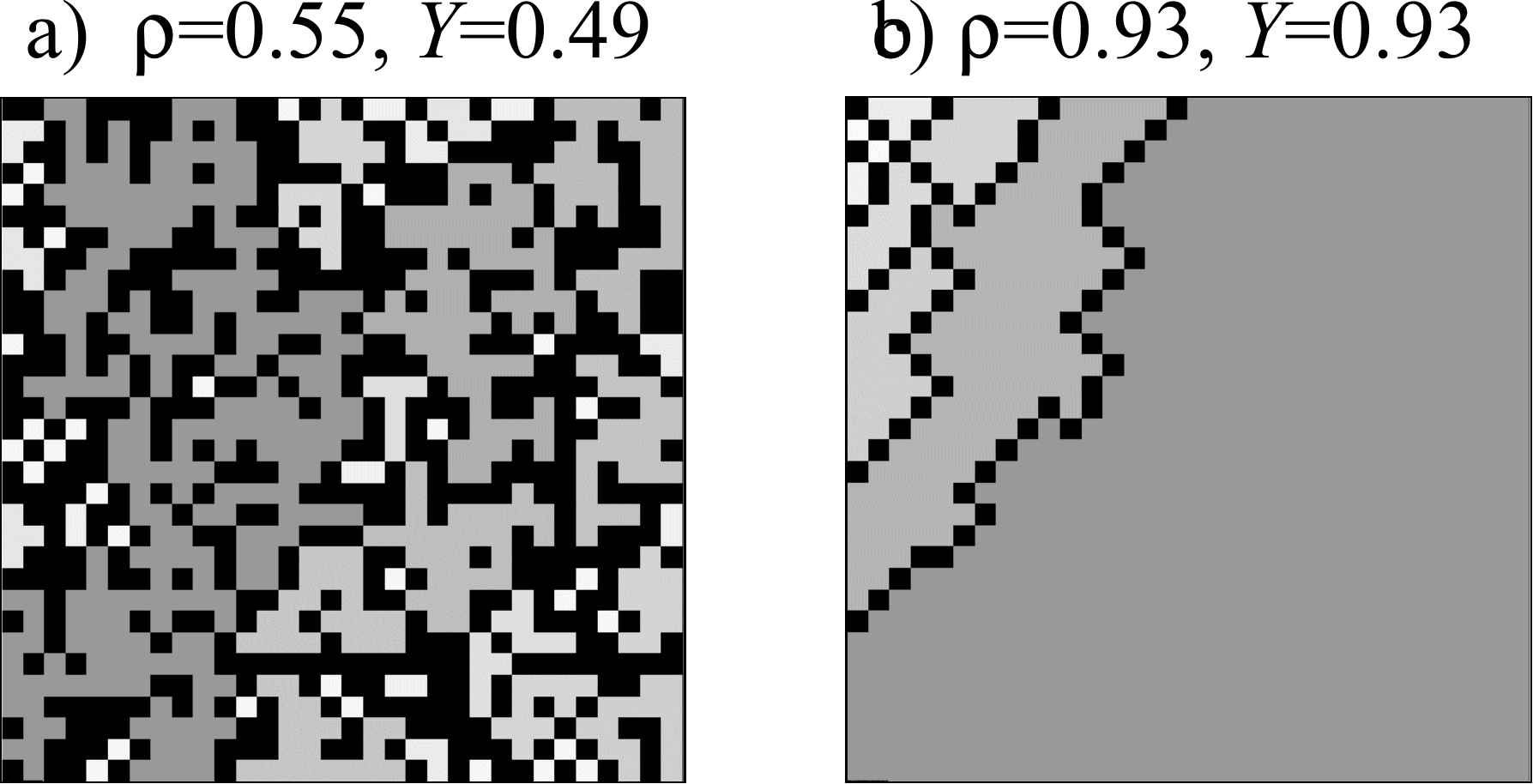}
}
\caption{Sample of percolation configuration on 
a $32\times 32$ lattice for (a) random percolation 
near $\rho_c$ and (b) a {\it HOT} state at maximal yield 
obtained by evolving lattice configurations. Unoccupied 
sites are black and clusters are gray, with darker 
shades indicating larger clusters. 
The designed lattice percolation was generated for the 
perturbation probability $f(\vec{r}) = f(r_1)f(r_2)$, 
where $f(x) = \exp(-\left(\frac{m_x+(x/N)}{\sigma_x}\right)^2)$, which 
were peaked at the upper left corner of the lattice 
[Courtesy of J.~Doyle \citep{carlson99a}].}
\label{fig_config}
\end{figure}
%%%%%%%%%%%%%%%%%%%%%%%%%%%%%%%%%%%%%%%%%%%%%%%%%%

%-----------------------------
\subsubsection{Fat tails and the generic HOT process}
%-----------------------------
It is not evident, at first sight, why the procedure of highly
optimized tolerance should lead to power-law scaling, or to fat 
tails in general. The emergence of power-law scaling from 
the {\it HOT} mechanism can however be understood by considering 
an abstract optimization setup as described by \citet{carlson99a}. 
The yield is defined as
\begin{equation}
Y(\rho) = \rho - \frac{1}{L^2}E[s],
\end{equation}
where $E[s]$ denotes the expectation value of event sizes
for a fixed distribution of perturbations $f(\vec{r})$. 
The yield $Y$ is maximal when a disruption triggers events
of minimal sizes.

For the case of optimized percolation, discussed in the
previous section, the event size $s$ was assumed to
be identical to the area $A(\vec{r})$ affected by a
disruption happening at $\vec{r}$. In a larger context
one may be interested not to minimize directly the
affected area $A$, but some importance measure $s$ of the
event, with the relevance $s$ of a given event
being a nonlinear function of the primary effect,
\begin{equation}\label{eq_hot_cost}
E[s] = \int f(\vec{r})s(\vec{r})\d\vec{r},
\qquad\quad s=\big(A(\vec{r})\big)^{\gamma}
~,
\end{equation}
where a polynomial dependence $s=A^\gamma$ has been assumed,
with $\gamma>0$. For the case of optimized percolation 
the yield $Y[\rho]$ is evaluated for fixed particle 
density $\rho$.  More generally, one can consider a 
constraint function $R(\vec{r})$ such that
\begin{equation}\label{eq_hot_constrain}
\int R(\vec{r})\d\vec{r} = \kappa
\end{equation}  
needs to be kept constant. Available resources are finite, 
$\kappa<\infty$, and need to be utilized optimally.
Real-world examples for resources are fire breaks 
preventing wildfires, routers and DNS servers 
preventing large failures of the Internet traffic and 
regulatory mechanisms preventing failure amplification 
in organisms. Allocating more resources to some location, 
to limit the size of events, will generically lead to a 
reduction in the size of the area affected by a disruption.  
One may thus assume that the area locally affected by an 
event is inversely related to the local density of resource 
allocation, that is, $A(\vec{r}) = 
(R(\vec{r}))^{-\beta}$, with $\beta$ being a positive 
constant related to the dimensionality of the system. 

The {\it HOT} state in this abstract system is obtained by 
minimizing the expected cost (Eq.~(\ref{eq_hot_cost})) subject 
to the constraint on available resources 
(Eq.~(\ref{eq_hot_constrain})), together with
$A = R^{-\beta}$. The optimal state is 
found by applying the variational principle and solving 
\begin{equation}
\delta\int\Big[f(\vec{r})\big(R(\vec{r})\big)^{-\gamma\beta}-
\lambda R(\vec{r})\Big]\d\vec{r} \equiv 0~,
\end{equation}   
where $\lambda$ is a Lagrange parameter. The variation,
relative to all possible resource distributions $R(\vec{r})$,
yields
\begin{equation}
f(\vec{x}) \sim \big(R(\vec{x})\big)^{\gamma\beta+1} 
\sim \big(A(\vec x)\big)^{-(\gamma + 1/\beta)} 
\sim \big(A(\vec x)\big)^{-\theta},
\qquad\quad
\theta = \gamma + 1/\beta~.
\end{equation}
This relation lead to $A\sim f^{-1/\theta}$, the larger the
event probability $f$, the smaller the affected area $A$.
The cumulative probability distribution $C(A)$ of observing an 
event which spreads over an area larger or equal than $A$, 
in the case of an optimal {\it HOT} state, becomes
\begin{equation}
C(A) = \int_{A(\vec{r})>A} f(\vec{r})\d\vec{r} = 
\int_{f(\vec{r}) < A^{-\gamma}}
f(\vec{r})\d\vec{r}~.
\end{equation} 
Although not all $f(\vec{r})$ will result in a scale-free 
scaling of event sizes, there is however a broad class 
of distributions leading to heavy tails in $C(A)$ and 
consequently in the distribution $P(A)$ of event areas. 
For example, in the one dimensional case 
an exponential, a Gaussian and a power-law distributed
$f(r)$ result in a heavy-tailed distribution of events.
One can show, in addition, that similar relations also 
hold for higher dimensional systems \cite{carlson99a}.
An example of a perturbation probability $f$ which does not result 
in heavy-tailed event sizes would be a uniform distribution 
or, alternatively, perturbations localized within a small 
finite region of the system.    

The above discussion of the {\it HOT} principle does not 
take into account the fact that real-world complex systems 
are, most of the time, part of dynamical environments, 
and that perturbations acting on the system will therefore not
be stationary, $f=f(\vec{r},t)$. The {\it HOT} principle 
can be generalized to the case of a time 
dependent distribution of disruptions $f(\vec{r}, t)$. 
A system can still be close to an optimal state in a
changing environment when constantly adapting to the 
changes and if the changes are sufficiently slow, that is,
if a separation of time scales exists \citep{zhou2000dynamics}. 
An adaptive {\it HOT} model was used 
by \citet{zhou2002mutation} to explore different scenarios 
for evolution and extinction, such as the effects of different 
habitats on the phenotype traits of organisms, the effects 
of various mutation rates on adaptation, fitness and diversity, 
and competition between generalist and specialist organism. 
In spite of using a very abstract and simple notion of organisms 
and populations, these studies were successful in capturing 
many features observed in biological and ecological systems 
\citep{zhou2005evolutionary}.

\section{Branching processes}
\label{sec:branching}

One speaks of an avalanche when a single event causes 
multiple subsequent events. Similar to a snowball rolling 
down a snowfield and creating other toppling snowballs. 
Avalanches will stop eventually, just as snowballs 
won't trundle down the hill forever. At the level of
the individual snowballs this corresponds to a
branching process---a given snowball may stop
rolling or nudge one or more downhill snowballs to
start rolling. The theory of random branching 
processes captures such dynamics of cascading events. 
First, we will discuss the classical
stochastic branching process and its relation to
{\it SOC}, branching models are critical when on the 
average the number of snowballs is conserved. 
Second, we will discuss vertex routing models 
for which local conservation is deterministic.

%%%%%%%%%%%%%%%%%%%%%%%%%%%%%%%%%%%%%%
\subsection{Stochastic branching}
%%%%%%%%%%%%%%%%%%%%%%%%%%%%%%%%%%%%%%

A branching or multiplicative process is formally defined 
as a Markov chain of positive integer valued random variables 
$\{Z_0, Z_1, \ldots\}$. One of the earliest application of the 
branching processes concerned the modelling of the evolution 
of family names, an approach known as the Galton-Watson 
process \citep{gros2010complex}. In this context $Z_n$ 
corresponds to the number of individuals in the $n$th generation 
with the same family name. More recently, the theory of branching 
processes was applied in estimating the critical exponents of 
sandpile dynamics, both for regular lattices \citep{alstrom1988mean}
and for scale-free networks \citep{goh2003sandpile}. In a
typical application branching processes are considered 
as mean-field approximations to the sandpile dynamics, 
obtained by neglecting correlations in the avalanche 
behavior \citep{zapperi1995self}.

%%%%%%%%%%%%%%%%%%%%%%%%%%%%%%%%%%%%%%%%%%%%%%%%
\begin{figure}
 \centerline{
\includegraphics[height=0.30\textwidth,angle=0]{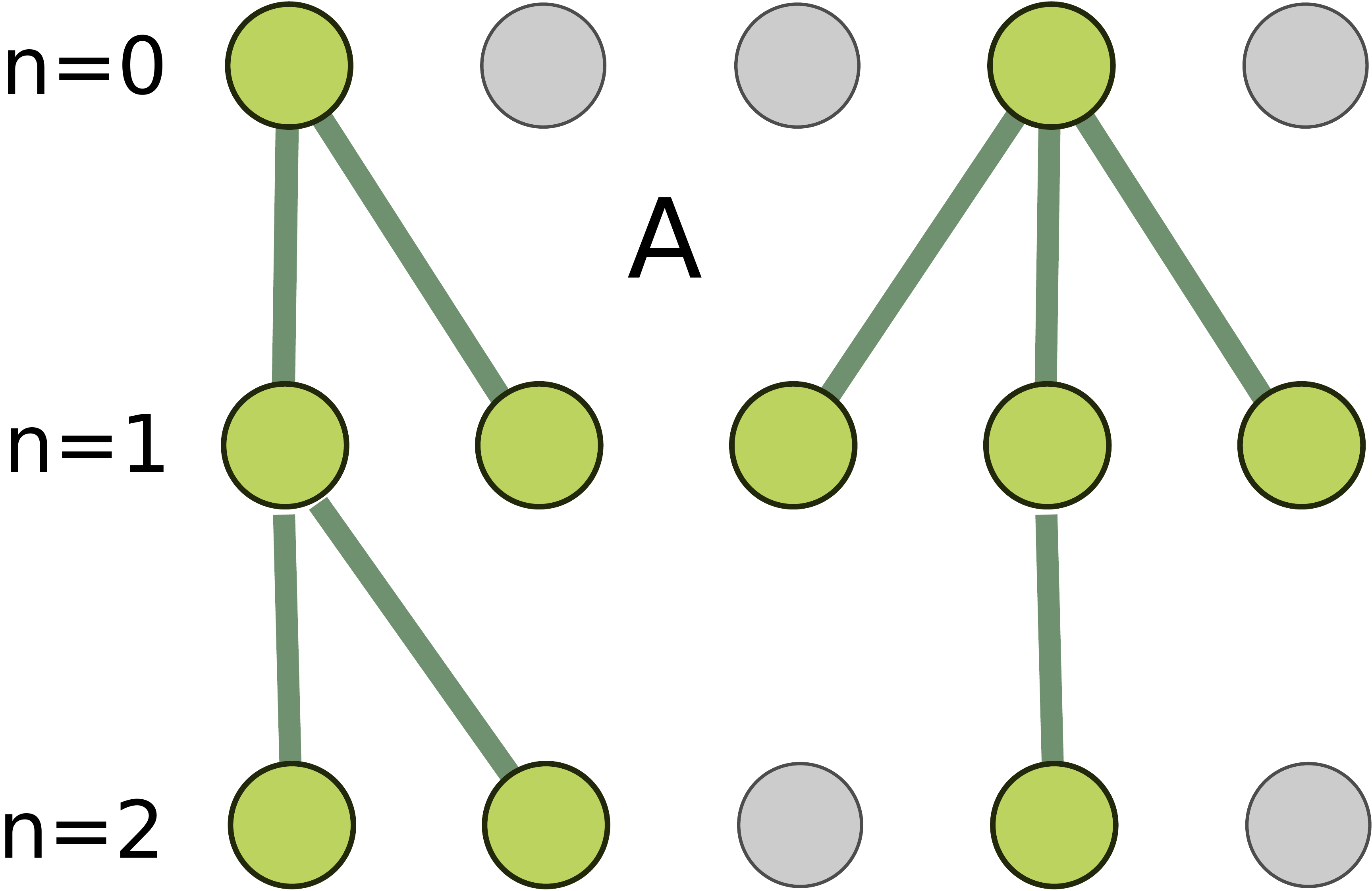}
\hspace{8ex}\ \ 
\includegraphics[height=0.30\textwidth,angle=0]{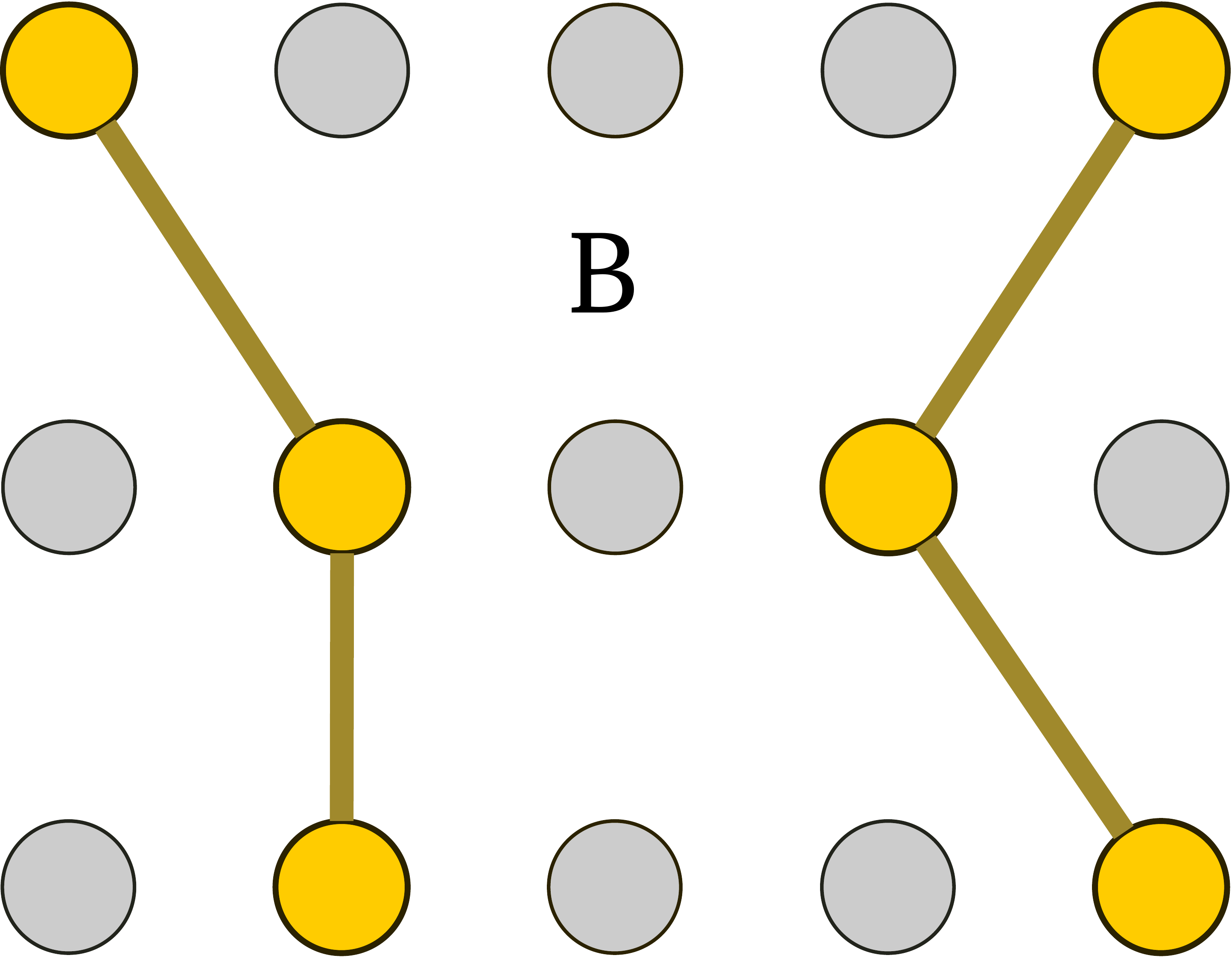}
}
\caption{Examples for branching processes (A, left) and 
routing processes (B, right), where $n$ denotes a time step.}
\label{fig_routing}
\end{figure}
%%%%%%%%%%%%%%%%%%%%%%%%%%%%%%%%%%%%%%%%%%%%%%%%

More abstractly, a random variable $Z_n$ represents the number 
of ``particles" present at iteration step $n$ generating 
a new generation of $Z_{n+1}$ descendents at step 
$n+1$ (see Fig.~\ref{fig_routing}). We denote with $p^{(n)}_k$ 
the probability that a single particle at time step $n$ 
generates $k$ offsprings at time step 
$n+1$ and with $P_n(Z_n)$ the probability of finding $Z_n$ particles
after $n$ iterations. One defines with
\begin{equation}
f_{n}(x) = \sum p^{(n)}_kx^k,
\qquad\quad
G_n(x) = \sum_{Z_n} P_n(Z_n) x^{Z_n}
\label{chap4:generating-functionals_defs}
\end{equation}
the corresponding generating functions $f_{n}(x)$ 
and $G_n(x)$ \citep{gros2010complex}.
A branching process may, in general, 
be time dependent, for a time-independent process 
$p^{(n)}_k\equiv p_k$ and $f_n(x)\equiv f(x)$. 
The recursion relation
\begin{equation}\label{eq_genfunc_particles}
G_n(x) = \sum_{Z_n} P_n(Z_n)x^{Z_n} 
= \sum_{Z_{n-1}}P_{n-1}(Z_{n-1})\left(f_{n-1}(x)\right)^{Z_{n-1}} 
= G_{n-1}(f_{n-1}(x))
\end{equation}
expresses the fact that branching processes are Markovian.
When using branching processes to study properties of {\it SOC} 
systems we are interested in the scaling of the cumulative 
number of offsprings $s= \sum Z_k$, corresponding to the
avalanche size (defined as the number of overall active 
sites), and in the duration $t$ of a branching process. 
An avalanche stops when no offsprings
are produced anymore, hence when $Z_{t}>0$ and $Z_{t+1} = 0$, 
which defines the duration $t$. 

The probability of having no particles left after
$n$ iterations is $q_n \equiv P_n(0) = G_n(0)$. One
defines with $q= \lim_{n \rightarrow \infty}q_n$ 
the overall extinction probability; a finite probability 
exists, for $q<1$, of observing infinitely long and infinitely 
large branching events. The regime $q<1$ is termed supercritical, 
while the critical and subcritical regimes are found when the
process extinction is certain, that is, $q=1$. The
extinction probability is hence a convenient measure
for characterizing the scaling regimes of branching processes.

The branching regimes are determined by the long term 
behavior of the average number of particles,
$$
E[Z_n] = \sum_{Z_n} P_n(Z_n) Z_n = G_n'(1)~.
$$ 
Defining with $\mu_{n} = \sum kp^{(n)}_k=f^{\prime}_{n}(1)$ 
the average number of offsprings generated by a single 
particle at time step $n$, one obtains the recursion relation
\begin{equation}
E[Z_n] = G_n^{\prime}(n)=f^{\prime}_{n-1}(1)G^{\prime}_{n-1}(1) = 
\mu_{n-1} E[Z_{n-1}]
=\mu_{n-1}\mu_{n-2}\cdot\cdot\cdot\mu_0 ~,
\label{chap3:E_recursion}
\end{equation}
when starting with a single particle, $E[Z_0]=1$.
Assuming that for large $n$ the expected number of particles 
scales as $E[Z_n] = e^{n\lambda}$, then for negative 
Lyapunov exponents $\lambda<0$ the expectation converges 
to zero, diverging on the other side for positive $\lambda>0$. 
Thus, $\lambda < 0$ is defined a subcritical branching 
process and $\lambda > 0$ the supercritical regime.
The Lyapunov exponent is given, through the
recursion relation (\ref{chap3:E_recursion}), as
\begin{equation}
\lambda =\lim_{n\to\infty}\left(\frac{1}{n}\ln E[Z_n]\right)
        =\lim_{n\to\infty}\left(\frac{1}{n}
\sum_{n=0}^{n-1} \ln \mu_n\right)~.
\label{chap3:lyapunov_lambda}
\end{equation}
The branching process is critical for $\lambda=0$.
For a time-independent branching process
one has $f_n(x) = f(x)$ and a fixed average number of
offsprings per particle, $\mu_n=\mu = f^{\prime}(1)$ for 
every $n$. Therefore, having $\mu = 1$ and $\ln\mu = 0$ 
at every iteration step is then a necessary condition
for the branching process to be critical. 
 
%%%%%%%%%%%%%%%%%%%%%%%%%%%%%%%%%%%%%%%%%%%%%%%%%%%%%
\begin{figure}
 \centerline{
\includegraphics[width=1\textwidth]{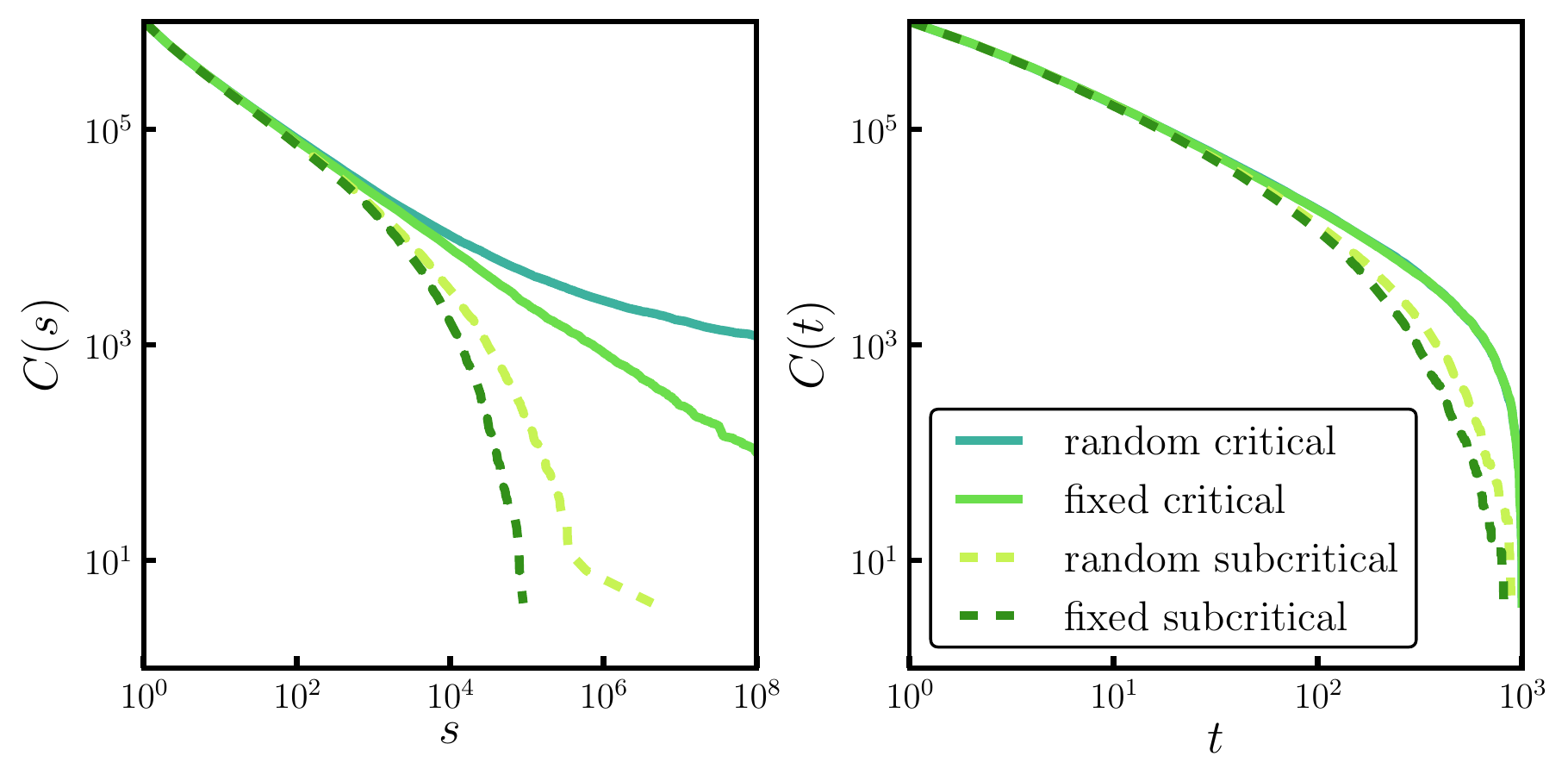}
}
\caption{Comparison of the complementary cumulative distributions
of sizes $s$ ($C(s)$, left) and durations $t$ ($C(t)$, right) 
for branching processes in fixed and random environments. 
The probability that a single particle generates 
$k$ offspring was set to a Poisson distribution $p_k = \mu_n^k e^{-\mu_n}/k!$.
At each time step one sets $\mu_n = e^{X_n}$, where
$X_n$ was drawn from a normal distribution 
$\mathcal{N}(\lambda, \sigma^2)$, with $\lambda = 0$ for the critical 
process, $\lambda = -0.01$ for the subcritical process, $\sigma = 0$ 
for fixed environment and $\sigma = 0.1$ for random environment.}
\label{fig_branch_changing}
\end{figure}
%%%%%%%%%%%%%%%%%%%%%%%%%%%%%%%%%%%%%%%%%%%%%%%%%%%%%

\citet{otter1949multiplicative} has demonstrated that 
in the case of fixed environments and a Poisson 
generating function $f(x)$ the tails of the distributions
$P(s)$ of avalanche sizes and durations $P(t)$ 
have the following scaling form:
\begin{equation}
P(s) \sim s^{-3/2}\mu^{s-1}e^{s(1-\mu)},
\qquad\quad
P(t)\sim t^{-2}\mu^{t-1}e^{t(1-\mu)}~.
\end{equation}   
The branching is critical for $\mu=1$, with the well-known
scaling exponents $3/2$ and $2$ for the avalanche size and
duration respectively.

The scaling behavior is more difficult to predict in 
the case of a changing or random environment. Consider 
an average number of offspring generated by a single particle 
which is given by $\mu_n = e^{X_n}$, where $X_n$ is drawn,
at each time step, from some probability distribution $\rho(x)$.
Again, the branching process is critical if $\lambda = 0$, that is, 
if $E_{\rho}[x] = 0$. Still, in contrast to fixed environment, 
the average number of particles $E[Z_n]$ fluctuates between infinity, 
$\lim_{n \rightarrow \infty}\sup \left(\ln E[Z_n] \right)= \infty$,
and zero, $\lim_{n \rightarrow \infty}\inf \left(\ln E[Z_n] \right)= -\infty$, 
where the supremum and infimum are taken over ensemble realizations 
\citep{vatutin2012total}. Furthermore, critical branching in random 
environments is a complex process and does not necessarily follow 
power-law scaling. \citet{vatutin2012total} has recently 
shown that, given a specific family of offspring generating 
function $f_n(x)$, the total size of the branching
process has logarithmic correction whereas the duration 
distribution still follows a typical power-law scaling.
In Fig.~\ref{fig_branch_changing} we present a comparison 
of the scaling behavior of critical and subcritical branching 
processes in fixed and random environments.

\begin{figure}[t]
\centerline{
\includegraphics[width=0.8\textwidth]{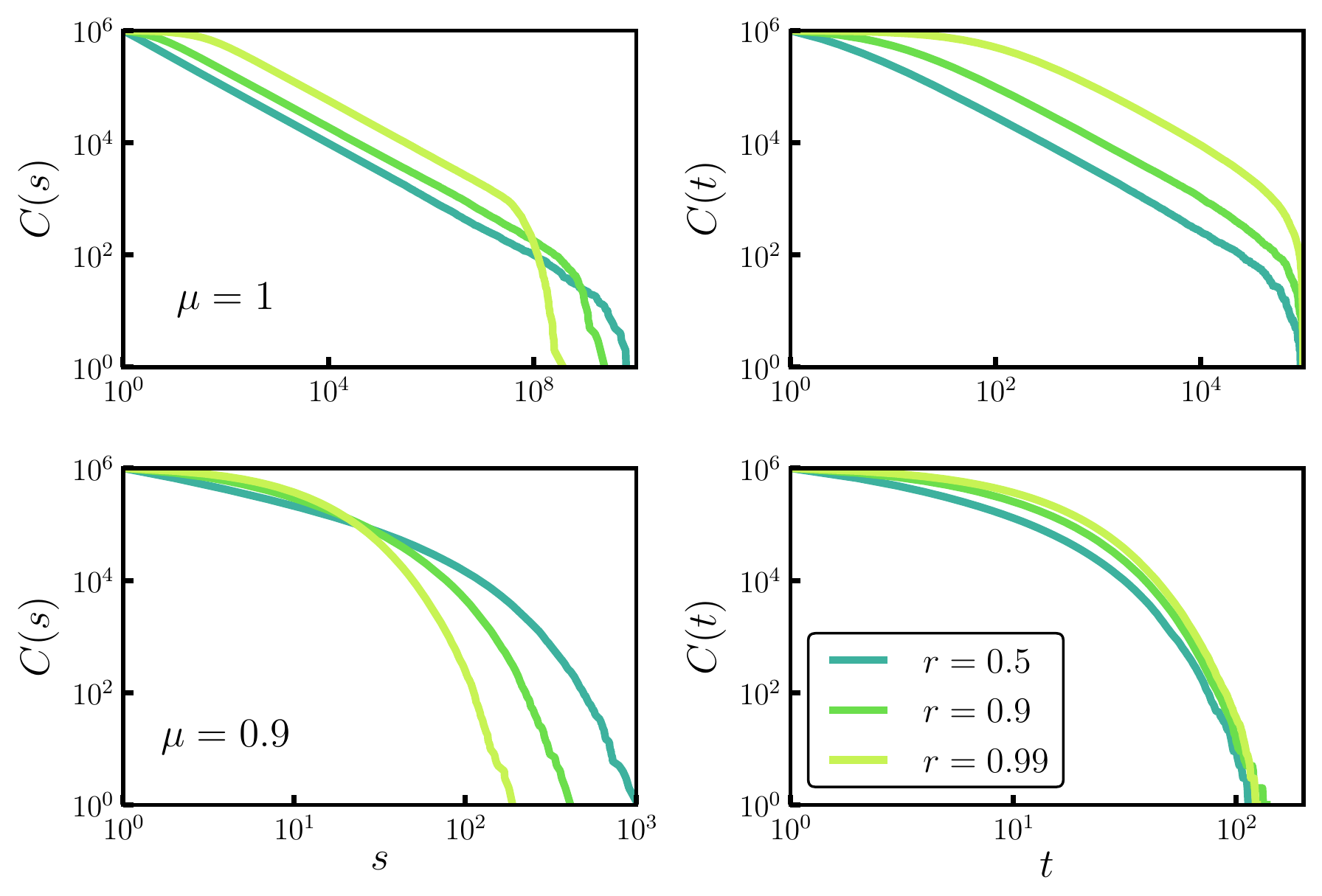}
}
\caption{The complementary cumulative distribution of 
sizes $C(s)$, and durations $C(t)$, of avalanches 
of a critical and subcritical branching process on a $d = 5$ 
dimensional lattice. The probability of activating a $j$th neighbor
of the $i$th active node is given as $p_{ij} = \alpha r(1-r)^{k-1}$, 
where $\alpha = \mu/\sum_{k=1}^{k=2d}p_{ij}$. Thus, each active node 
on average activates $\mu$ neighbors. For critical branching (top) 
$\mu = 1$, for subcritical branching (bottom) $\mu = 0.9$~. 
Increasing $r$ leads to an increase in the probability of 
activating only the first neighbor, hence in the limit 
$r \rightarrow 1$ only one node is active in each time step,
the process becomes deterministic.}
\label{fig:branch_directed}
\end{figure}

When mapping a real-world phenomenon to a branching process,
it is assumed that the phenomenon investigated propagates 
probabilistically. For example, when considering the propagation 
of activity on a finite network, each of the neighbors of 
an active node may be activated with some probability, 
say $p_{ij}$. Thus, the probability that the $i$th node 
will activate a certain number of neighboring nodes 
is given by the following generating function:
\begin{equation}
 f^{(i)}(x) = \prod_{j=1}^{k_i}(1-p_{ij}+p_{ij}x)~,
\end{equation}
where the degree $k_i$ denotes the total number of 
neighbors of the $i$th node. On the average the $i$th 
node will activate $\mu^{(i)} = \sum_j p_{ij}$ 
neighbors. This branching dynamics leads to correlation
effects due to loops in the network structure. In the
simplest approximation one neglects correlation effects 
and the avalanche propagation will be critical when 
every site activates, on the average, one node,
$\mu^{(i)} = 1$. In Fig.~\ref{fig:branch_directed} 
we present the critical scaling behavior of avalanche 
size and durations as we switch from the case when there 
is equal probability of activating any of the neighboring 
nodes ($p_{ij} = p_i = 1/k$) to the case where the 
activation  one of the neighbors 
($p_{ij} \rightarrow 1$ when $j=j^*$ and 
$p_{ij} \rightarrow 0$ when $j \neq j^*$).

%%%%%%%%%%%%%%%%%%%%%%%%%%%%%%%%%%%%%%%%%%%
%%%%%%%%%%%%%%%%%%%%%%%%%%%%%%%%%%%%%%%%%%%

This probabilistic description of branching process on a network 
is useful for mapping the behavior of a real-world phenomena, 
when the exact state of the whole physical system is unknown, that 
is when at any moment only a small subset of the complete system 
is studied. Even a deterministic process will appear stochastic if 
there are hidden, non-observable variables and dependencies of 
the current state on the exact history, viz if the process is
non-Markovian. For example the activation of a network node may lead 
to the activation of the same set of nodes whenever the same activation 
history is repeated. Neglecting memory effects can lead to 
the conclusion that neighboring nodes are activated in probabilistic 
manner. Using a random branching process for modelling is, in this
case, equivalent to an average of the observed activations, 
over sampled system states. 

In the next section we will discuss the scaling behavior of a special 
case of branching processes, such that $\mu = 1$ and $p_{ij} = 1$ 
for some $j=j^*$. These conditions are satisfied when the activation 
of a single node leads with certainty to the activation of exactly 
one of its neighbors. We call this limiting case of a branching 
process a routing process \citep{markovic2009vertex}. 

%%%%%%%%%%%%%%%%%%%%%%%%%%%%%%%%%%%%%%
\subsection{Vertex routing models}
%%%%%%%%%%%%%%%%%%%%%%%%%%%%%%%%%%%%%%
 
A routing process can be considered as a specialization of
random branching, see Fig.\ \ref{fig_routing}. For random
branching the probability $p_{ij}$ of activating the $j$th 
neighboring node is equal for all neighbors, 
that is $p_{ij} = 1/k_i$ for every $j=1,\ldots,k_i$, 
where the degree $k_i$ is the number of neighbors of the 
$i$th node. For a routing process, in contrast, only a 
single neighbor is activated.  An example of a system 
exhibiting routing-type behavior is a winner-take-all 
neural network \citep{gros2007neural,gros2009cognitive}, 
where at any time only a single neuron may be active, or, 
alternatively, only a single clique of neurons becomes 
active suppressing the activity of all other competing 
cliques \citep{gros2007neural}. One may also view 
routing processes as the routing of information packages 
and study in this context the notion of information 
centrality \citep{markovic2009vertex}, which is defined 
as the number of information channels passing 
through a single node.

Here we discuss the relation of vertex routing to scaling in
critical dynamical systems. Routing models are critical by
construction with the routing process being conserved. The 
type of vertex routing models considered here are exactly solvable
and allow to study an interesting question: Does the scaling
of an intrinsic feature, e.g.\ of a certain property of the
attractors, coincide with what an external observer would 
find when probing the system? Vertex routing models allow for 
a precise investigation of this issue and one finds that the 
process of observing a complex dynamical system may introduce 
a systematic bias alternating the resulting scaling behavior. 
For vertex routing models one finds that the observed scaling 
differs from the intrinsic scaling and that this disjunction 
has two roots. On one hand the observation is biased by the 
size of the basins of attraction and, on the other hand, the 
intrinsic attractor statistics is highly non-trivial in the sense
that a relative small number of attractors dominates phase 
space, in spite of the existence of a very large number 
of small attractors. 

\subsubsection{Markovian and non-Markovian routing dynamics}

\begin{figure}[t]
\centerline{
\includegraphics[width=0.35\textwidth,angle=0]{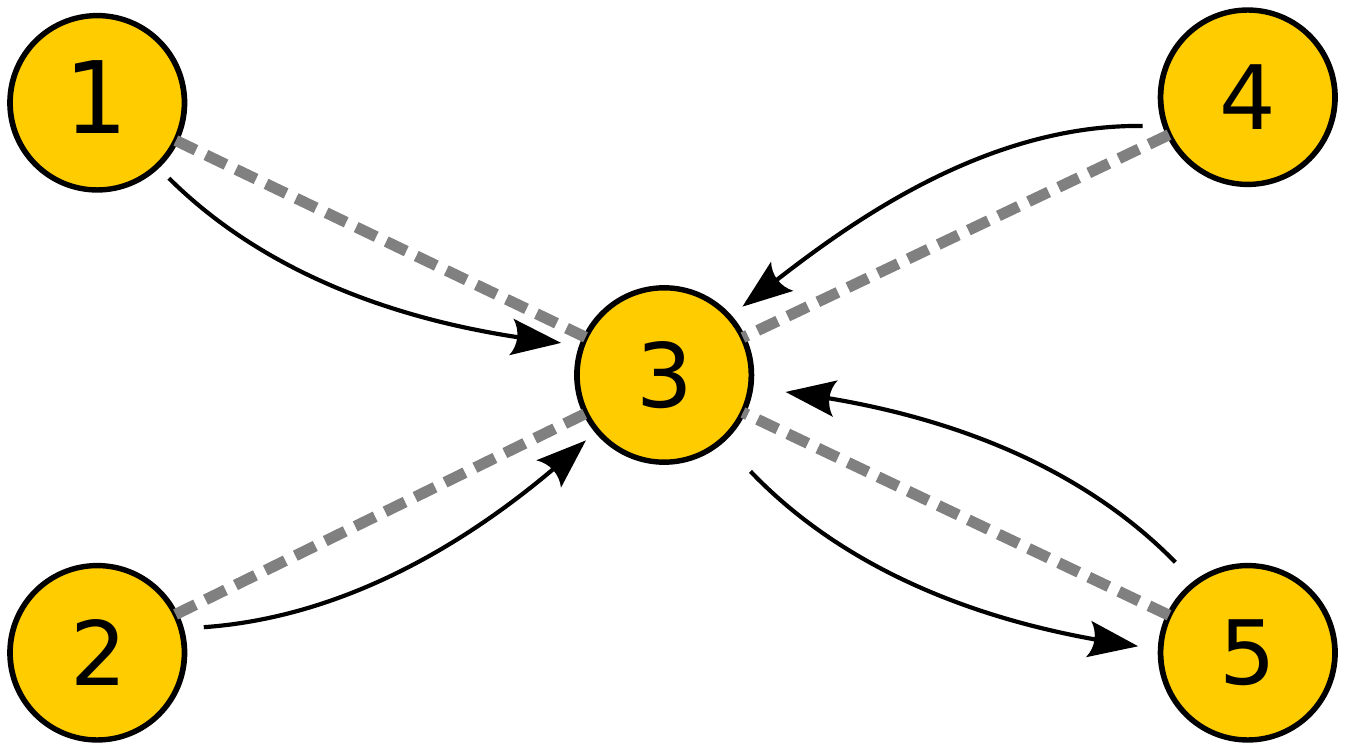}
\hspace{6ex}\ \ 
\includegraphics[width=0.35\textwidth,angle=0]{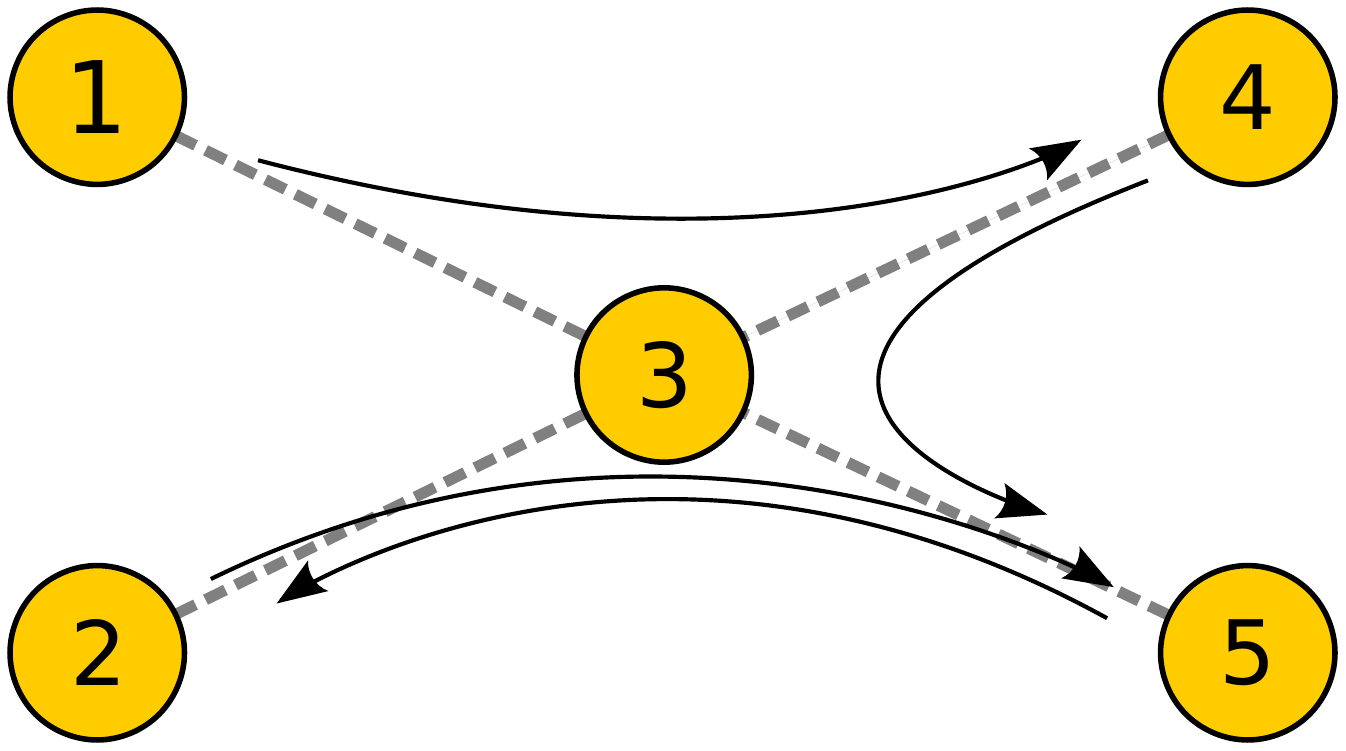}
}
\caption{Examples of routing process.
Left: For the Markovian case, $m=0$. An information 
package is always routed to vertex 5 independently 
of where it came from. Right: For a one-step memory, $m=1$. 
Information packages arriving at vertex 3 from the vertices 
4 and 2 are routed to vertex 5, while packages arriving 
from vertex 1 and vertex 5 are routed to vertex 4 and 2 
respectively. }
\label{fig_noderouting}
\end{figure}

We discuss here routing on complete networks, {\it i.e.}\ networks
which are fully connected, and consider the routing
process as the transmission of an information package,
which may represent any preserved physical quantity.
In general, routing of the information package to one 
of the neighboring nodes may depend on the routing history,
that is, on the set of previously visited (activated) nodes. We 
denote with $m$ the depth of the routing memory retained. The
routing is then Markovian if $m=0$ and non-Markovian otherwise.
An illustration of a basic routing transition
is presented in Fig.~\ref{fig_noderouting} for $m=0$ and $m=1$.

Let us denote with $v_t$ a node active at time step $t$, 
where $v_t \in V=\{1,\ldots,N\}$ with $N$ denoting the
network size.  Which of the $N-1$ neighbors of the node $v_t$ 
will become activated in the next time step $t+1$
will depend, through the transition probability 
$P(v_{t+1}=j|v_t,\ldots,v_{t-m})=p_{j|v_t,\ldots,v_{t-m}}\in \{0,1\}$, 
on the set of the $m$ previously visited nodes. The routing process is
considered conserved whenever $\sum_{j}p_{j|v_t,\ldots,v_{t-m}}=1$. 
For example, given some routing history in a network 
of $N=20$ nodes, say $v_t=3$, $v_{t-1} = 4,\ \ldots, v_{t-m} = 15$, 
there would be only one possible successor vertex, say 
$v_{t+1}=8$, and all other $N-1$ nodes would be unreachable,
given the specified routing history.

%%%%%%%%%%%%%%%%%%%%%%%%%%%%%%%%%%%%%%%%%%%%%%%%%%%%%%%%5
%%%%%%%%%%%%%%%%%%%%%%%%%%%%%%%%%%%%%%%%%%%%%%%%%%%%%%%%5

A sequence of $m+1$ vertices can be seen as a point 
in the (enlarged) phase space of routing histories with
$p_{j|v_t,\ldots,v_{t-m}}$ defining the adjacency matrix 
on the directed graph of phase space elements. To give an 
example, a point $[v_{m+1},\ldots,v_{1}]$ 
of the enlarged phase space is connected to some other point 
$[v_{m+2},\ldots,v_{2}]$ if $P_{v_{m+2}|v_{m+1},\ldots,v_1}=1$, 
where $v_i \in V$. The volume of the enlarged phase space, 
given as the total number of containing elements, 
is $\Omega = NK^{m}$ where $K=N-1$, for the case of a
fully connected network. 

\subsubsection{Intrinsic properties vs.\ external observation}

One usually considers as ``intrinsic'' a property
of a model when evaluated with quenched statistics,
hence when all parameters, like connectivities, 
transition probabilities, etc., are selected initially
and then kept constant \citep{gros2010complex}. 
An external observer has however no direct access to 
the internal properties of the system. An unbiased observer
will try to sample phase space homogeneously and then
follow the flow of the dynamics, evaluating the properties of
the attractors discovered this way. Doing so, the likelihood
to end up in a given attractor is proportional to the
size of its basin of attraction. The dynamics of the
observational process is equivalent to generate the
transition matrix ``on-the-fly'', viz to a random sampling
of the routing table $p_{j|v_t,\ldots,v_{t-m}}$.
Both types of dynamics can be evaluated exactly for
vertex routing models.

\paragraph{Intrinsic attractor statistics}
We first consider quenched dynamics, the transition 
probabilities $p_{v_{t+1}|v_t,\ldots,v_{t-m}}$ 
are fixed at the start and not selected during the 
simulation. A routing process initiated from a randomly 
selected point in phase space will eventually settle 
into a cyclic attractor. The ensemble averaged
distribution of cycle lengths is obtained when using
the set of all possible realizations of the routing 
tables, created by randomly selecting the values for the
transition probability, 
$p_{v_{t+1}|v_t,\ldots,v_{t-m}} \in \{0,1\}$,
while maintaining following conditions
\begin{displaymath}
\sum_{v_{t+1}}p_{v_{t+1}|v_t,\ldots,v_{t-m}}=1,\qquad  
\sum_{v_{t+1}, v_{k<t+1}}p_{v_{t+1}|v_t,\ldots,v_{t-m}} = K ~,
\end{displaymath}
where $K=N-1$ is the coordination number. The average number 
of cycles of length $L$, when the routing is dependent on 
the $m$ previous time steps, and for a network 
with $N$ nodes, is given by \citep{kruskal1954expected,markovic2012critical}
\begin{equation}
\label{eq_cyclength}
\langle C_m \rangle(L,N)=
\frac{N}{LK}\frac{(K^{m+1})!}{K^{(m+1)(L-1-m)}(K^{m+1}+m+1-L)!}~.
\end{equation}
The relation (\ref{eq_cyclength}) is, for finite networks with
$N<\infty$, an approximation for the non-Markovian case with
$m>0$, as it does not take into account corrections from self 
intersecting cycles, {\it i.e.} cycles in which a given node 
of the network is visited more then once. 
\citet{beck1989scaling} studied this model for the Markovian
case, in analogy to random maps, mainly in the context of 
simulating chaotic systems on finite precision CPUs 
(central processing unit of computer hardware).

%%%%%%%%%%%%%%%%%%%%%%%%%%%%%
\begin{table}[b]

\centering
\begin{tabular}{l|l|l|l}
%\cline{3-4}
\multicolumn{2}{c|}{}
 & \ quenched \ & \ random \ \\
\hline
\multirow{2}{*}{\ $m=1$\ }
    & \ $\displaystyle \langle n \rangle$ \ & \ $\log(N)$         \ & \ -- \ \\
    & \ $\langle L \rangle$   \ & \ $N/\log(N)$       \ & \ $N$ \ \\
%   & \ median \ & \ $N^{0.8}/\log(N)$ \ & \ $N$ \ \\
\hline
\multirow{2}{*}{\ $m=0$\ }
  & \  $\langle n \rangle$ \ & \ $\log(N)$          \ & \ --       \ \\
  & \ $\langle L \rangle$   \ & \ $\sqrt{N}/\log(N)$ \ & \ $\sqrt{N}$ \ \\
% & \ median \ & \ $N^{0.4}/\log(N)$ \ & \ $\sqrt{N}$ \ \\
\hline
\end{tabular}
\caption{ \label{tbl_exponents}
Scaling with the number of vertices $N$,
for the number of cycles $\langle n \rangle$ and for the mean 
cycle length $\langle L \rangle$ for history independent process ($m=0$) 
and the history dependent process ($m=1$), and 
for the two probing methods, quenched sampling and 
random sampling.}
\end{table} 
%%%%%%%%%%%%%%%%%%%%%%%%%%%%%%%%%

One can show that there is, in the limit of large networks, 
an equivalence between increasing the network size and increasing 
the memory dependence. This relation can be seen from the following 
memory dependent scaling relation
\begin{equation}
\langle C_{m+\tau} \rangle(L,N) \propto \langle C_m \rangle(L,N^\prime),
\qquad\quad
N^\prime \approx 1+(N-1)^{1+\frac{\tau}{m+1}}~,
\end{equation}
to leading order (for large $N$). Obviously, 
when $m=0$ we get $N^\prime-1 \approx (N-1)^{\tau+1}$, 
thus each 
additional step of history dependence effectively increases 
exponentially the phase space volume. On the other hand, in the 
limit $m\rightarrow\infty$ we obtain $N^\prime = N$, any additional 
memory step in the system with already long history dependence 
will not drastically change the total number of cycles.

%%%%%%%%%%%%%%%%%%%%%%%%%%%%%%%%%%%%%%%%
\begin{figure}[t]
\centerline{
\includegraphics[width=0.8\textwidth]{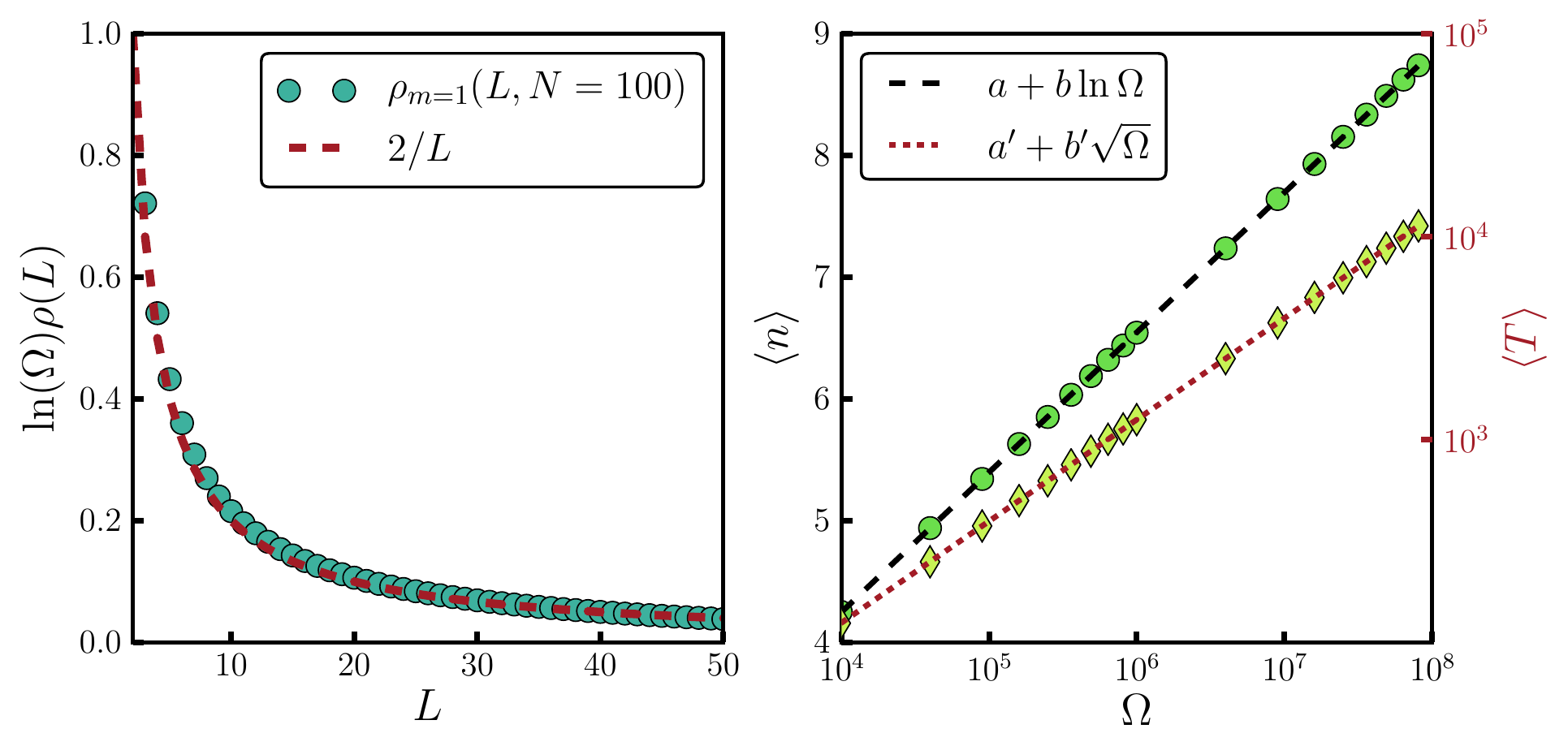}
           }
\caption{Left: The cycle length distributions 
$\rho_{m}(L,N)$, rescaled by $\log(\Omega)$,
for the vertex routing model ($N$: network size,
$L$ cycle length). The dashed line, $2/L$, represents 
the large-$N$ and small-$L$ limiting behavior.
Right:
As a function of phase space volume $\Omega$,
the average total number of cycles 
$\langle n \rangle$ (circles, linear scale - left
axis) and the expected total cycle length 
$\langle T \rangle$ (diamonds, logarithmic scale
- right axis).
The dashed line is fit using $a+b\ln\Omega$ 
($a = -0.345(3)$, $b =0.4988(2)$), and the doted line 
a fit using $a^{\prime} + b^{\prime}\sqrt{\Omega}$ 
($a^{\prime} = -0.3311(5)$, $b^{\prime} = 1.25331 \pm 2\cdot10^{-7}$).
The coefficient of determination is $R^2 = 1.0$ in both cases, 
within the numerical precision.
}
\label{fig_cycle_length_distribution}
\end{figure}
%%%%%%%%%%%%%%%%%%%%%%%%%%%%%%%%%%%%%%%%% 

\begin{figure}[t]
\centerline{
\includegraphics[width=0.8\textwidth]{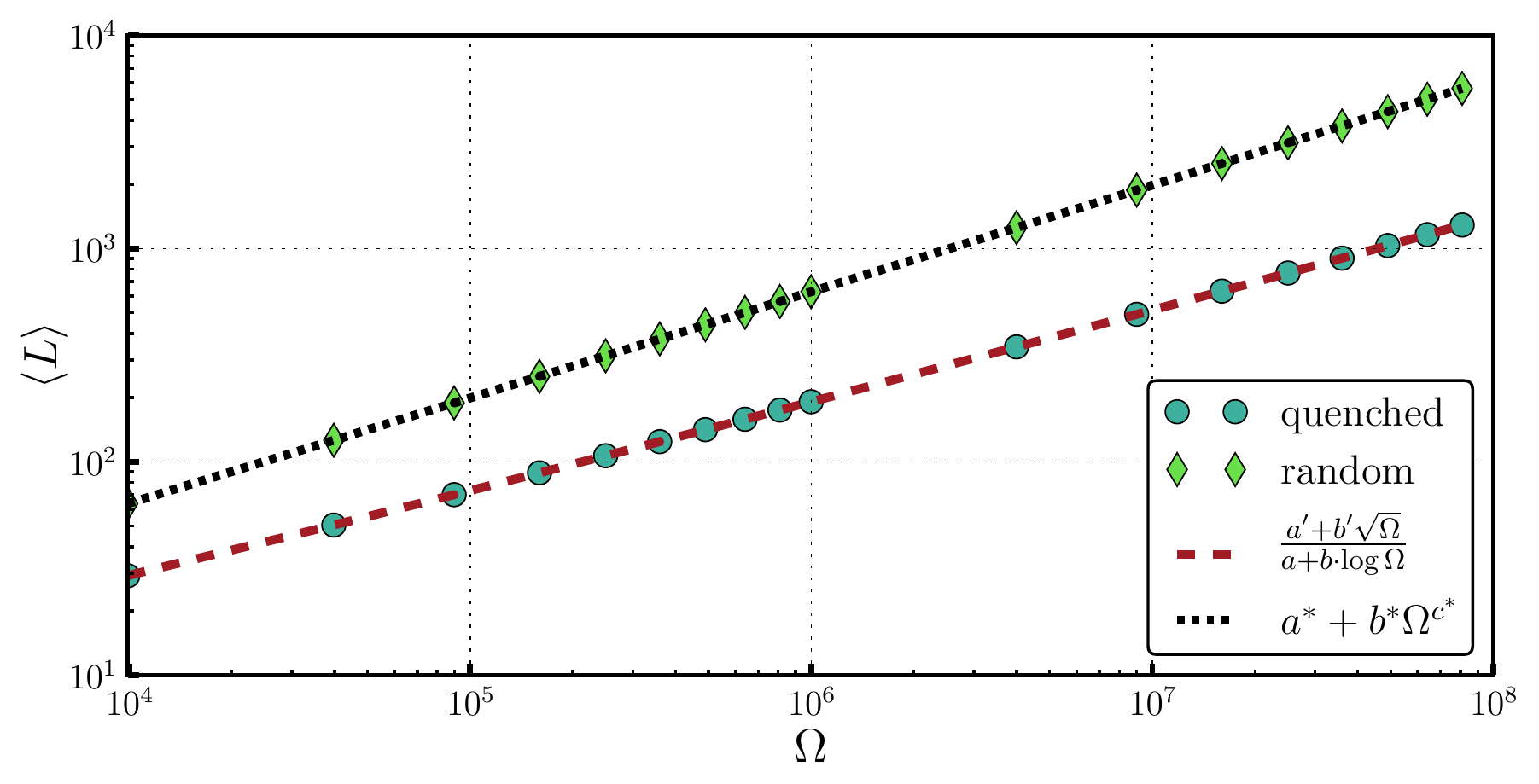}
}
\caption{The mean cycle length $\langle L\rangle$ for the 
vertex routing with the quenched dynamics (blue circles) 
and the vertex routing with random sampling 
(green diamonds), as a function of the phase space volume 
$\Omega$; log-log plot. The dashed line is the fit 
of the form ${a^{\prime}+b^{\prime}\sqrt{\Omega}}/{a+b\ln\Omega}$,
for  the parameters see Fig.~\ref{fig_cycle_length_distribution}.
The doted line is a fit of the form $a^* + b^* \Omega^{c^{*}}$, 
with $a^{*} = 1.3319(3)$, $b^{*} =  0.62666\pm 2 \cdot 10^{-6}$, 
$c^{*} = 0.5 \pm 9\cdot10^{-8}$. The coefficient of 
determination is $R^2 = 1.0$ in both cases, within the 
numerical precision. 
}
\label{fig_scaling_mean}
\end{figure}

The analytic expression (\ref{eq_cyclength}) for the 
cycle-length distribution can be evaluated numerically
for very large network sizes $N$, or alternatively as
a function of phase space volume $\Omega = NK^m$.
The total number of cycles 
$\langle n_m \rangle(\Omega) = \sum_L \langle C_m \rangle(L, \Omega)$
present in the system shows logarithmic scaling as a 
function of the phase space volume $\Omega$,
as shown in Fig.~\ref{fig_cycle_length_distribution}. 
The growth is hence slower than any polynomial of the number 
of vertices $N$, which is in contrast to critical Kauffman 
models, where it grows faster then any power of $N$ 
\citep{drossel2005number, samuelsson2003superpolynomial}. 
A numerical evaluation of the total cycle length, defined as 
$\langle T_m \rangle_{\Omega} = 
\sum_L L\langle C_m \rangle_{\omega}(L)$, 
shows power-law scaling with phase space volume, namely
as $\sim\sqrt\Omega$. 
Thus, the mean cycle length scales as 
\begin{equation}
\langle L_m \rangle_{\Omega} = 
\frac {\langle T_m \rangle_{\omega}} {\langle n_m \rangle_{\Omega}}
= \frac{a^{\prime}+b^{\prime}\sqrt{\Omega}}{a+b\ln\Omega}~,
\end{equation}
as shown in Fig.~\ref{fig_scaling_mean}.
%
%%%%%%%%%%%%%%%%%%%%%%%%%%%%%%%%%%%%%%%%%
%%%%%%%%%%%%%%%%%%%%%%%%%%%%%%%%%%%%%%%%%
%
The probability $\rho_m(L,N)$ of finding, for a network 
with $N$ nodes, an attractor with cycle length $L$
is obtained by normalizing the expression (\ref{eq_cyclength}).
One can show that the rescaled distribution $\log(\Omega)\rho_m(L,N)$ 
has the form  $2e^{-L^2/2K^{m+1}}/L$, for small cycle lengths $L$,
falling off like
\begin{equation}
\log(\Omega)\rho_m(L,N) \propto \frac{K^{(m+1)(M-\frac{1}{2})}}{M!}~,
\end{equation}
for large $L\rightarrow K\Omega/N+1$, where $M = K^{m+1}+1-L$.

\paragraph{Observed attractor statistics}
Instead of considering quenched routing dynamics, 
one can sample stochastically the space of all possible 
realizations of routing dynamics \cite{gros2013observing}. 
In practice this means 
that at each time step one randomly selects the next 
element in the sequence of routing transitions.
Algorithmically this is equivalent of starting at a
random point in phase space and then following the flow. 
This is actually the very procedure carried out when probing 
a dynamical system from the outside. A cycle is found when
previously visited phase space elements is 
visited for a second time.

Starting from a single element of phase space, the activation propagates 
until the trajectory reaches the same element for the second time. The 
probability of such a trajectory having a path length $s$, is given 
by 
\begin{equation}
p_s = \frac{(s-1)(K^{m+1})!}{K^{(m+1)s)}(K^{m+1}-s+1)!}~. 
\end{equation}
In a path of length $s$, the observed cycle will have a
length $L \leq s$. Thus, the joint probability of observing 
a cycle of length $L$ within a path of length $s$ is given by
\begin{equation}
p(L,s) = \frac{\Theta(s-L)\Theta(L-2)(K^{m+1})!}{K^{(m+1)s}(K^{m+1}-s+1)!}~,
\end{equation}
with $\Theta$ being the Heaviside step function.
Finally, one obtains the probability $\tilde{\rho}_m(L,N)$
(with $\tilde\rho_m$ denoting random dynamics and
$\rho_m$ quenched dynamics) of observing a cycle of length $L$ 
as a sum over all possible path lengths, that is
\begin{equation}
\tilde{\rho}_m(L,N)=\Theta(L-2)\sum_{s=L}^{K^{m+1}+1}
\frac{(K^{m+1})!}{K^{(m+1)s}(K^{m+1}-s+1)!}~.
\end{equation}
Interestingly, the mean cycle length scales as $\sqrt{\Omega}$ 
when using random sampling as a method for probing the system 
of routing transition elements. The comparison of the respective
scaling behaviors, as a function of the network size and for $m=0,1$,
is given in Table \ref{tbl_exponents}. There are two implications
\cite{gros2013observing}.
\begin{itemize}
\item The results for the vertex routing model indicate
      that one needs to account for the procedure used 
      to probe the scaling behavior of a complex system.
\item Certain properties of critical dynamical systems, like the
      number of attractors, may not show power-law scaling, even
      at criticality.
\end{itemize}
Vertex routing models and random boolean networks are,
furthermore, in different classes. The scaling relations
shown in Table \ref{tbl_exponents} do not translate into
the ones for the Kauffman net 
\citep{drossel2005number, samuelsson2003superpolynomial}
when rescaling the dependence of the phase space volume 
$\Omega$ from $N(N-1)$ (valid for the $m=1$ routing model) 
to $2^N$, as valid for the Kauffman net.

\section{Modelling experimental data} 
\label{sec:exp}

A mathematical model of real-world phenomena 
should both replicate the phenomena and capture 
the structure and the function of the described physical 
system. One may divide theory models 
as ``descriptive" or ``explanatory'' \citep{willinger2002scaling}. 
A descriptive model tries to reproduce the statistical properties 
of the phenomena in question, while containing often 
unrealistic and simplistic assumptions about the 
structure of the modeled system. Thus, not attempting
to explain the underlying generative mechanism of the 
phenomena of interest. In contrast, an explanatory model 
would reproduce both the phenomena while capturing the 
known structural and functional properties of the 
system modeled. It is, however, difficult to
actually  prove that a given model is ``correct''.
When modeling systems which are very complex, one
has necessarily to resort to some simplifying assumptions
and to neglect certain experimental aspects seen as
secondary; and to concentrate on the primary
aspect on interest, e.g.\ the power-law scaling of
certain observables. Our discussion here will hence not
be able to give definite answers. \citet{willinger2002scaling} 
has pointed out in this context, that although descriptive 
models may provide an initial description for the possible 
causes of the phenomenon studied, a correct prediction 
of the dynamical behavior would require a consistent 
explanatory model for which the various assumptions 
incorporated into the model have been verified. 
Thus, we would like to understand whether {\it SOC} 
models provide an adequate explanatory description for
various real-world phenomena and, if not, which extensions 
of current models are required or what would be an 
alternative explanatory model.

In the following sections we will give a short review of 
the some of the known statistical properties of the empirical 
time series of earthquake magnitudes, solar flares 
intensities and sizes of neuronal avalanches and compare 
experimental avalanche statistics with theory predictions,
mostly for dissipative {\it SOC} models. We will also point 
out plausible alternative mechanisms leading to power-law scaling 
of event sizes without requiring a critical regime. 

%---------------------------------------
\subsection{Earthquakes and Solar flares}
%---------------------------------------
Solar flares are large energy releases on the surface of the Sun 
and they are observed as a sudden brightening of a region on 
Sun's surface. As the distribution of peak intensities of solar 
flares follows a power-law scaling, \citet{lu1991avalanches} 
proposed {\it SOC} for a generative mechanism of flares in the 
solar corona. Looking at the total flare energy, which represents 
the size of an avalanche $s$, one finds that it follows a 
power-law scaling with an exponent $\tau_s \in [1.6,1.75]$ 
\citep{crosby1993frequency, clauset2009power}.

Similarly, \citet{sornette1989self} have initially suggested that the 
scaling behavior of earthquakes magnitudes would correspond to 
that of the {\it SOC} systems, a proposition motivated by the well 
known Gutenberg-Richter and Omori laws. The Omori law describes 
the empirical evidence that the frequency $f(t)$ of earthquake 
aftershocks decays, as function of time $t$ passed since the 
earthquake, as $1/t$, whereas the Gutenberg-Richter law 
states that the probability of observing an earthquake of 
magnitude of at least $M$ scales as $10^{-bM}$, where $b$ is 
a positive constant. The size of an avalanche $s$ is taken to
be proportional to the scalar seismic moment, and its relation to 
the earthquake magnitude as $M = \frac{3}{2}\log_{10}(s)$ \citep{kagan2002modern}. 
Hence, the probability $P(s)$ of finding an event of size $s$ 
follows a power-law scaling, that is $P(s) \sim s^{-\tau_s}$.
The scaling exponent $\tau_s$ falls in the range $[1.6, 1.7]$,
independent of the region and of the depth of the earthquakes 
\citep{kagan2002modern, clauset2009power}, with values closer 
to the mean field prediction of $\tau_s = 3/2$ also 
being discussed \citep{kagan2010earthquake}. Note, that 
similar scaling laws are also observed in the scaling properties 
of solar flares, suggesting a common interpretation of 
these two phenomena \citep{de2006universality}.

%%%%%%%%%%%%%%%%%%%%%%%%%%%%%%%%%%%%%%%%%
\begin{figure}[t]
\centerline{
\includegraphics[width=1.0\textwidth]{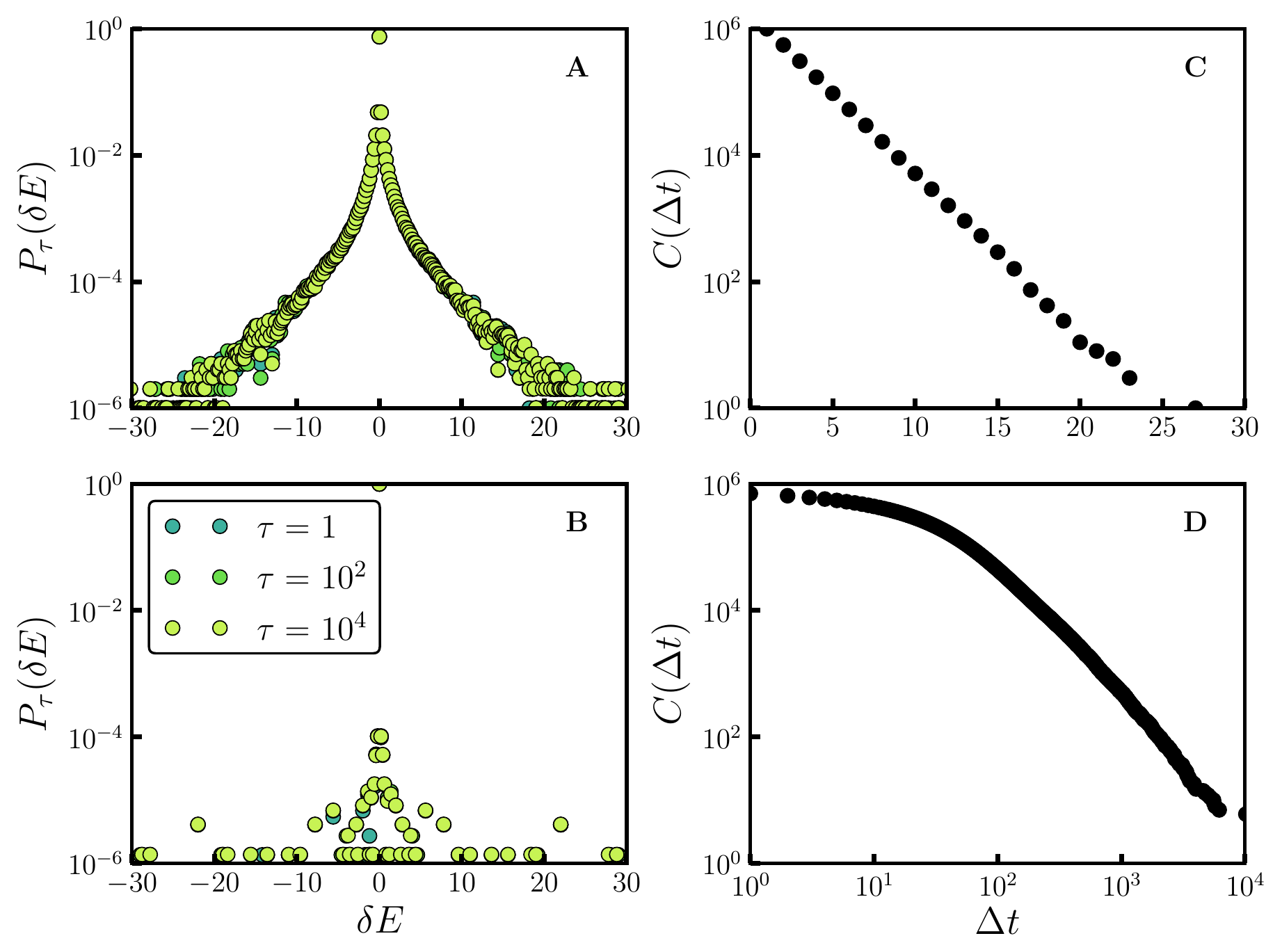}
}
\caption{Probability distributions for released energy 
fluctuations $P_{\tau}(\delta E)$ (see Eq.\ref{eq_fluc_energy}), 
obtained from (A) the avalanches generated by the BTW 
sandpile model (see Sect.~\ref{sec:soc}),
(B) the data set of earthquakes in North California in a period 
1968 - 2012 (earthquakes of magnitude $M \geq 1$ in the 
NCEDC Earthquake Catalog). Overlapping data for various
inter-event time scales $\tau$ indicate self-similarity.
The respective CCDF (complementary cumulative probability 
distributions) of waiting times estimated 
(C) from the time series generated by the BTW sandpile mode and 
(D) from the time series of earthquakes.
}
\label{fig_fluc_energy}
\end{figure}
%%%%%%%%%%%%%%%%%%%%%%%%%%%%%%%%%%%%%%%%%

The statistics of the released energy fluctuations, or the 
so called ``returns'', is an important quantity 
characterizing self-similarity of a stochastic process, 
and a good yardstick for controlling the 
quality of modeling efforts. The quantity
\begin{equation}
\delta E = \frac{s(t+\tau)-s(t)}{\sigma_{\tau}},
\qquad\quad
\sigma_{\tau}^2 = \left\langle(s(t+\tau)-s(t))^2\right\rangle
\label{eq_fluc_energy}
\end{equation}
corresponds to the relative difference in the size of 
avalanches released at times $t$ and $t+\tau$ respectively. 
One may evaluate, for a fixed inter-avalanche time $\tau$, 
the distribution $P_\tau(\delta E)$ measuring the probability 
of finding an fluctuation $\delta E$ in the released energy. 

Real-world and {\it SOC} avalanches may differ with
respect to the statistics of the returns. The 
distribution $P_{\tau}(\delta E)$ is invariant with respect
to a change of the time scale $\tau$, for classical {\it SOC} systems, 
that is, $P_{\tau}(\delta E) = P_{\tau'}(\delta E)$ for any 
$\tau' \ne \tau$. Experimentally observed energy fluctuations 
change---in the case of turbulent phenomena---with 
the inter-event time scale $\tau$, exhibiting 
multifractal scaling \citep{carbone2002extent}. This 
observation led to the conclusion that classical {\it SOC} 
models cannot produce the higher order statistics typical 
for turbulent flows, which are however captured properly 
by models describing the energy cascades in turbulence 
\citep{boffetta1999power, freeman2000power}. For the
case of earthquakes, interestingly though, \citet{caruso2007analysis} 
pointed out that the distribution $P_{\tau}(\delta E)$ of energy 
fluctuations is independent on the scale $\tau$, thus the time 
series is self-similar, as shown in Fig.~\ref{fig_fluc_energy}.

Another important quantity for characterizing a time series of
experimentally observed events is the waiting time distribution 
(WTD); the distribution of durations of quiet periods between 
events. The WTD observed for earthquakes and solar flares 
differs markably from the one produced by classical {\it SOC} systems,
with the empirical time series showing a power-law distributed 
WTD and with the SOC waiting time distribution closely following an 
exponential distribution, as typical for a Poisson process 
charactering a memoryless time series \citep{boffetta1999power, 
freeman2000power, yang2004earthquakes, davidsen2004seismic, wheatland2009waiting}.
\citet{sanchez2002waiting} demonstrated that a modified
sandpile model can produce a scale-invariant WTD, and
multifractal scaling for the energy fluctuations.
In addition, \citet{paczuski2005interoccurrence} showed, 
that the WTD follows a power-law when one considers 
the time scale of avalanches instead of the time scale of the
external drive and putting a threshold to the minimal recorded
intensity, at any point in time. Setting a signal threshold is an 
usual experimental procedure to distinguish between small 
events and background noise. Furthermore, \citet{sattin2006self} 
demonstrated that one obtains, when the external drive is 
spatially correlated, both power-law scaling for the WTD and 
multifractal scaling for the energy fluctuations \citep{sattin2006self, 
charbonneau2007avalanche}. Thus, certain constrains to the driving 
force in {\it SOC} theories can generate some of the behavior 
observed in the empirical data.

Nevertheless, some concerns remain. When predicting the 
occurrence of an event of a certain size the distribution 
of waiting times is not as important as the correlations 
between waiting times. The predictability of a time series can 
be quantified by estimating the index of long-range time 
dependence, also known as the Hurst exponent $H$ 
\citep{samorodnitsky2007long}. For $H=1/2$ the time series 
is uncorrelated and unpredictable; this is exactly the 
value of the Hurst exponent obtained in different {\it SOC} 
models \citep{caruso2007analysis}---even in the 
presence of spatial correlation in the external driving 
force. In contrast, the estimates for the Hurst exponent 
for the time series of earthquakes and solar flares
indicate the presence of a long-term memory in the empirical 
data \citep{lennartz2008long, paczuski2005interoccurrence}, 
that is, $H \in (1/2,1]$. These long-term correlations suggest
that large events are more likely to be followed by events of 
similar or larger magnitude, possibly allowing for the prediction 
of intense events. For example, specific patterns have been 
observed in the seismic activity data preceding the main 
event, thus opening a venue for predicting large earthquakes 
\citep{evison1977fluctuations, johansen2000new, manshour2009turbulent}.

\citet{jagla2010realistic} introduced a modified {\it OFC}
earthquake model, see Sect.~\ref{subsec:dissipativesoc},
and proposed a solution for this inconsistency between 
theoretical and experimental results. The modifications 
to the original {\it OFC} model consist of implementing 
structural relaxation and random threshold values for 
each node of the lattice---resembling the spatial 
inhomogeneity of real earthquake faults 
\citep{kawamura2012statistical}. The relaxation 
mechanism equalizes the stress levels among 
neighboring nodes and works on the
time scales of the driving forces---essentially infinitely 
slower then the time scale of avalanche topplings.
The avalanches generated by this model follow 
a power-law scaling, with exponents independent on 
the dissipation levels; unlike the standard OFC 
model with inhomogeneities (see \ref{subsec:dissipativesoc}). 
Furthermore, the simulated time series is spatially and 
temporally correlated and exhibits patterns of aftershocks 
like the one observed in earthquakes and solar 
flares. Aftershocks are triggered by the relaxation 
mechanism after the main shocks---initiated by the 
external drive---due to the non-uniform distribution 
of thresholds.  

A few questions still remain. Is this modified OFC model 
robust in the presence of non-uniform interactions between
neighboring nodes \citep{zhang2009analysis}? Is the modified 
OFC model robust in the presence of complex network structures?
An interesting issue since there are indications that the underlying 
network of earthquake epicenters has scale-free and small world 
structure \citep{baiesi2004scale, abe2004small}. Finally, 
is the mechanism of structural relaxations universally 
applicable to other physical systems that show {\it SOC} 
dynamics or are system specific modifications required? 
If the required modifications to dissipative {\it SOC} 
models---in the presence of inhomogeneities---are system 
specific then the {\it SOC} behavior would start to depend
on the exact dynamical constrains and local interaction 
rules, thus the universal properties of such regimes would
be lost.  

%-----------------------------------------------------
\subsubsection{Tuned versus self-organized criticality}
%-----------------------------------------------------

When studying naturally occurring phenomena, like
solar flares and earthquakes, one cannot control
experimental conditions and their effect on the 
behavior of the system. Small-scale experimental 
studies of power-law phenomena \citep{zaiser2006scale},
in which the experimental conditions are carefully 
controlled, might provide important insights for our 
understanding of the power-law behavior observed in 
their large-scale counterparts.

\citet{friedman2012statistics} analyzed the scaling behavior 
of fractures in metallic nanocrystals induced by an externally 
applied, slowly increasing, stress. A fracture or a slip occurs 
when the local stress level, within the crystal, exceeds the 
local threshold stress, with the slips generated by the fast 
release of pinned deformations. The process stops when the 
loose segments  get repinned or annihilated, thus forming 
an avalanche. The avalanches are typically of length 
scales which are large with respect to the microscopic
length scales. The distributions of slip sizes $s$, measured in 
different materials, follow a power-law , $P(s)\sim s^{-1.5}$, over 
several orders of magnitude and fall on a same scaling function. 
Interestingly, the size of the largest expected event $s_{max}$ scales 
with the strength of the externally induced stress $f$, as $s_{max}(f) = (f_c-f)^{-2}$, 
which diverges only for $f=f_c$ \citep{zaiser2007slip,friedman2012statistics}. 
The results for the statistics of slip-avalanches in
nanocrystals obtained by \citet{friedman2012statistics} 
have been analyzed within a molecular-field approximation
for a micromechanical model for deformations in solids
\citep{dahmen2009micromechanical}. Within this model there is a second-order
phase transition between brittle and hardening crystals
(becoming respectively more/less susceptible to stress in
the wake of a slip), thus scale-free avalanche statistics is observed.

In contrast, within {\it SOC} framework, the maximal size of an 
avalanche $s_{max}$ depends only on the system size and diverges 
in the thermodynamic limit independent on the other system 
parameters. Thus, to relate critical like behavior to a {\it SOC} 
or SOqC mechanism, one should demonstrate that no other parameters 
except system size influence the scaling. In other words, one 
should exclude tuned criticality as possible explanation.
For example, the power-law scaling of earthquakes might be caused by 
near-critical stress levels in earth crust, which are just a transient 
state typical for the current geological era and not an attracting 
state, as would be the case in self-organized critical process.  
Unfortunately, this kind of hypothesis is difficult to test, as one 
cannot control the environmental parameters generating the earthquakes. 

%%%%%%%%%%%%%%%%%%%%%%%%%%%%%%%%%%%%%%%%%
\begin{figure}[t]
\centerline{
\includegraphics[width=0.8\textwidth]{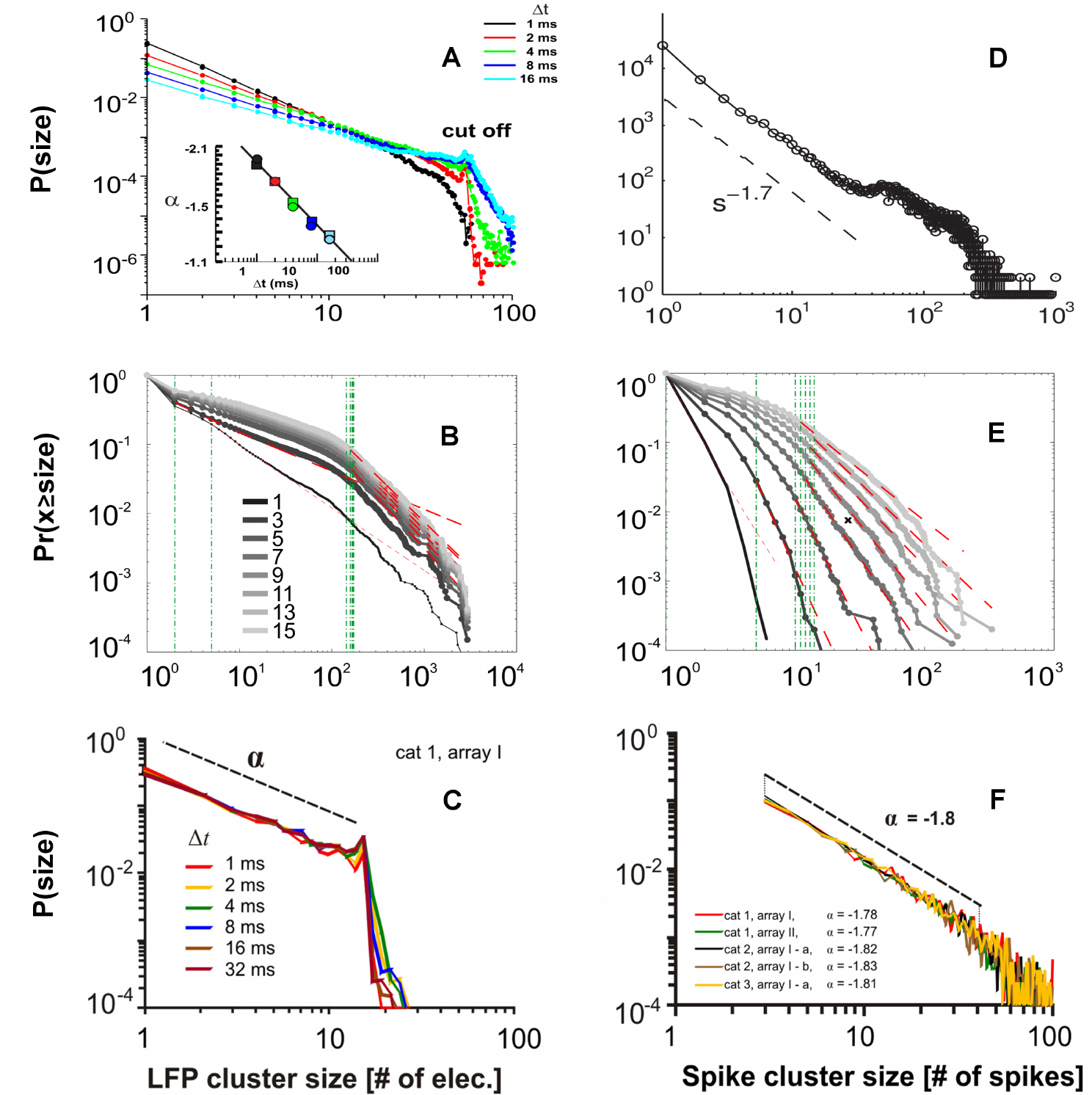}
}
\caption{Distribution of sizes of neuronal avalanches estimated 
from the LFPs (left column) and from neural spike recordings 
(right column). (A) (adapted from \citet{beggs2003neuronal}) 
and (D) (adapted from \citet{friedman2012universal}) 
show the data obtained from in vitro recordings in acute slices 
and organotypic cultures. Examples of data obtained from in vivo 
neural activity in humans (B,E) (adapted from \citet{dehghani2012avalanche}) 
and cats (C,F) (adapted from \citet{hahn2010neuronal}).
}
\label{fig_neural_avalanches}
\end{figure}
%%%%%%%%%%%%%%%%%%%%%%%%%%%%%%%%%%%%%%%%%

%---------------------------------------
\subsection{Neuronal avalanches}
%---------------------------------------

Neuronal avalanches are sequences of bursts of neural activity 
which, separated by quiet periods, spread across 
the neural tissue. Since the introduction 
of {\it SOC} theory it has been hypothesized that the
brain operates in the critical dynamical regime,
as many features of neural spiking activity resemble 
the properties of sandpile models, namely the sudden 
release of energy (action potential) and the transmission 
of released energy to neighboring nodes (interaction of 
neurons mediated by neurotransmitters or ion diffusion).
One of the first experimental evidences supporting this hypothesis 
was given by \citet{beggs2003neuronal}. They investigated 
the spontaneous neural activity measured in organotypic 
cultures (tissue which, removed from an organ, continues 
to develop as it would have done in the body) and in acute 
slices of rat cortex, observing power-law scaling of neuronal 
avalanches as extracted from the recordings of local field 
potentials. Similar evidences were later obtained from in vivo 
neural activity in humans \citep{ribeiro2010spike},
monkeys \citep{petermann2009spontaneous}, cats \citep{hahn2010neuronal}
and also from high-resolution data measured in cultured slices 
of cortical tissue extracted from living rats \citep{friedman2012universal}.
In Fig.~\ref{fig_neural_avalanches}  we presented the distribution 
of sizes of neuronal avalanches adapted from various studies.

A local field potential (LFP) represents the recorded 
voltage generated by the sum of all currents on the 
surface of the small electrode embedded within the neuronal 
tissue. These currents reflect the dendritic activity within 
a small volume surrounding the electrode. The neuronal 
avalanches are constructed from the sequence of negative 
peaks of the LFPs propagating across multiple electrodes, 
because negative voltage peaks are correlated to synchronized 
spiking activity of nearby neurons \citep{beggs2003neuronal, 
kelly2010local}. To distinguish between the troughs of LFPs 
from the troughs generated by the background noise, one has to define  
a threshold value for the recorded voltage. Only signals dropping 
below the threshold are considered in the definition of an avalanche. 
One calls an electrode active if the value of LFP on that 
electrode is below the threshold value. After identifying 
the relevant signals, the data is divided into time bins and the 
neuronal avalanche is defined as the sequence of recorded activity. 
An avalanche starts when at least a single electrode is 
active and ends when the signal is below threshold on all electrodes 
for at least one time bin. The avalanche duration is determined as 
the elapsed time between the first and the last bin; the size of the 
avalanche can be chosen either as the total number of active 
electrodes or as the absolute sum of LFP amplitudes over 
all active electrodes during the avalanche duration.

\citet{beggs2003neuronal} found that the 
avalanche size follows a power law with exponent close 
to $-3/2$ (see Fig.~\ref{fig_neural_avalanches} A), 
with the avalanche duration following a power 
law with an exponent close to $-2$. These values for the
critical exponents are, interestingly, identical with the 
mean-field results for critical branching processes in
fixed environments (see Sect.~\ref{sec:branching}). Note 
that the experimentally observed scaling behavior,
thus the values of the exponents, of the neuronal avalanches 
will depend on the choice of the threshold value and on the 
selected width of the time bins \citep{priesemann2013neuronal}. 
Still, these values can be fixed if one takes into account certain 
properties of white noise signals and the propagation speed
of action potentials along the neural cell membrane 
\citep{beggs2003neuronal}. 

The initial work of \citet{beggs2003neuronal} 
lacked rigorous statistical estimates of the scaling 
laws, the confirmation of similar scaling behavior for in vivo 
recording, and evidence for a critical state going
beyond the distribution of avalanche sizes and the $1/f$ 
scaling of the power spectrum (both necessary signatures for 
a critical state). A consensus on the dynamical state of 
neural activity is still missing even with experiments repeated 
and a refined data analysis including the previously missing 
factors. A central problem is the recording of neural activity 
in vivo with sufficiently high resolution, and the variations 
of the statistical properties of recorded activity between 
subjects and species \citep{petermann2009spontaneous, 
priesemann2009subsampling}. 

\citet{touboul2010can} showed, in a study performed on 
awake cats, for which the LPFs were measured with 8 channel 
multi-electrode arrays, that an exponential distribution 
is a better fit to the avalanche size distribution then 
a power-law distribution. \citet{dehghani2012avalanche} 
reached a similar conclusion by analyzing avalanches from 
recordings from the cerebral cortex of cat, monkey and human,
both made during wakefulness and during sleep 
(See Fig.~\ref{fig_neural_avalanches} 
B and E). He concluded 
that the optimal fit of the avalanche distributions is 
actually a double-exponential distribution\footnote{This results 
have been questioned, possibly being affected by the existence 
of a cutoff for large avalanches \citep{priesemann2012}, with
the number of recording electrodes limiting the maximal observable 
size of neuronal avalanches.}.
In contrast to the studies of \citet{touboul2010can} and 
\citet{dehghani2012avalanche}, several investigations 
found evidence for power-law distributed neuronal avalanches. 
\citet{petermann2009spontaneous} argued for scale invariant features 
of the cortical activity recorded from awake monkeys. 
Similarly, \citet{klaus2011statistical} recently showed that a power 
law is the best fit for the neuronal avalanches recorded both 
{\it in vivo} and {\it in vitro}. 

One of the possible explanation, for these opposing 
experimental results, may be traced back to the small 
number of recording electrodes used in some experiments, 
which may lead to a sub-sampling of the local neural activity. 
\citet{priesemann2009subsampling} argued that
critical processes can appear subcritical in the scaling 
behavior if the activity is averaged over a small number of 
recording elements, relatively to the total number of elements 
which are actually generating the critical phenomena.
Still, one should note that the analysis of peaks in LFP signals 
is an rather indirect measure of the neural activity patterns. 
\citet{touboul2010can} argued that simple thresholding of a stochastic process 
can generate an apparent power-law behavior, and that the use of LFP 
recordings for identifying the scaling properties of neuronal 
avalanches may hence be problematic. Furthermore, if one takes the 
positive peaks of the LFP signal, which are not related to 
spiking activity, instead of the negative peaks, applying 
the same procedure to estimate neural avalanches, one finds 
similar scaling behavior as for the negative peaks.
Thus, \citet{touboul2010can} and \citet{dehghani2012avalanche} 
proposed that the observed scaling behavior may be a consequence 
of a thresholding procedure, and not a reflection of an 
underlying critical or near critical state. They stressed the
point that one should investigate the scaling behavior of
the avalanches estimated both from the negative and from the
positive LFP peaks, with criticality being of possible relevance only
if the respective scaling behaviors would differ qualitatively.

%I have removed this part!
% Nevertheless, 
% \citet{dehghani2012avalanche} used for recordings a number 
% of electrodes comparable or even larger then the number 
% used in opposing studies and still observed exponential 
% scaling of avalanche sizes. It would be surprising that 
% changing the size of multi-electrode arrays by an order 
% of magnitude influences the avalanche statistics, as the 
% number of electrodes is still several orders of magnitude 
% smaller then the average number of neurons present 
% in the small patch of the surrounding neural tissue 
% \citep{skoglund1996heterogeneity}.

Beside estimating neuronal avalanches indirectly from the 
propagation of LFPs, one can also directly record neural 
spikes. For example, \citet{hahn2010neuronal} recorded 
spontaneous neural activity of adult cats under anesthesia 
and beside LFPs they also measured neural spikes. 
For both cases, they've found evidence of power-law 
distributed neuronal avalanches (see C and F subplots of 
Fig.~\ref{fig_neural_avalanches}). Also, \citet{ribeiro2010spike},
observed power-law distributed neuronal avalanches recorded 
from the cerebral cortex and from the hippocampus of rats; in awake, 
asleep and anesthetized animals. These results are puzzling as 
one would expect sleep and awake states to be characterized by 
distinct dynamical regimes and by different responses to external 
stimuli \citep{landsness2011electrophysiological}. Nevertheless, 
controversy persist even regarding the direct measurements of spiking 
activity, as \citet{dehghani2012avalanche} reported absence of 
power-law distributed avalanches.

In a recent study, where they recorded neural spikes in cultured cortical 
slices with high density multi-electrode arrays, \citet{friedman2012universal}
showed that the average shapes of neuronal avalanches
of different durations collapse to a single curve 
under an appropriate scaling transform, a strong evidence 
for a critical regime which even allows for the determination 
and the comparison of the dynamical universality class 
\citep{kuntz2000noise}. Interestingly though, out of ten 
samples of organotypic cultures used in this study, only two of 
them showed clear evidence for critical neuronal avalanches 
(see Fig.~\ref{fig_neural_avalanches} D). The other samples 
showed subcritical or supercritical behavior. This suggest, 
that self-organization of cortical networks to a critical 
state may not be a generic property, but that it might depend
on environmental conditions, on the interaction 
between different mechanisms of neural plasticity or on the current 
functional properties of global brain networks \citep{priesemann2013neuronal}. 

%-----------------------------------------------------
\subsubsection{The origins of neuronal power laws}
%-----------------------------------------------------

The underlying causes for the observed neural power laws
are still under debate. On the experimental side, to
give an example, the $1/f$ scaling of the power spectrum 
of the recorded LFPs could be ascribed to biophysical filtering 
effects of the extracellular media on the recorded signal 
\citep{bedard2006does, bedard2009macroscopic, el2009network}. 
\citet{touboul2010can} noted, in addition, that power-law scaling 
of peak events may arise from a thresholded stochastic 
process, a plausible model for the generic neural dynamics,
which is however devoid of any connection to criticality or 
self-organization.

A basic precondition for the brain to retain functionality
is, in agreement with experimental results, that the level 
of the average cortical activity remains within a certain 
range, neither exploding over time nor dying out. Mapping 
bounded neural dynamics to a branching process hence cannot 
result in neither a subcritical (with the neural activity 
becoming eventually extinct) nor in a supercritical (with 
an exploding neural activity) regime. This line of argument 
is valid if the majority of neural activity studied is 
stimulated internally and not induced by external sensory 
inputs. This is the case for the upper cortical layers, 
which are responsible for the intra-cortical communication. 
Interestingly, these upper cortical layers are also mostly the 
ones for which evidence for neuronal avalanches has been reported,
together with a critical branching ratio \citep{plenz2012neuronal}.

A support for the {\it SOC} causes of neuronal avalanches 
comes from several theoretical studies using networks of spiking 
neurons. These artificial neural networks are related to dissipative 
{\it SOC} models and as such require a fine tuning of the external 
drive, which initiates the neuronal spikes, relative to the number 
of neurons in the network \citep{bonachela2010self}. 
Nevertheless, one can still consider dissipative models 
as very close approximations to true {\it SOC} behavior 
observed in conserved sandpile models, as discussed in 
section \ref{sec:soc}. More importantly, these neural network models, 
although replicating very closely the experimentally observed 
statistical properties of neuronal avalanches, achieve critical 
behavior only for networks consisting of purely excitatory 
neurons \citep{levina2007dynamical} and the introduction of 
biologically realistic levels of inhibition breaks the 
power-law scaling of neuronal avalanches 
\citep{millman2010self, de2012dragon}. 
The experimental observations \citep{beggs2003neuronal} indicates 
that a network of excitatory neurons operates in a supercritical 
regime, which allows for a fast transfer of information, whereas 
inhibition has the role of stopping large neuronal avalanches 
and to localize information processing. 
Rather then spontaneously emerging from separation of time 
scales between external driving and internal dissipative 
mechanisms, the critical behavior in cortical networks seems to
be reached through various plasticity mechanisms, whenever 
such a dynamical regime is optimal for given environmental conditions. 
This kind of reasoning is closer to the {\it HOT} theory 
(see Sect.~\ref{subsec:hot}), which states that power-law scaling 
emerges through design aimed at optimal functioning in uncertain 
environments. For the case of the brain, and in general for entire 
organisms, this design is thought to emerge through natural selection.

Finally, in order to understand why a critical behavior 
of neuronal avalanches may be computationally favorable,
and hence be selected through Darwinian evolution,
one should look for the conditions under which the 
critical state may constitute an optimal working regime. 
An analysis of information retention and 
information transmission in simple models of branching 
processes on complex networks has shown that critical 
regimes offer certain advantages, when considering 
the computational performance of the network \citep{plenz2012neuronal}. 
\citet{beggs2003neuronal} showed that the information 
transmission between input and output layers 
of a network is maximal in the critical branching regime, whereas 
\citet{haldeman2005critical} found that the critical 
state is optimal for information retention. Also, \citet{kinouchi2006optimal} 
demonstrated that the critical regime is related to a maximal 
sensitivity of a neural network to the variations
in the input activity. This interesting characteristic of the 
critical regime may be explained by fact that the dynamical regime 
at the border of a second-order phase transition shares in part the 
properties of the two phases. The activity in the frozen
state would be, in this view, related to nonlinear computations
with the activity in the chaotic state favorable for fast 
information transmission and parallelization of 
computational processes \citep{rossello2012neural}.       
Thus, it is plausible that cortical areas organize into distinct
dynamical states, depending on the required functionality; 
the critical regime might be an attracting dynamical state for 
computations needing the features of both states, that is, a 
large flexibility in information processing. 
Still, it is important to extend this simple models in a 
way which captures neural variability, adaptability and evolutionary 
design in order to reevaluate the hypothesis discussed 
above in biologically realistic setups.

%---------------------------------------
\subsection{Beyond power laws - dragon kings}
%---------------------------------------

We will conclude this section with a short discussion 
of the emerging topic of "life beyond power laws",
which deals with an intriguing perspective regarding 
the possible origins of large catastrophic events. 
\citet{sornette2009dragon} and \citet{sornette2012dragon} 
pointed out that there is growing evidence indicating 
that extremely large events often transcend the heavy-tailed 
scaling regularly observed by the bulk of the data sets. 
These outliers were named "dragon kings", in order to stress 
their unique and diverse generating mechanisms, and their 
extreme size, which is typically off the charts. 

The generating mechanisms of dragon kings are believed to differ 
from the ones generating the smaller events, such as the various 
mechanisms discussed in this review. Furthermore, they are diverse 
and system dependent, having however several common properties.
For a dragon king to emerge an additional amplification mechanism 
is required, a mechanism which may not be present at all times
in the system. The system then undergoes a temporary phase transition, 
or bifurcation, leading to a qualitative new state and possibly 
to large-scale events. These kinds of transitions may be caused 
by a sudden increase in coupling strength of interacting components, 
leading to increased positive feedback, and possibly to a synchronized 
regime, spanning across a large fraction of the  system 
\citep{sornette2012dragon}. Interestingly, certain precursors 
typically precede a dragon king event, thus predicting an incoming 
catastrophe may be possible in certain cases. 
\citet{johansen2000critical} used the existence of a 
log-periodic precursors as an indicator for an impending 
material failure, and \citet{sornette2001significance} 
applied similar methods for the prediction of bursts of 
financial bubbles, that is, market crashes. 

Dragon kings are rare events, although more frequent then
what would be expected when using only the distribution 
of smaller events as a reference. Unfortunately, these
features make them difficult to identify and to differentiate
between dragon kings and a regular large-scale events.
Tools and methods used for the identification of dragon kings
often depend on the particularities of the system in 
question \citep{sornette2009dragon, 
sornette2012dragon}. Only recently had \citet{pisarenko2012robust} 
proposed a robust statistical test able to identify anomalies 
in the tails of power-law or exponential distributions, even 
when only a few dozens of observations are available.

Evidences for the existence of dragon kings have been found 
in numerous phenomena, such as various extreme weather 
phenomena, material rupture events, the distributions 
of financial runs of losses, in the statistics of epileptic 
seizures in humans and animal models, and many others 
\citep{sornette2009dragon,sornette2012dragon}. Still, in 
several cases the evidence of the dragon kings existence 
is inconclusive \citep{sornette2012dragon}. For example,  
it is still debated whether genuine dragon kings exist in 
the distribution of earthquake magnitudes.

\section{Conclusions}

The concept of self-organized criticality ({\it SOC}) 
is an intensely studied and discussed mechanism
for generating power-law distributed quantities. 
This theory has been proposed as an explanation 
for power-law scaling observed in various real-world 
phenomena. We have focused here on several well-studied 
phenomena, notably earthquakes, solar flares, 
and neuronal avalanches; just a three out of a plethora of 
phenomena exhibiting fat tails. Given the amount of existing 
empirical data, it is important to understand to which 
extent the theory of {\it SOC} contributes to an understanding 
of the underlying causes of the observed power-law behavior 
in real-world complex dynamical systems.

The current experimental evidence is still inconclusive 
with respect to a possible causal relation of the emergent 
power laws to an underlying self-organized critical state. 
In any case, extensions of the original sandpile 
model, such as dissipative models like the 
{\it OFC} earthquake model, are essential for replicating the 
fat-tailed avalanche statistics which are temporally and 
spatially correlated, a key property of many real-world data 
sets. Furthermore, a satisfactory description for real-world 
systems would also need to account for the observed inter-event 
correlations, which by themselves are key to improved predictions 
of catastrophic events.   
  
An alternative for an underlying self-organized critical 
state is the concept of highly organized tolerance 
({\it HOT}), which does not require a critical dynamical 
state for generating distributions with heavy tails. The 
theory of {\it HOT} proposes an explanation for the 
emergence of scale invariance in artificial and natural 
systems as a consequence of system design, where the design 
aims to achieve an optimized and robust performance in uncertain environments. 
For the case of living organisms this robust design may 
plausibly emerge through natural selection, and also result 
as such from a self-organizing process, albeit on longer 
time scales.

In this context, an interesting and hitherto open research question 
regards the relation between self-organization
and criticality in general. Essentially all proposed models
for generating scale-invariant observables are based on
self-organizing processes, some of which lead to critical
states, while others do not. For example, any dynamical system,
which retains its average activity homeostatically within 
certain bounds, as it is done in various cortical areas, is 
statistically equivalent to a self-organized critical branching 
process, and hence scale invariant. Balancing different types 
of drives, such as external driving and internal dissipation, may 
lead, on the other hand, to a self-organized, non-critical and
heavy-tailed state, a route proposed by the coherent noise model.

A further complication concerning this discussion is added by the
circumstance that critical dynamical systems may not actually be 
intrinsically scale invariant, which is in contrast to
thermodynamic critical systems. We discussed property in 
the context of vertex routing models. Another important aspect
regards the process of probing a complex dynamical system,
which is normally done by a stochastic sampling of phase
space and then following the dynamical flow. The measurement
process may actually have a qualitative effect on the 
resulting scaling properties of observables, an effect 
which has been worked out in detail for the
case of vertex routing models. Both effects can be traced back
to a highly non-trivial statistics of the attractors which 
might emerge in a critical dynamical system.

On the experimental side, power-law regimes are
routinely observed in both physical and biological 
systems. Considering the functional aspect, critical 
dynamical states have been argued to be advantageous 
for non-linear sensory processing and self-sustained 
neural computation \citep{moretti2013griffiths}, which 
are crucial characteristic biological neural networks. 
Living organisms are the product of self-organizing 
processes and it is therefore likely -- 
considering the functional advantages of 
critical regimes -- that the observed heavy-tailed
distributions will result from self-organizing 
principles. The {\it SOC} mechanism would imply 
that an underlying critical state, if realized, 
would be based on a very specific
generating mechanism namely the separation of time
scales between a fast internal dissipation (which may 
occur either at the boundary, for conserved sandpile 
models, or locally, for dissipative {\it SOC} models) and a
slow external driving, as exemplified by absorbing
state transitions. It may, however, also be the case
that the underlying state is non-critical and is either 
the product of various regulatory mechanisms (like 
homeostatic plasticity), as proposed within the 
{\it HOT} theory, or the result of balancing 
external driving and internal dissipation occurring
on similar time scales, as within the coherent 
noise model.

An important aspect regards the modeling of experimental
data. Estimating the dynamical state of an avalanche-like 
phenomenon, such as neuronal avalanches, by mapping it to a 
branching process, to obtain an estimate of the respective 
branching parameter, comes with several difficulties. 
The value of the estimated branching parameter will depend 
on the assumed charachteristics of the environment, {\it e.g.}
is the environment fixed or changing over time. 
Thus, the modeling assumptions will influence the
conclusion regarding the character of the avalanche 
dynamics \citep{taylor2013identification, hartley2013identification}. 
In addition, it is still unknown to which extent
history dependent branching, that is, the memory of the
system, influences the scaling behavior of avalanche 
sizes and durations. These difficulties may lead to wrongly 
identifying critical systems as non-critical, and vice versa.

Finally, in spite of the evidence that quite different 
physical systems exhibit dynamical properties akin to 
the one observed in various sandpile models, there is 
no convincing proof that the generative mechanism for 
power-law scaling, as proposed by {\it SOC}, constitutes
the true causal explanation. A substantial controversy 
regarding the interpretation of empirical data still persists, 
and the resolve of this controversy will, together with 
novel approaches for experimental setups and data analysis, 
require measurements with higher resolution. 

On a final note, what one actually considers a
self-organized process is to a certain extent a question 
of semantics. It is possible, in many circumstances, to 
tune a system towards a critical point. There is general 
agreement that the underlying process can be considered 
self-organized whenever this tuning process occurs through internal 
drives on time scales shorter than (or comparable to) the 
experimental time scale. The tuning of internal 
parameters may however also result from processes
acting on much longer time scales, like, for example, 
Kauffman's notion of ``life at the edge of criticality'',
as a consequence of Darwinian selection.
In both cases the dynamical state will never be,
for real-world systems, exactly at the critical point,
but fluctuating around it, albeit on very long time 
scales.

\section*{Acknowledgement}

Authors thank, in no particular order: Alain Destexhe, 
Nima Dehghani, Didier Sornette, Viola Priesemann, 
Juan Antonio Bonachela Fajardo and Dietmar Plenz, for 
helpful discussions, comments, and suggestions. 

\bibliographystyle{plainnat}
\bibliography{powerLaws}

\end{document}